\documentclass[remotesensing,article,accept,moreauthors,pdftex,10pt,a4paper]{mdpi} 
\newcommand{\blue}{\textcolor{black}}

\newcommand{\sigmab} {{\boldsymbol{\sigma}}}
\newcommand{\nub} {{\boldsymbol{\nub}}}

\newcommand{\varthetab} {{\boldsymbol{\vartheta}}}

\newcommand{\phib} {\boldsymbol{\phi}}

\newcommand{\etab} {\boldsymbol{\eta}}

\newcommand{\Cmat} {\mathbf{C}}

\newcommand{\Kmat} {\textbf{K}}

\newcommand{\Imat} {\textbf{I}}

\newcommand{\cvec} {\textbf{c}}

\newcommand{\avec} {\textbf{a}}

\newcommand{\svec} {\textbf{s}}

\newcommand{\uvec} {\textbf{u}}

\newcommand{\rvec} {\textbf{r}}

\newcommand{\RMSPE} {\textit{RMSPE}}
\newcommand{\MAPE} {\textit{MAPE}}
\newcommand{\MPE} {\textit{MPE}}

\renewcommand{\P}{\mathbf{P}}

\renewcommand{\S}{\mathbf{S}}

\newcommand{\x}{\mathbf{x}}
\newcommand{\Yvec}{\mathbf{Y}}

\newcommand{\Zvec}{\mathbf{Z}}

\newcommand{\cov}{\mathrm{cov}}

\newcommand{\Gau}{\mathrm{Gau}}

\newcommand{\E}{\textrm{E}}

\newcommand{\var}{\textrm{var}}

\newcommand{\corr}{\textrm{corr}}

\firstpage{1} 
\articlenumber{155}
\doinum{10.3390/rs10010155}
\pubvolume{10}
\pubyear{2018}
\copyrightyear{2018}
\history{Received: 9 November 2017; Accepted: 17 January 2018; Published: 22 January 2018}

\Title{On Statistical Approaches to Generate Level 3 Products from Satellite Remote Sensing Retrievals}

\Author{Andrew Zammit-Mangion *\orcidA{}, Noel Cressie\orcidA{} and Clint Shumack}

\AuthorNames{Andrew Zammit-Mangion, Noel Cressie and Clint Shumack}

\address[1]{%
National Institute for Applied Statistics Research Australia, University of Wollongong, Wollongong, NSW 2522, Australia; \blue{azm@uow.edu.au (A.Z.-M.);} ncressie@uow.edu.au (N.C.); cshumack@uow.edu.au (C.S.)}

\corres{Correspondence: azm@uow.edu.au; Tel.: +61-02-4221-5112}

\abstract{Satellite remote sensing of trace gases such as carbon dioxide (CO$_2$) has increased our ability to observe and understand Earth's climate. However, these remote sensing data, specifically~Level 2 retrievals, tend to be irregular in space and time, and hence, spatio-temporal prediction is required to infer values at any location and time point. Such inferences are not only required to answer important questions about our climate, but they are also needed for validating the satellite instrument, since~Level 2 retrievals are generally not co-located with ground-based remote sensing instruments. Here, we discuss statistical approaches to construct Level 3 products from Level 2 retrievals, placing~particular emphasis on the strengths and potential pitfalls when using statistical prediction in this context. Following this discussion, we use a spatio-temporal statistical modelling framework known as fixed rank kriging (FRK) to obtain global predictions and prediction standard errors of column-averaged carbon dioxide based on Version 7r and Version 8r retrievals from the Orbiting Carbon Observatory-2 (OCO-2) satellite. The FRK predictions allow us to validate statistically the Level 2 retrievals globally even though the data are at locations and at time points that do not coincide with validation data. Importantly, the validation takes into account the prediction uncertainty, which~is dependent both on the temporally-varying density of observations around the ground-based measurement sites and on the spatio-temporal high-frequency components of the trace gas field that are not explicitly modelled. Here, for validation of remotely-sensed CO$_2$ data, we use observations from the Total Carbon Column Observing Network. We demonstrate that the resulting FRK product based on Version 8r compares better with TCCON data than that based on Version 7r, in~terms of both prediction accuracy and uncertainty quantification.}


\begin{document}


\section{Introduction}\label{sec:intro}

Level 2 retrievals from satellite remote sensing instruments are typically retrieved irregularly in space and time. Hence, in order to validate these retrievals against either ground-based remote-sensing data (e.g., \cite{Chevallier_2017}) or atmospheric transport model output (e.g., \citep{Tiwari_2006,Hammerling_2012}), some form of {gap-filling}, or~{spatial prediction}, is required. A wide variety of approaches has been proposed that are either deterministic (e.g., geographic co-location methodologies; see \citep{Inoue_2013, Butz_2011}) or statistical (e.g., kriging; see \citep{Katzfuss_2011,Zeng_2017}). Statistical~techniques tend to be more computationally intensive, on the one hand, but on the other hand, they allow for uncertainty quantification. This is indispensable when validating satellite remote sensing products to ground-based measurements or to other transport-model outputs, as it puts into context the magnitudes of any observed discrepancies. They have also been shown to be more accurate in practice, in a mean-squared-error sense, than their deterministic counterparts \citep{Nguyen_2014b}. 

A large variety of statistical techniques has been proposed to produce Level 3 maps from satellite remote sensing Level 2 retrievals. These include spatial kriging (e.g., \citep{Hammerling_2012, Jing_2014}), spatial block kriging \citep{Tadic_2015} and spatio-temporal kriging (e.g., \cite{Zeng_2017}). These studies \cite{Hammerling_2012, Jing_2014,Tadic_2015,Zeng_2017} generate predictions and prediction standard errors based on a subset of retrievals and, hence, are all variants of the {local} kriging procedure proposed by Haas \citep{Haas_1995}. Local methods are advantageous for two key reasons: (i) they are much faster than global kriging methods that require consideration of all data points simultaneously; and (ii) they allow for a straightforward ad hoc consideration of non-stationarity, where the spatial properties of the field being studied vary in both space and time. However, as we shall show, local methods need to be used with care, specifically only in situations when the signal-to-noise ratio (SNR) is sufficiently high, so that basing a prediction and a prediction standard error on a hand-picked neighbourhood of retrievals is justified. That care is required even when the local neighbourhoods are large. A theoretical problem with local kriging is that it is based on many local geostatistical models that are not implied by a single spatial stochastic process. Thus, covariances between the process at two locations can be different for different local models, resulting in an incoherent probability structure. In particular, the~modelled covariances of the local spatial processes may together yield invalid covariance matrices.

Another class of statistical techniques that are designed to work with large datasets is based on dimensionality reduction. Of these, one of the most popular is fixed rank kriging (FRK), first~proposed by Cressie and Johanesson \cite{Cressie_2008} and later applied to generate remote sensing products in a variety of contexts, both in space (e.g., \citep{Nguyen_2012}) and in space-time (e.g., \citep{Katzfuss_2011, Nguyen_2014a}). FRK is based on a coherent probability structure, and the advantage of dimensionality reduction is the reduced complexity in computing the predictions and prediction standard errors. The reduced-rank model used in FRK is designed to produce smooth (in the sense of second-order derivatives of small magnitude) predictors that may appear to be very different from, say, transport-model outputs. However, we shall show that this smoothness is not necessarily detrimental if the goal is optimality in the mean-squared-error sense.

Several authors have produced Level 3 products by combining information from the retrievals with transport-model output. In some cases, parameters appearing in the kriging weights (or~covariance function) are estimated from transport-model output (e.g., \citep{Alkhaled_2008,Chevallier_2017}). In other cases, the Level 3 maps are a by-product of an inversion scheme, for example in a 4DVAR framework (e.g.,~\citep{Engelen_2009}). In this article, we are mostly concerned with the ``vanilla case'' in which one generates a Level 3 product directly from the retrievals, without consideration of other sources of information. In our discussion of \blue{Level-3} product generation, we also go one step further and assume that the covariance function of the process is known. This simplification is made so that we can focus on the issue of prediction and how various model assumptions and local methods affect the quality of the inferences we can make. In practice, we work with an estimated covariance function, as is standard practice in geostatistics \cite{Cressie_1993} (Chapter~3).

This article is divided into two parts. 
The first part, Section \ref{sec:pedag}, is largely tutorial in nature. It~considers several methods for prediction adopted in the literature, and it gives insight into their merits and drawbacks. It does this through a simulation experiment that is used as a running example. Specifically, Section \ref{sec:pedag} explores how kriging \blue{for the process, and for the observations}, leads to different interpretations of the Level 3 products; it discusses the smoothness of the reduced-rank predictor \blue{FRK}; and it explores where and when \blue{reduced-rank predictors} should be used as opposed to local kriging methods. 
The second part, Section \ref{sec:OCO-2}, illustrates the use of FRK in generating Level 3 products from Orbiting Carbon Observatory-2 (OCO-2) Level 2 data in the Version 7r and Version~8r Lite Files \cite{Eldering_2017,OCO2v7r,OCO2v8r}. These products are then validated against data from the Total Carbon Column Observing Network (TCCON, \cite{Wunch_2011_TCCON,Wunch_2017_TCCON}). The paper concludes with Section \ref{sec:discussion} giving a summary and a brief discussion of other recently-developed spatio-temporal prediction methods that are ideally suited for large heterogeneous remote sensing datasets such as those produced by OCO-2.

\section{Spatio-Temporal Prediction from Retrievals}\label{sec:pedag}

From our sampling of the recent literature, kriging remains the most widely-used `statistical interpolation' method for producing Level 3 products containing maps of both predictions and prediction standard errors from Level 2 satellite remote sensing retrievals. However, there are many variants of kriging that can be, and have been, used to derive these products. In this section, we discuss a number of these variants, and we explore the circumstances under which some approaches are more appropriate than others. For ease of exposition, throughout this section, we focus on the prediction aspect and assume that mean and covariance parameters are known.

\subsection{Observation Space vs. Process Space}\label{sec:spaces}

An issue that is not often discussed, from both a theoretical and practical perspective, is whether the Level 3 product is a prediction of the retrievals (i.e., in {observation space}) or of the underlying process (i.e., in {process space}). While the former is a prediction of what a retrieval would be if it were to be done at some point in space and time, the latter is a prediction of the actual process (e.g., the true column-averaged CO$_2$) at that point in space and time. Currently, little distinction appears to be made between the two variants, yet the two types of predictions are intrinsically different. Which is used affects the conclusions one should draw when validating satellite remote sensing instruments.

 Consider a field of interest, say column-averaged CO$_2$, that is indexed by both space and time. Assume the field is random, and denote it by $Y$. Then, $Y(\svec;t)$ is a random variable representing the field evaluated at some location on the sphere $\svec$ at time $t$. Consider now a remote sensing retrieval of this field at location $\svec_i$ and $t_i$, which we denote as $Z(\svec_i;t_i)$, for $i = 1,\dots,m$. A typical model relating these retrievals to the field is the linear additive model: 
$$
Z(\svec_i;t_i) = Y(\svec_i;t_i) + \epsilon_i,
$$
for $i = 1,\dots,m$, where $\{\epsilon_i\}$ is a set of independent measurement errors with variance $\sigma^2_{\epsilon}$, typically~assumed to be Gaussian with a mean that is usually set to zero, but does not need to be zero. 

Consider the vector of retrievals $\Zvec \equiv (Z(\svec_1;t_1),\dots,Z(\svec_m;t_m))'.$ Kriging, or optimal linear spatio-temporal prediction, in {observation space} refers to predicting $Z(s_j^*;t_j^*), j = 1,\dots,N,$ from the retrievals $\Zvec$, where $(s_j^*; t_j^*)$ is a prediction space-time location that for simplicity we assume does not coincide with an observation location. Clearly, this can only be done if the measurement-error variance $\sigma^2_{\epsilon}$ at the prediction location is known, since the kriging equations explicitly involve $\var(Z(\svec_j^*;t_j^*)) = \var(Y(\svec_j^*;t_j^*)) + \sigma^2_{\epsilon}$. Kriging in {process space}, also known as `filtered kriging,' refers~to predicting $Y(s_j^*;t_j^*), j = 1,\dots,N,$ from the observations $\Zvec$ \cite{Cressie_2011} (Section~4.1.2). In both cases, {optimal}~should be taken in the sense that the predicted quantities minimise their respective mean squared prediction errors under the model assumptions and predictor constraints. Other predictors that are optimal in a different sense are sometimes more appropriate (e.g., \cite{Zhang_2008}). For the purposes of Level 3 products, minimising the mean-squared prediction error results in kriging predictors and is generally considered suitable for mapping.  

Recall that here, all parameters in the statistical model for $Y$ are assumed known and fixed. Then, it is straightforward to show that, at unobserved locations, the conditional expectations, $\E(Y(s_j^*;t_j^*) \mid \Zvec)$ and $\E(Z(s_j^*;t_j^*) \mid \Zvec)$, are the optimal (unbiased) linear predictors in process space and observation space, respectively, and that they are identical.
 However, the prediction variances are different due to the extra variability present in observation space: For unobserved locations,
$$\textrm{var}(Z(s_j^*;t_j^*) \mid \Zvec) = \textrm{var}(Y(s_j^*;t_j^*) \mid \Zvec) + \sigma^2_{\epsilon}, \quad j = 1,\dots,N.$$

Consequently, the questions that Level 3 products in observation space answer may be very different from those that Level 3 products in process space answer. Specifically, the former provide uncertainty measures on retrievals, while the latter provide uncertainty measures on the underlying process itself. This distinction is particularly important in validation studies of satellite remote sensing~retrievals.

{Example 1:} The implications of carrying out comparisons in an inappropriate space can be seen from a simple simulation experiment that we shall use throughout this section. Let $Y$ be a spatial process with mean zero and covariance function $C_Y(\svec_1,\svec_2)\equiv \sigma^2_YR_Y(\svec_1,\svec_2)$, where $R_Y(\svec_1,\svec_2) \equiv e^{-\|\svec_2 - \svec_1\| / \tau}$ is a correlation function (in~this case, a stationary exponential correlation function), $\tau$ is the e-folding length scale at which the correlation drops to $e^{-1}$ and $\sigma^2_Y$ is the process variance.

Because we are working from a simulation, we know the covariance function, as well as the mean. Consider the spatial domain $D \equiv [0,1]\times[0,1]$, and let $\tau = 0.15$ and $\sigma^2_Y=1$. Assume further that we have 1000 measurements of the process randomly placed in $D$ and that the measurement-error variance $\sigma^2_\epsilon = 1$ is constant and known. These measurements play the role of `retrievals.' For prediction locations, we consider a 100 $\times$ 100 grid on $D$. Predictions and prediction errors on this grid play the role of the Level 3 product. We also consider 200 random diagnostic locations on $D$, which serve as validation in both the process space and the observation space. The true field is simulated from $Y$ under Gaussian assumptions, and it plays the role of the scientific quantity of interest, such as the output from a forward transport model.

\begin{figure}[H]
	\centering
	\includegraphics[width=2.4in]{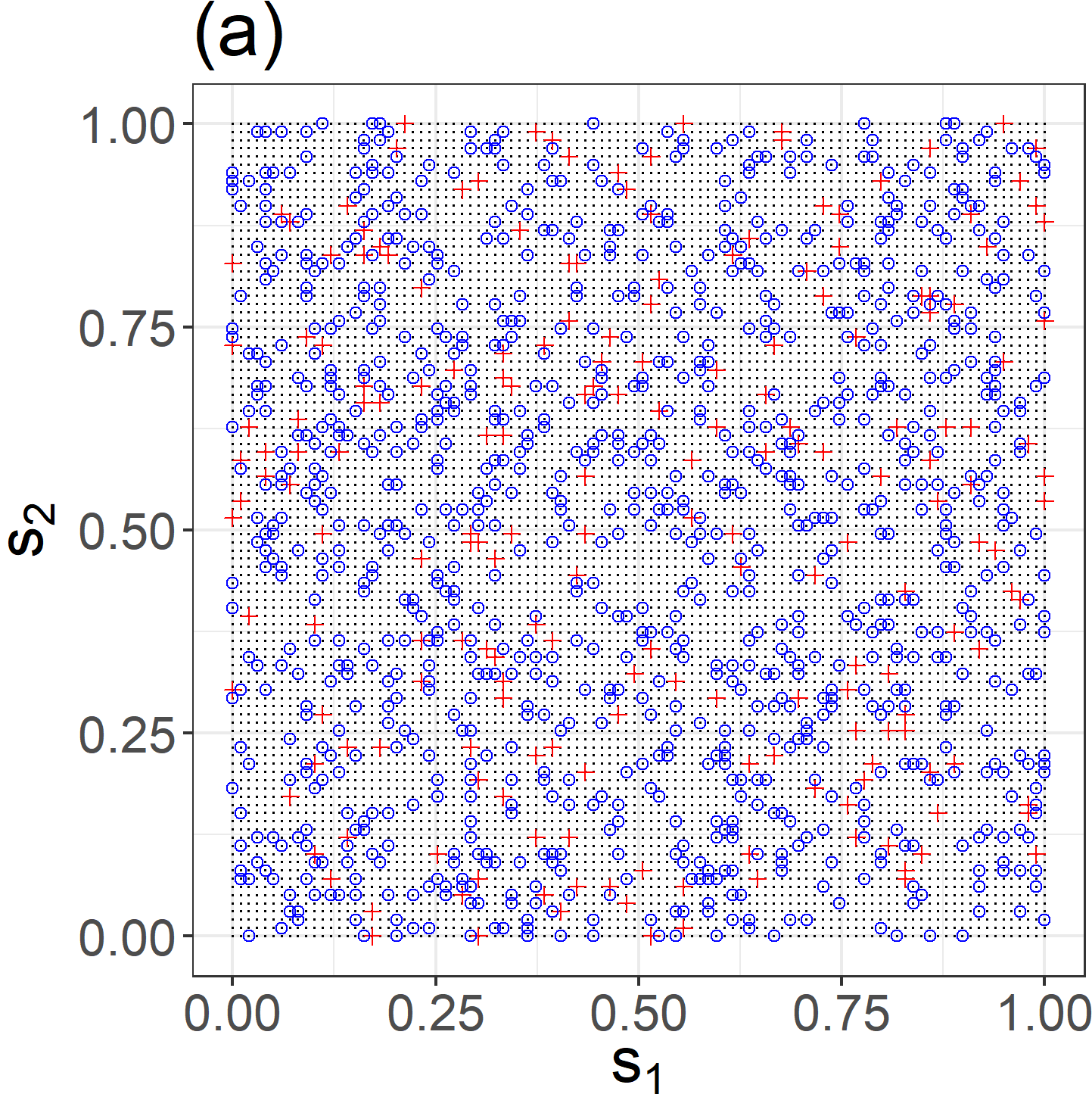}
   \includegraphics[width=2.8in]{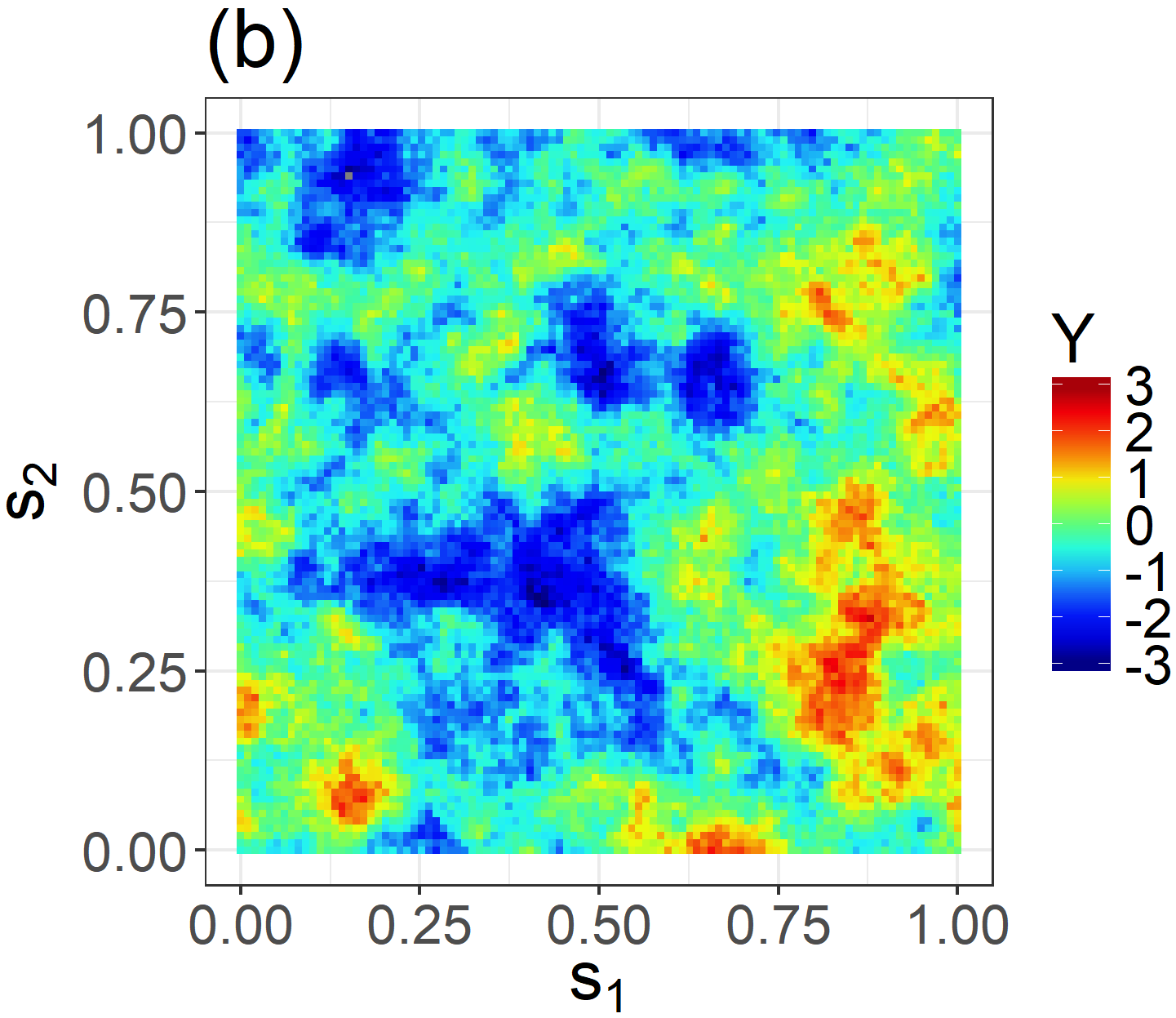}
   \includegraphics[width=2.8in]{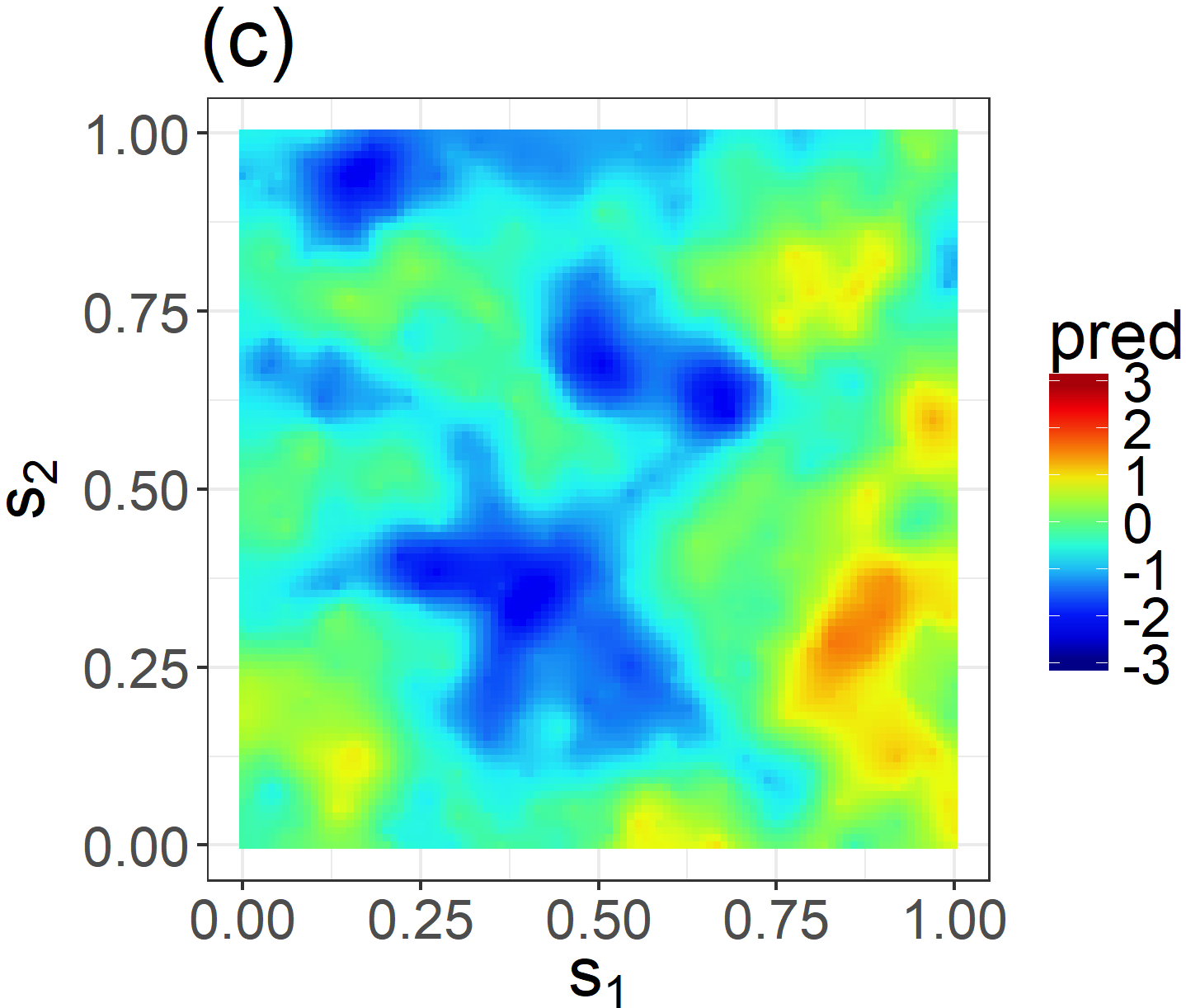}
   \includegraphics[width=2.8in]{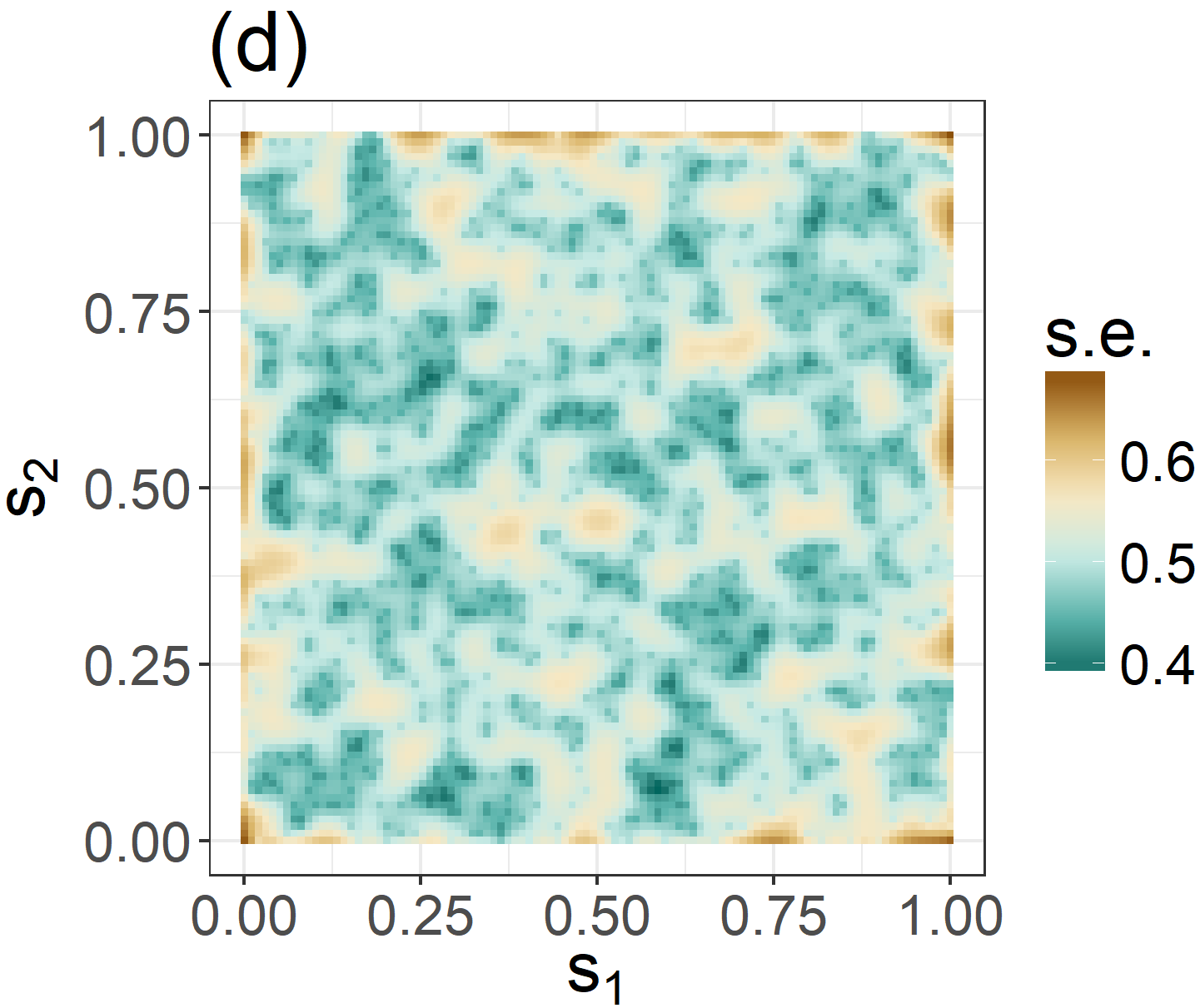}
	\caption{Simple kriging in a simulation experiment (see the text in Section \ref{sec:spaces} for details). (\textbf{a})~Experimental setup: the blue circles are the retrieval locations; the red crosses are the validation (unobserved) locations; and the black dots define the prediction grid at which the Level 3 maps are constructed. (\textbf{b}) The `true' unobserved process of interest $Y$. (\textbf{c}) The simple kriging predictor \blue{(pred)} in process space on the prediction grid from the simulated retrievals. (\textbf{d}) The prediction standard errors \blue{(s.e.)} obtained as part of simple kriging on the prediction grid. \label{fig:exp1}}  
\end{figure}

Figure \ref{fig:exp1}a--d shows the experimental setup, the simulated (true) field, the optimal prediction in process space (which at unobserved locations is the same as the optimal prediction in observation space) and the prediction standard error in process space, respectively. The prediction standard error is seen to be high in regions of sparse measurements, and the predictor satisfactorily reproduces the unobserved true signal, as would be expected from an optimal spatial predictor.

Level 3 products based on statistical techniques always come in pairs, with one map showing a prediction and a companion map showing the prediction standard error. The first map on its own can only be used to derive discrepancy diagnostics for the prediction, such as the mean prediction error (MPE), the mean absolute prediction error (MAPE) or the root-mean-squared prediction error (RMSPE), where here, ``mean'' refers to the ``sample mean'' or ``average'' given in the definitions below. The MPE is designed to give an an estimate of the predictor `bias', but it does not reveal anything about the variance of the predictor. Although the MAPE increases with both the predictor bias and variance, it does so in a highly nonlinear fashion; it is commonly used because of its intuitive interpretation. On~the other hand, the MSPE is an estimate of the sum of the squared bias and the predictor variance, and it is commonly viewed (or its square root, as we do here) as a \blue{``gold standard''} when comparing~predictors.

 Denote the process values at $N_d$ diagnostic locations as $\Yvec_d$ and the optimal predictor at those locations as $\hat\Yvec_d$. {MPE}, {MAPE} and {RMSPE} are then given by:
\begin{align*}
 \MPE &= \frac{1}{N_d}\sum_{k=1}^{N_d}(Y_{d,k} - \hat{Y}_{d,k}), \\
 \MAPE &= \frac{1}{N_d}\sum_{k=1}^{N_d}|Y_{d,k} - \hat{Y}_{d,k}|, \\
 \RMSPE &= \sqrt{\frac{1}{N_d}\sum_{k=1}^{N_d}(Y_{d,k} - \hat{Y}_{d,k})^2}, 
\end{align*}
respectively. The process values, $\Yvec_d$, to which the predictor is compared, are typically obtained from data that are ignored when constructing the predictor or data from another instrument. In this example, we know the true (simulated) process values and thus set $\Yvec_d$ to the true values. For example, the~{RMSPE} over the $N_d = 200$ diagnostic locations was 0.4995.

MPE, MAPE and RMSPE are valuable, but diagnostics that treat uncertainty quantification require the second map, which shows the prediction standard errors. One such diagnostic that is very important is the {coverage}, which can help the analyst assess whether the prediction standard errors are correct. If uncertainty is quantified correctly, $x$\% of (independent) validation data should fall within the $x$\% prediction interval. Usually a large prediction interval (e.g., $x = 90$ or $95$) is chosen since this is likely to be more useful to the analyst than a smaller prediction interval. However, any prediction interval, or set of intervals, may be chosen for diagnostic purposes. Clearly, using the correct prediction standard error when constructing the interval is required to determine the correct coverage. In Section~\ref{sec:coverage}, we further develop this idea for the situation where measurements are biased.

For nominal 90\% coverage, Table \ref{tab:coverage} shows the empirical coverage when validating against (i) the true field (which in this case plays the role of the output from a perfect CO$_2$ transport model in process space) and (ii) left-out retrievals in observation space.

\begin{table}[H]
	\caption{Coverage diagnostics when validating against the `true' (simulated) process values and the left-out retrievals using prediction intervals constructed in (i) process space and (ii) observation space. The nominal coverage is 90\%, obtained by assuming that the prediction standard errors are correct.\label{tab:coverage}}
	\centering
	\begin{tabular}{ccc}
\toprule
 \textbf{Validating against} & \textbf{\begin{tabular}[c]{@{}c@{}}Empirical Coverage in\\ Process Space (Nominal Is 90\%) \end{tabular}} & \textbf{\begin{tabular}[c]{@{}c@{}}Empirical Coverage in \\Observation Space (Nominal Is 90\%) \end{tabular}} \\
\midrule
 (Simulated) process & 0.89 & 1.00 \\
 Left-out retrievals & 0.51 & 0.93 \\
\bottomrule
 \end{tabular}
\end{table}

The table shows the misinterpretations that are possible: Using prediction standard errors in process space to construct the intervals when validating against other retrievals would give the impression that our predictions are too `optimistic' (from the table, only 51\% of the left-out retrievals fell into the constructed 90\% prediction interval). Using prediction standard errors in observation space to construct the intervals when validating against a process realisation (output from a transport model, say) would give us the impression that our inferences on the process are too `pessimistic' (from the table, all of the process values fell into the constructed 90\% prediction interval).

These results are illustrative of one particular diagnostic: indeed, there is a whole suite of probabilistic diagnostics that require consideration of prediction standard errors (see \cite{Gneiting_2014} for a detailed review). In all these cases, incorrect prediction intervals will lead to incorrect \blue{probabilistic} diagnostics and hence incorrect conclusions. Therefore, it is crucial to know whether the Level 3 products generated from statistical techniques that are used in an analysis are defined for process space or for observation~space. 

\subsection{Level 3 Maps Generated Using Statistical Techniques Will Appear Smooth}\label{sec:smooth}

A common criticism of kriging predictions is that they appear to be too smooth when compared to what is expected from the process. However, an attempt to produce a predictor that looks `realistic' will likely yield a suboptimal predictor (as we show below, it is the method of conditional simulation that should be used instead to produce `realistic' outputs). For example, the~covariance-matching-constrained kriging of \citep{Aldworth_2003} will have larger mean squared error than the usual kriging predictor. Note that the optimal prediction in Figure \ref{fig:exp1}c is smoother than the true, unobserved, process in Figure \ref{fig:exp1}b, and this can be established theoretically (see below). Kriging is not meant to yield the underlying process, but it is meant to give correct coverage with optimal prediction intervals, which makes the ``too smooth'' criticism a moot point. The purpose of this section is to show that the optimal predictor of $Y(\svec^*_j;t^*_j)$ in the mean-squared-error sense (which is the conditional expectation $E(Y(\svec^*_j;t^*_j)|\Zvec)$), is a data smoother and that a smooth predictor is not necessarily a poor predictor.

Consider a setup similar to that of Section \ref{sec:spaces}, and define the second-order difference quotient of the process as,
$$
D_Y(\svec_1^*,\svec_2^*,\svec_3^*) \equiv \frac{1}{h^2}(Y(\svec_1^*) - 2Y(\svec_2^*) + Y(\svec_3^*)),
$$
where $\svec_1^* + (h,0)' = \svec_2^* = \svec_3^* - (h,0)'$ is a finite-difference approximation of the field's curvature in the direction of the first coordinate at location $\svec_2^*$. 
Now, under Gaussian and mean-zero assumptions, it is straightforward to show that $D_Y(\svec_1^*,\svec_2^*,\svec_3^*)$ is normally distributed with mean zero and variance: 
$$\sigma^2_{D_Y}\equiv \frac{\sigma^2_Y}{h^4}(6 - 4R_Y(\svec_1^*,\svec_2^*) - 4R_Y(\svec_2^*,\svec_3^*) + 2R_Y(\svec_1^*,\svec_3^*)),$$
where $R_Y(\svec,\rvec) \equiv \corr(Y(\svec),Y(\rvec))$. Since $|D_Y(\svec_1^*,\svec_2^*,\svec_3^*)|$ is half-normal, we have that: 
$$
E(|D_Y(\svec_1^*,\svec_2^*,\svec_3^*)|) = \sqrt{\frac{2\sigma^2_{D_Y}}{\pi}},
$$
from standard properties of the half-normal distribution \cite{Johnson_1994} (\blue{Chapter 13,} Section~10.1). 

Now, consider the second-order difference quotient of the optimal predictor, $\E(Y(\svec) \mid \Zvec)$,
$$
D_{E(Y|Z)}(\svec_1^*,\svec_2^*,\svec_3^*) = \frac{1}{h^2}[E(Y(\svec_1^*)|\Zvec) - 2E(Y(\svec_2^*)|\Zvec) + E(Y(\svec_3^*)\mid \Zvec)].
$$

Then, $D_{E(Y|Z)}(\svec_1^*,\svec_2^*,\svec_3^*)$ has exactly mean zero for linear spatial trends and approximately mean zero for smooth spatial trends and small $h$. Importantly, it can be shown to have variance equal to $\sigma^2_{D_Y} - c$, where \blue{$0 \le c \le \sigma^2_{D_Y}$}; see Appendix \ref{app:A1}. Hence,
\vspace{12pt}
$$
E(|D_{E(Y|Z)}(\svec_1^*,\svec_2^*,\svec_3^*)|) =\sqrt{\frac{2(\sigma^2_{D_Y}-c)}{\pi}} \blue{~\le~} E(|D_Y(\svec_1^*,\svec_2^*,\svec_3^*)|).
$$

We can therefore expect, on average, that the optimal predictor is smoother (in the sense that it will have smaller absolute second-order derivatives) than the process. The degree of smoothness is a function of the SNR, an aspect that we explore further in Section \ref{sec:local}. 
If rougher maps are sought based on the same spatial covariance function, the analyst has no option but to relinquish the concept of mean-squared optimality (see \cite{Aldworth_2003,Zhang_2008} for further discussion).

This result highlights one important property of the Level 3 products, which in turn allows for their correct interpretation as summaries of random quantities. Specifically, the quantities $\{Y(\svec^*_j;t^*_j) : j = 1,\dots,N\}$ are dependent random variables and not deterministic scalars. Their conditional expectation with respect to the observed data $\Zvec$ (the predictor) illustrates just one aspect of these random variables. Since one cannot visually determine a Level 3 product's characteristics from the conditional expectation alone, it is often preferable to generate a few {conditional simulations} \cite{Cressie_1993} (Section~3.6.2) of the process conditional on $\Zvec$, to reveal whether the spatial structure is correctly captured or not in the product. 

\begin{figure}[H]
	\centering
  	\includegraphics[width=2in]{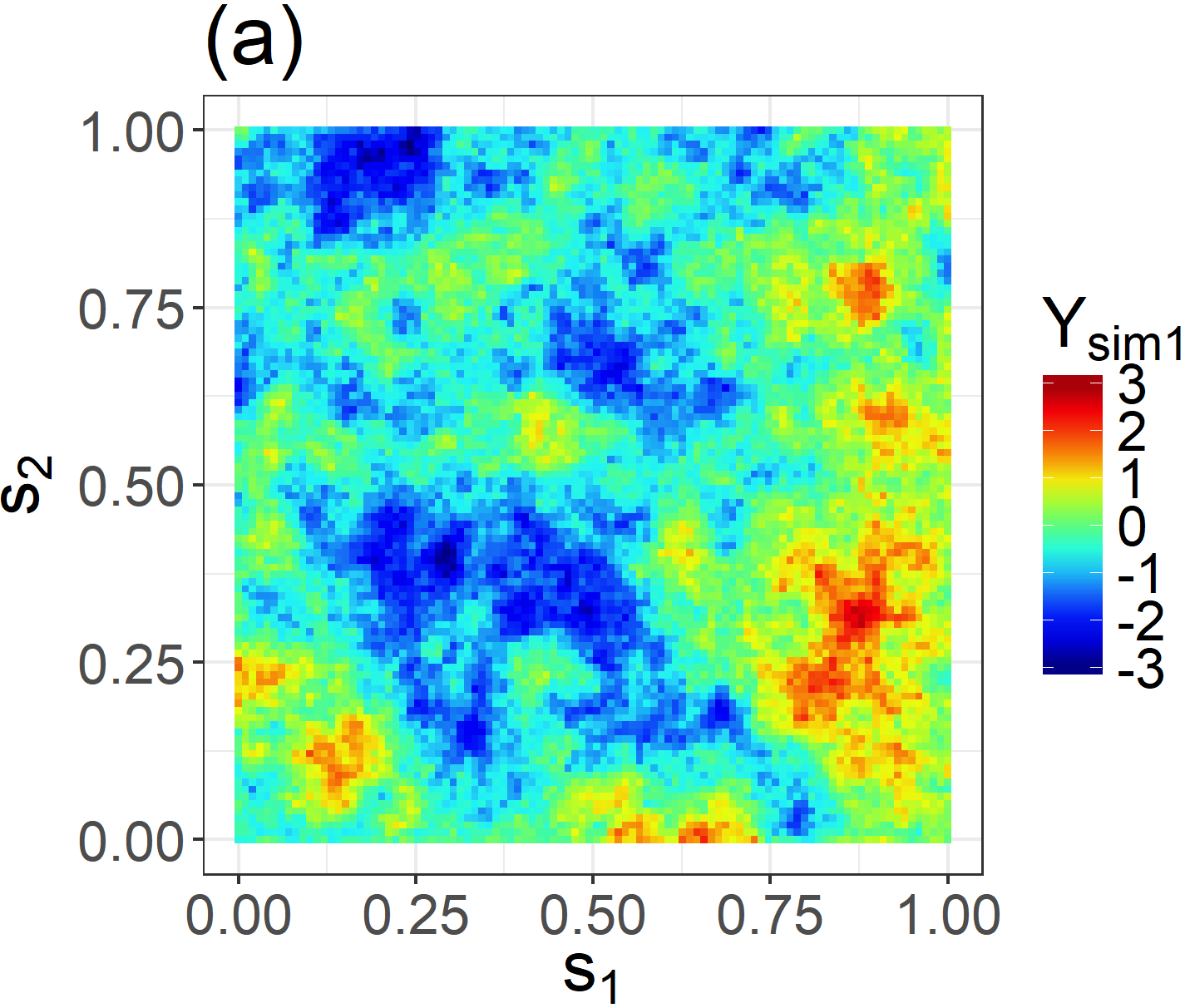}
   \includegraphics[width=2in]{./E1_field}
   \includegraphics[width=2in]{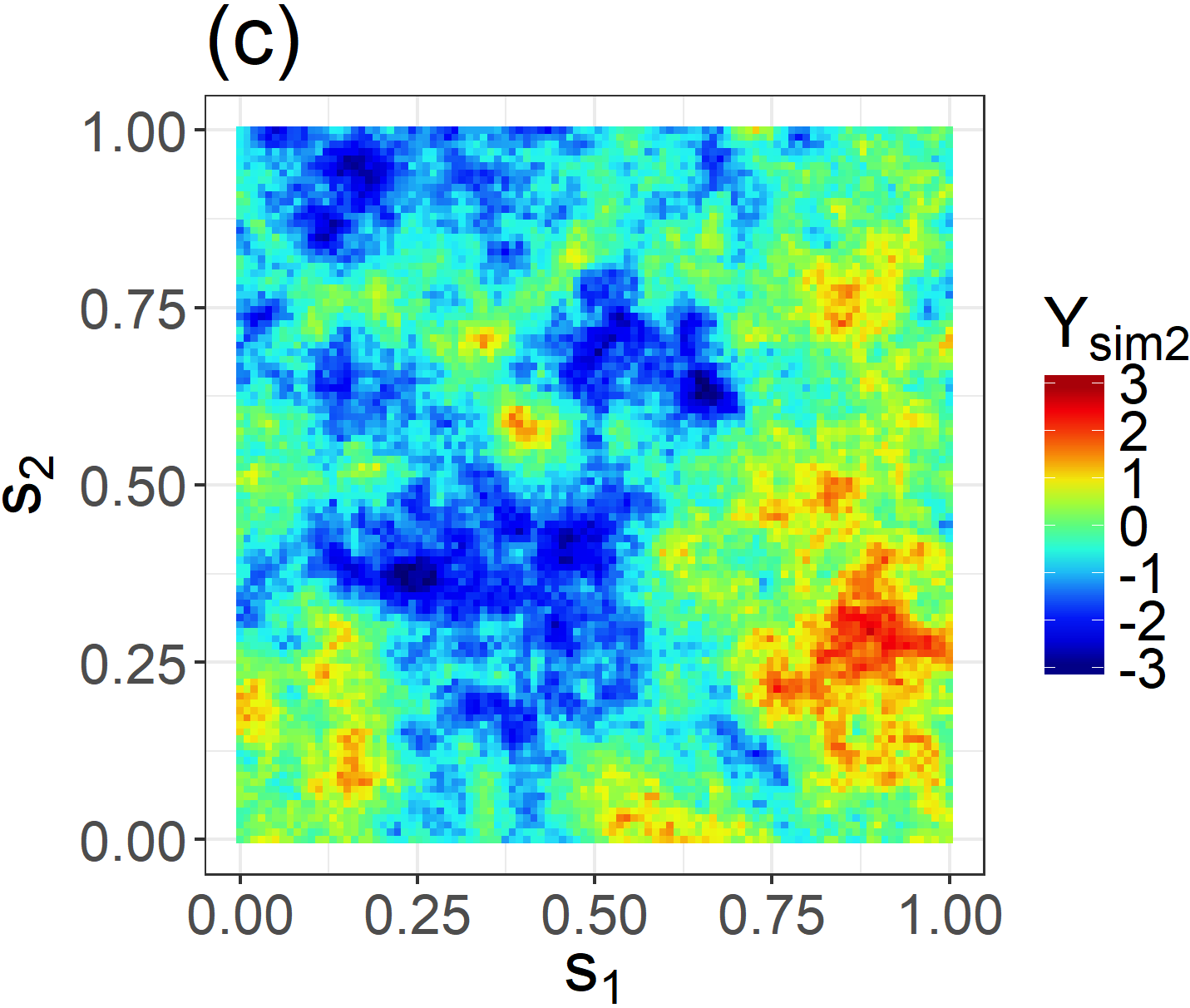}
	\caption{Conditional simulations from the Level 3 product (see the main text for details). (\textbf{a}) First conditional simulation; (\textbf{b}) the true process (identical to Figure \ref{fig:exp1}b); (\textbf{c}) second conditional simulation.\label{fig:exp1B}}
\end{figure}

{Example 2:} In Figure~\ref{fig:exp1B}a--c, we show two conditional simulations, together with the true process $Y$, from~the simulation experiment of Example 1, where the true mean and covariance-function parameters are known. The~conditional simulations reveal that the statistical Level 3 product encodes reasonable spatial structure, despite the optimal predictor being `smooth.' 

\vspace{-6pt}


\subsection{Fixed Rank Kriging}\label{sec:FRK}

The rather complex structures evident in Figure \ref{fig:exp1}b are `smoothed out' in the optimal prediction of Figure~\ref{fig:exp1}c, for reasons elucidated in Section \ref{sec:smooth}. This smoothing action implies that there is likely to be a simpler, lower-dimensional model that could be used to satisfactorily carry out pointwise inferences in both process space and measurement space. A suite of reduced-rank methods, an example of which is FRK \cite{Cressie_2008,Wikle_2010,Zammit_2017}, is designed to do just that, and these yield optimal predictors based on large datasets at a fraction of the computational cost of traditional kriging. FRK has often been used to produce Level~3 products from Level 2 retrievals (e.g., \cite{Cressie_2008,Nguyen_2012,Zammit_2017}).

FRK assumes that the process $Y$ can be decomposed into a weighted sum of $r$ pre-specified basis functions, where the weights are random. To simplify the specification of the random field $Y$, we~assume that $\E(Y(\cdot ; \cdot)) = 0$. Then:
$$
Y(\svec;t) = \sum_{i=1}^r\phi_i(\svec;t)\eta_i + \zeta(\svec;t);\quad \svec \in D, t > 0,
$$
\noindent which is a linear random effects model, where the random vector of coefficients $\etab \equiv (\eta_1,\dots,\eta_r)'$ have mean zero and covariance matrix $\Kmat$ and where the fine-scale variation term $\zeta(\svec;t)$ plays the role of the `nugget' (without a measurement-error component) in geostatistical models. From this basis-function representation, we immediately have:
$$C_Y(\svec,\rvec;t,u) \equiv \cov(Y(\svec;t),Y(\rvec;u)) = \phib(\svec;t)'\Kmat\phib(\rvec;u) + \var(\zeta(\svec;t))I(\rvec = \svec; u = t),$$
where $\phib(\svec;t) \equiv (\phi_1(\svec;t),\dots,\phi_r(\svec;t))'$ and $I(\cdot)$ is the indicator function.

The choice of basis functions $\{\phi_i\}$ is important. At the very least, they should be able to accurately reconstruct the optimal predictor under more general assumptions about $\cov(Y(\svec;t),Y(\rvec;u))$, and~ideally, they should also be able to provide a good representation of the prediction-standard-error map. 
Hence, it is common to set $r$ as large as computationally feasible, which usually comes at the cost of imposed model constraints through a structured $\Kmat$ (e.g., \cite{Lindgren_2011,Nychka_2015}). 

All excess variation in the process that cannot be explained by the spatio-temporal components $\{\phi_i(\cdot;\cdot)\eta_i\}$ is absorbed by the white-noise component $\zeta(\cdot;\cdot)$ that ensures that the total pointwise prediction standard error is accurately quantified. However, this component does not contribute to improved prediction itself, since it is spatially uncorrelated. The matrix $\Kmat$ and $\var(\zeta(\cdot;\cdot))$ are parameters of the spatio-temporal covariances. In general, they cannot be assumed known; in what follows, we shall use the method of maximum likelihood to estimate them from the data (see \cite{Zammit_2017}).

\begin{figure}[H]
	\centering
	\includegraphics[width=1.9in]{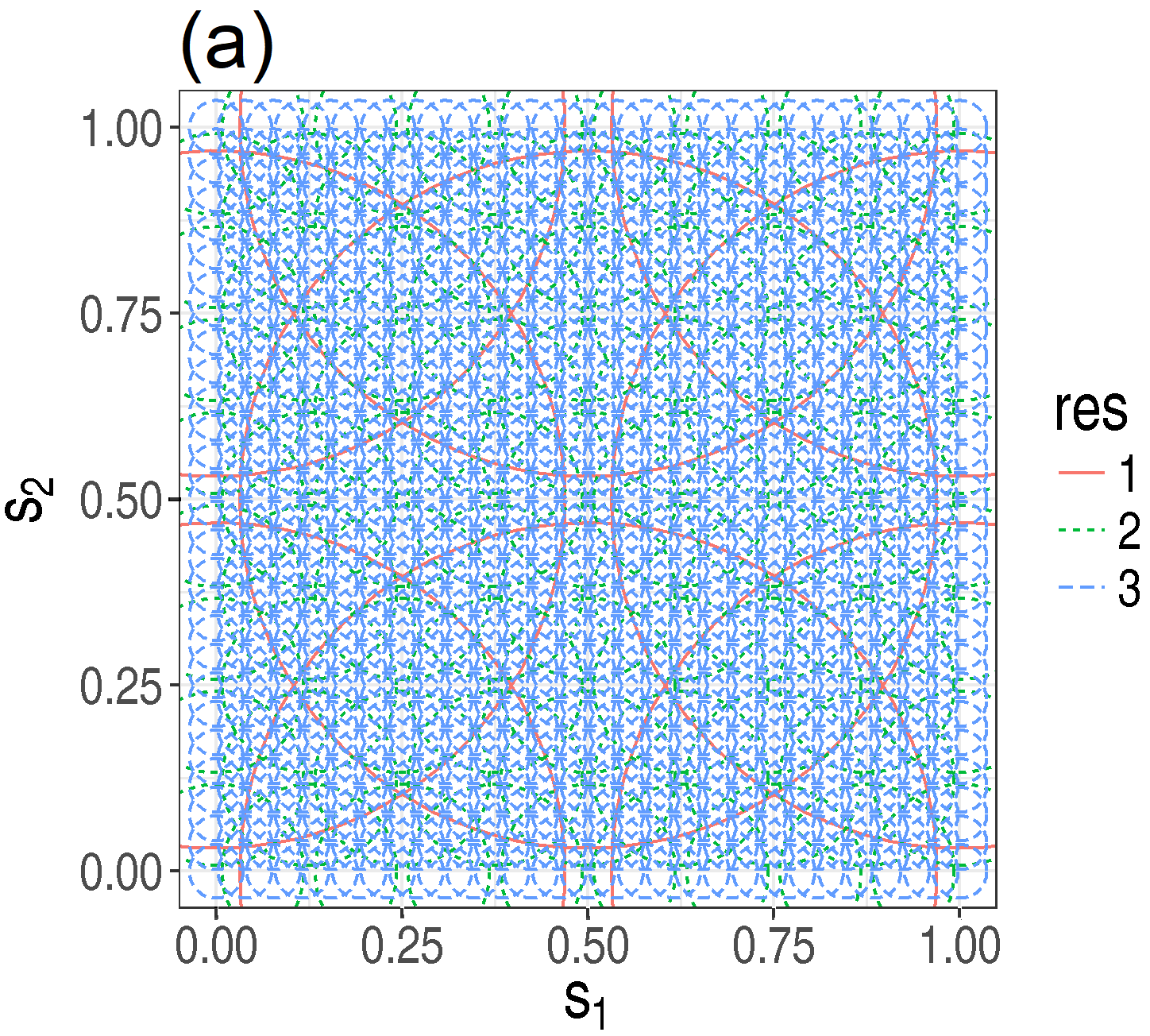}
	\includegraphics[width=2in]{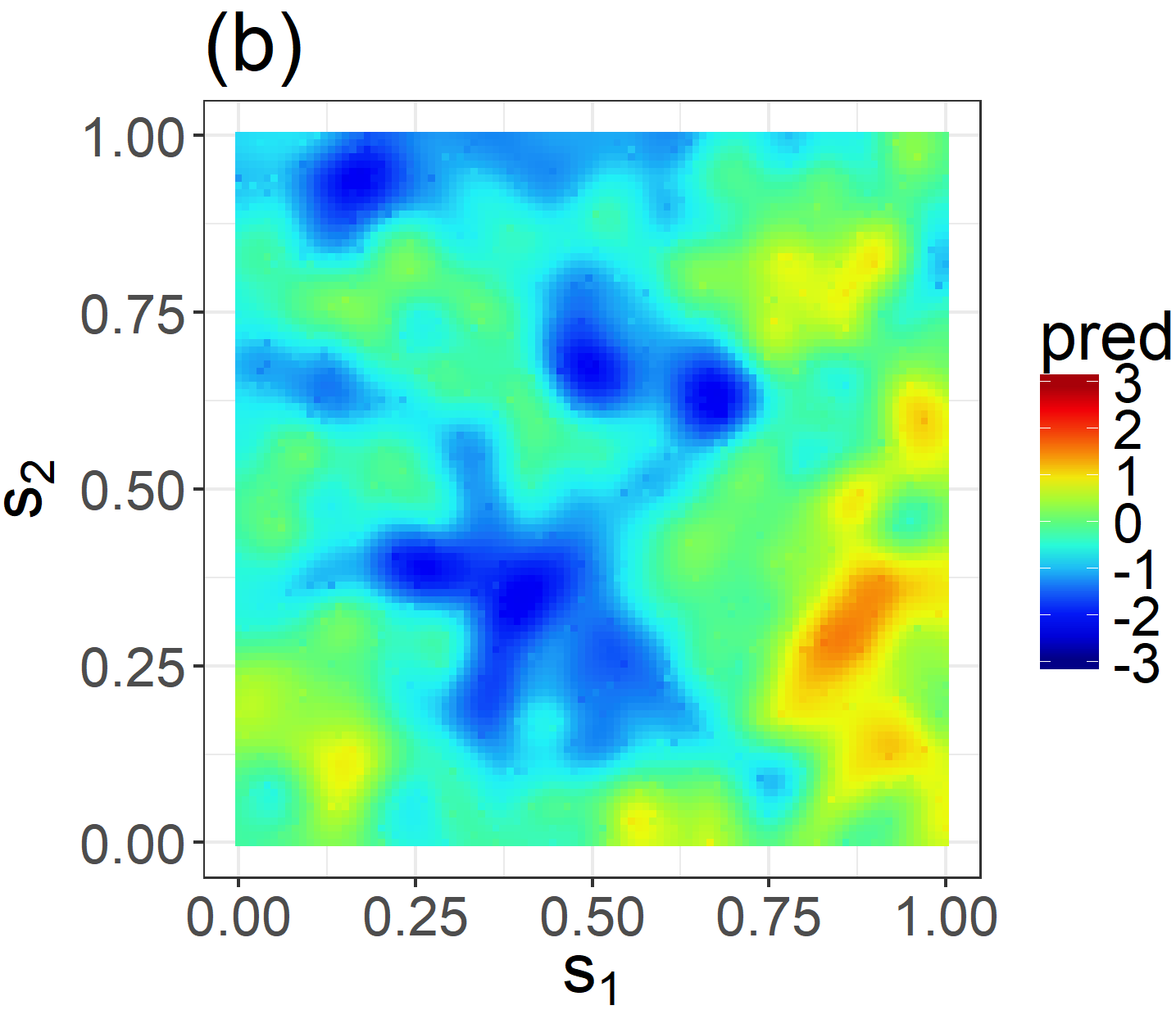}
	\includegraphics[width=2in]{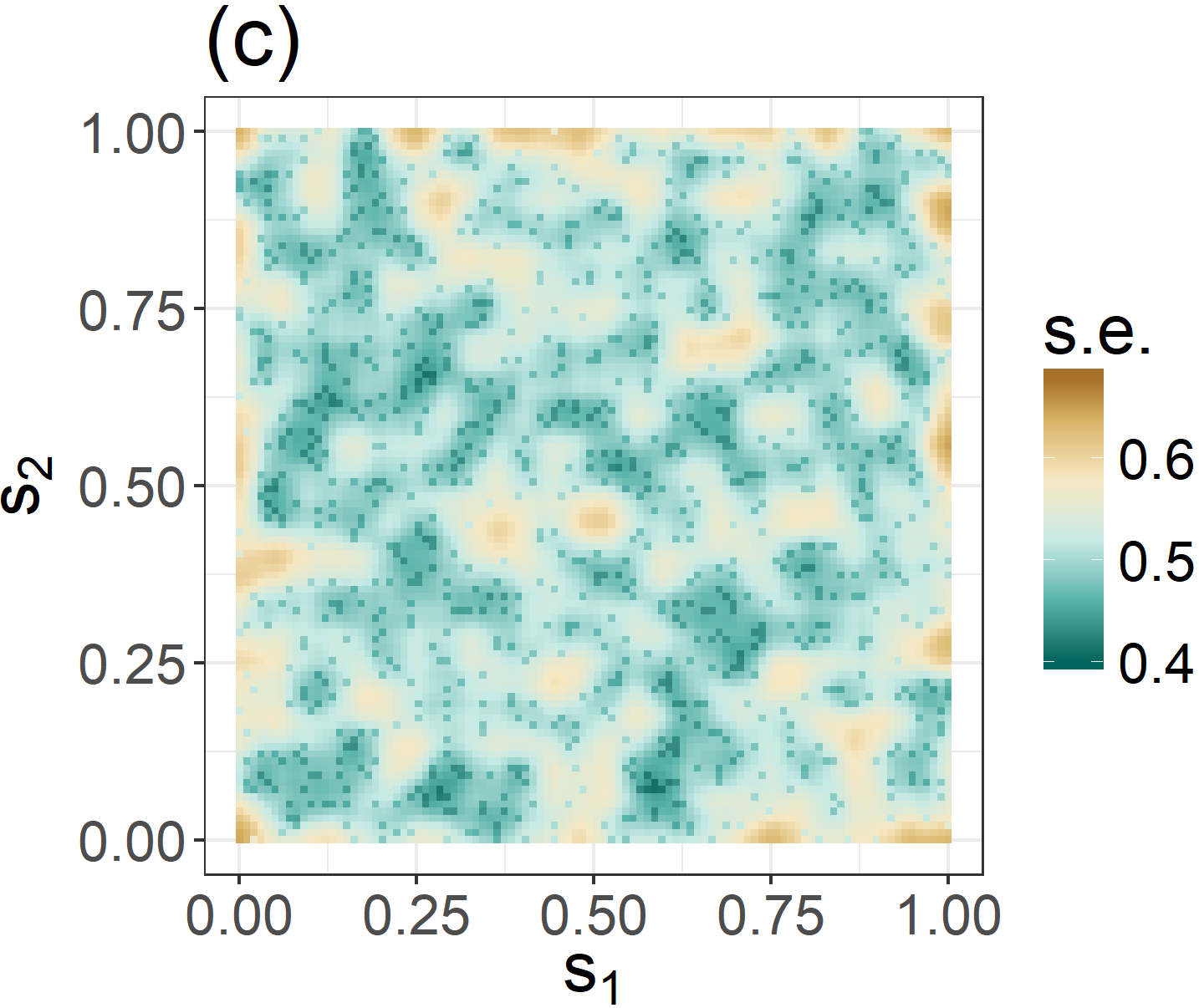}
	\caption{Fixed rank kriging (FRK) in a simulation experiment (see the text of Section \ref{sec:FRK} for details). (\textbf{a}) Arrangement of multi-resolution basis functions: the red solid line, green dotted line and blue dashed line denote the first, second and third resolution of the basis functions, respectively. (\textbf{b}) The prediction \blue{(pred)} following FRK on the prediction grid using the simulated retrievals. (\textbf{c}) The prediction standard errors \blue{(s.e.)} following FRK on the prediction grid using the simulated retrievals. \label{fig:exp2}}
\end{figure}

{Example 3:} To illustrate FRK, consider the spatial-only example of Section \ref{sec:spaces}. The basis functions $\{\phi_i\}$ were chosen to be multi-resolution bisquares of the form,
 \begin{equation*}
b(\svec,\rvec) \equiv \left\{\begin{array}{ll} \{1 - (\|\rvec- \svec\|/A)^2\}^2; &\| \rvec -\svec\| \le A \\ 
0; & \textrm{otherwise}, \end{array} \right. 
\end{equation*}
where $A > 0$ is the aperture. The bisquare basis functions were regularly distributed in the domain; see~Figure~\ref{fig:exp2}a. The matrix $\Kmat$ was restricted to be block diagonal by resolution, with each block a covariance matrix derived from an exponential covariance function; see \cite{Zammit_2017} for details. The FRK predictions and prediction standard errors using these basis functions and known measurement-error variance are given in Figure \ref{fig:exp2}b,c, respectively. Notice that these have a practically identical appearance to those in Figure \ref{fig:exp1}c,d, respectively, despite $\Kmat$ and the variance of the process $\zeta$ being estimated from the data. The RMSPE at the diagnostic locations using FRK was 0.4996 (compared to 0.4995 using the optimal predictor), while the coverages were identical to those obtained using classic kriging in Table~\ref{tab:coverage}.

\begin{figure}[H]
	\centering
	\includegraphics[width=2in]{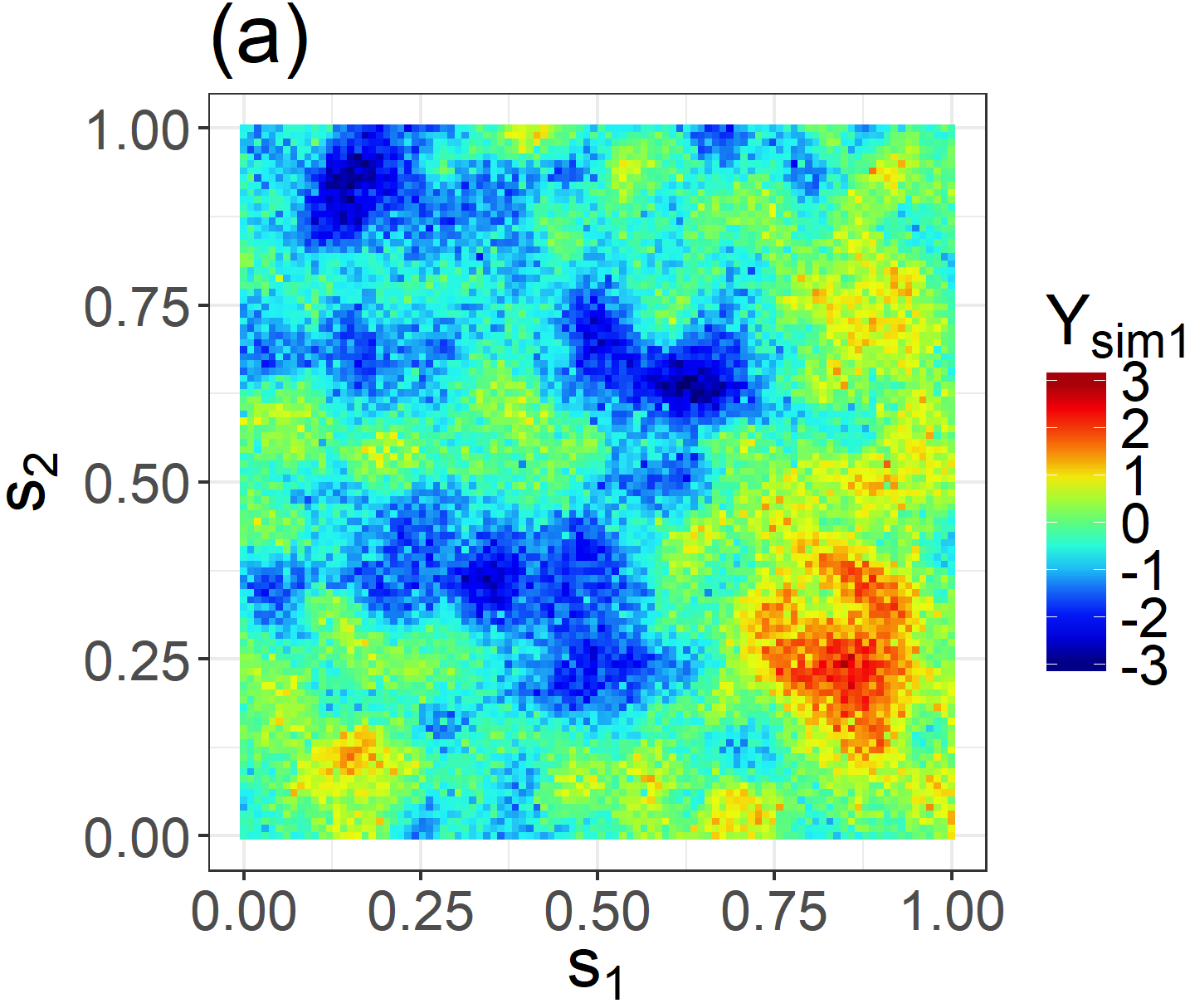}
	\includegraphics[width=2in]{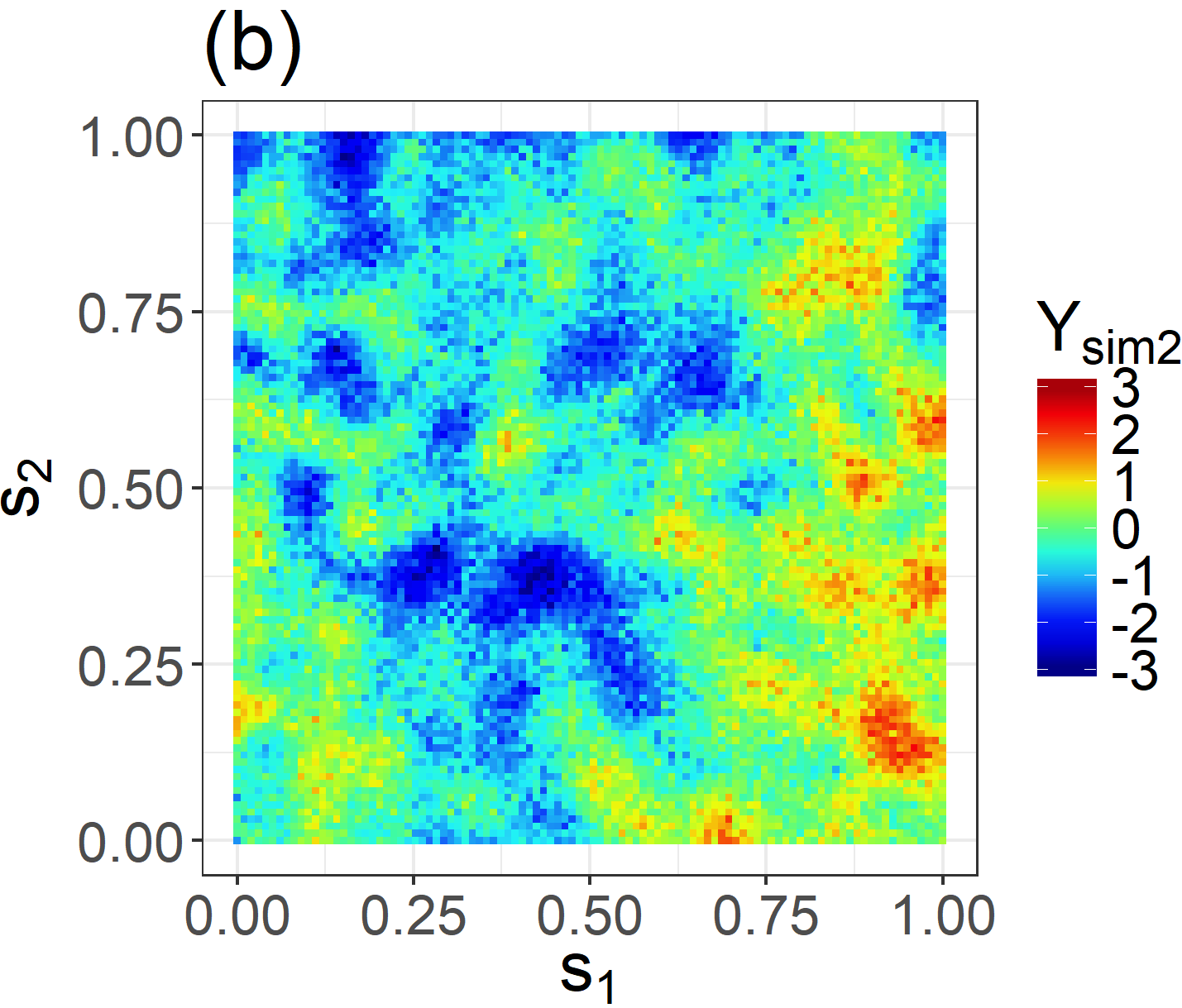}
	\includegraphics[width=1.7in]{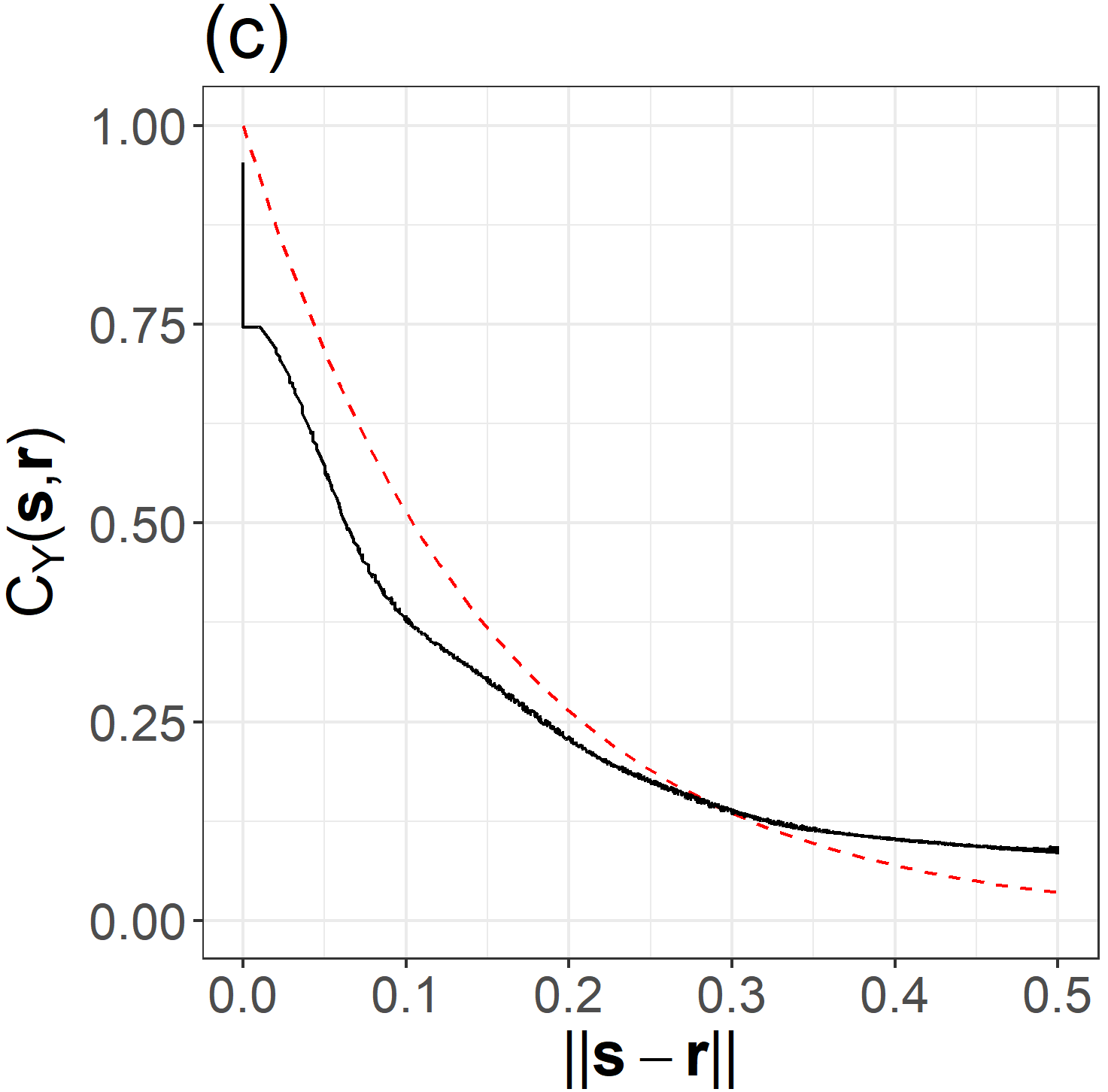}
	\caption{Conditional simulations and fitted covariance function using FRK (see~the text of Section \ref{sec:FRK} for details). (\textbf{a}) A conditional (on the observations) simulation from the fitted model that yields FRK; (\textbf{b}) a second conditional simulation from the same model; (\textbf{c}) true stationary covariance function (red dashed line) and estimated covariance function, averaged suitably over space, that was used in FRK (black solid line). \label{fig:exp2B}}
\end{figure}

Despite the FRK prediction and coverage being virtually indistinguishable from that of the classic kriging prediction, conditional simulations now are visually different from each other; see Figure~\ref{fig:exp2B}a,b. The reason for this is that while the basis functions are able to adequately reproduce the optimal predictor and prediction standard error surfaces, they are less suited to reconstruct the sharp gradient in the covariance function close to the origin. Specifically, the reduced-rank covariance function, $\sum_{i=1}^r\sum_{j=1}^r\phi_i(\svec)\phi_j(\rvec)K_{ij}$, is not able to reproduce the drop that appears in the exponential covariance function for $\svec$ close to $\rvec$, which plays a big role in the `rough' behaviour of exponential fields shown in Figure~\ref{fig:exp1B} \cite{Cressie_1993} (Section~2.3.1). Instead, a smooth covariance function is fit, and the excess variation is absorbed by $\zeta$, which is reflected in the sudden jump at the origin in Figure \ref{fig:exp2B}c, which~in turn is reflected in `speckles' in the sample paths. Technical details on the quality of the approximation in terms of the Kullback--Leibler divergence are given in \cite{Stein_2014}. 

Example 3 clearly shows that the sample paths from conditional simulation based on FRK are generally too speckled because of the assumption of an uncorrelated $\zeta$. Alternative models that assume a fine-scale correlated structure can remedy this \cite{Ma_2017}. The claim that FRK constrains the predictive surface to be smoother than the optimal predictor is not true in general. Indeed, if one is only concerned with predictions, FRK can even yield the optimal predictor {exactly} in our example when the true covariance function is known. In this case, one needs as many basis functions as data points. Specifically, for the simulation study of Example 3, one needs to set the vector $\phib(\cdot)' = \cvec(\cdot)'\Kmat^{-1}$ and $\Kmat = (\Cmat + \sigma^2_\epsilon\Imat)$, where $\cvec(\cdot)' = (C_Y(\cdot,\svec_i): i = 1,\dots, m)$, and $\Cmat = (C_Y(\svec_i,\svec_j):i,j = 1,\dots,m)$ (see~Appendix \ref{app:A2} for a proof).

\vspace{-6pt}

Using basis functions other than these leads to a surface that is suboptimal for a process with covariance function $C_Y(\cdot,\cdot)$; however, in Example 3, we demonstrated nearly perfect performance with $r < m$ and the use of a set of basis functions that was not optimised for the task. Hence, over-smoothing with FRK is the consequence of either a poor choice (or an insufficient number) of basis functions, or a poor choice for $\Kmat$; but, it is not a consequence of the use of a reduced-rank method in itself. Spectral considerations, such as those considered in \cite{Zammit_2012}, can be used to ensure that an adequate set of basis functions is chosen. Finally, the last decade has seen a surge in reduced-rank methods based on dimensionality reduction that can handle orders of magnitude more basis functions than FRK by putting additional constraints on $\Kmat$ or $\Kmat^{-1}$ (e.g., \cite{Lindgren_2011,Nychka_2015,Katzfuss_2017}). They are briefly discussed in Section \ref{sec:discussion}. 

In summary, FRK can yield the optimal predictor, the correct coverage and a good approximation to the true, underlying, covariance function even when this is not specified {a priori}. This flexibility is important, since a specific choice of covariance function considerably affects predictions and prediction standard errors. In addition, it can be used with large datasets (hundreds of thousands to millions compared to a few thousand as with classic kriging), and it can also handle structured non-stationarity. Indeed, FRK using the entire dataset only becomes less desirable from a practical standpoint when the system has a high SNR. This is where {local} kriging methods come into their own; these are discussed~next.

\subsection{Fixed-Window and Moving-Window Local Space-Time Kriging}\label{sec:FixedWindow}

It is generally problematic to consider satellite remote sensing datasets in their entirety when carrying out spatio-temporal prediction, for two reasons. The first reason is computational: When~carrying out optimal prediction, the computational cost of prediction always increases with dataset size. Further, there is little to be gained by using data far away from the space-time prediction location, due to the use of covariances that decay with space and time separation. The second reason is modelling: It requires effort to define and fit spatio-temporal models with non-stationary covariance functions at fine, medium, and large scales. Using dynamic models (e.g., \cite{Wikle_1999,Stroud_2001}), one can alleviate both the computational and the modelling problems somewhat. However, even then, fixed-lag (i.e.,~local) smoothing is generally used for datasets that span large temporal scales.

One way to circumvent these difficulties is to generate Level 3 products by considering spatial snapshots of retrievals, that is, retrievals that fall into temporal bins of fixed width. When using spatial snapshots the bin width is clearly important, and one needs `to balance the competing goals of including as many observations as possible, while avoiding time periods over which the [$\dots$] field itself would change substantially' \cite{Tadic_2015}. Unfortunately, this statement is difficult to operationalise, and~it results in the use of different bin widths by different users. For example, \blue{both six-day bins \cite{Hammerling_2012,Tadic_2015} and monthly bins \cite{Watanabe_2015} have been used to produce spatial maps from the Greenhouse gases Observing SATellite (GOSAT) retrievals.} 

Unless a spatio-temporal field is highly correlated in time, spatial-only kriging of spatio-temporal data is likely to yield poorer predictions than spatio-temporal kriging: Retrievals that are spatially close but on different satellite orbits that are days apart, may be reflections of completely different process values that will be treated as `similar' when doing spatial-only prediction. From a modelling point of view, when temporally binning data, implicitly one is considering a separable spatio-temporal covariance function of the form $C_Y^{(t)}(t,t')C_Y^{(s)}(h)$, where the spatial covariance function $C_Y^{(s)}(h)$ is modelled from the data but the temporal covariance function $C_Y^{(t)}(t,t')$ is just an indicator function, namely $C_Y^{(t)}(t,t') = 1$ if $t,t'$ are in the same temporal bin, and $= 0$ otherwise. This choice of $C_Y^{(t)}(t,t')$ says that two data in different bins are independent, which is likely to be a very poor representation of reality. Clearly, this is a strong assumption, and any class of continuous temporal covariance functions (such as the exponential) is likely to be an improvement over this choice.

\begin{figure}[H]
	\centering
	\includegraphics[width=3.4in]{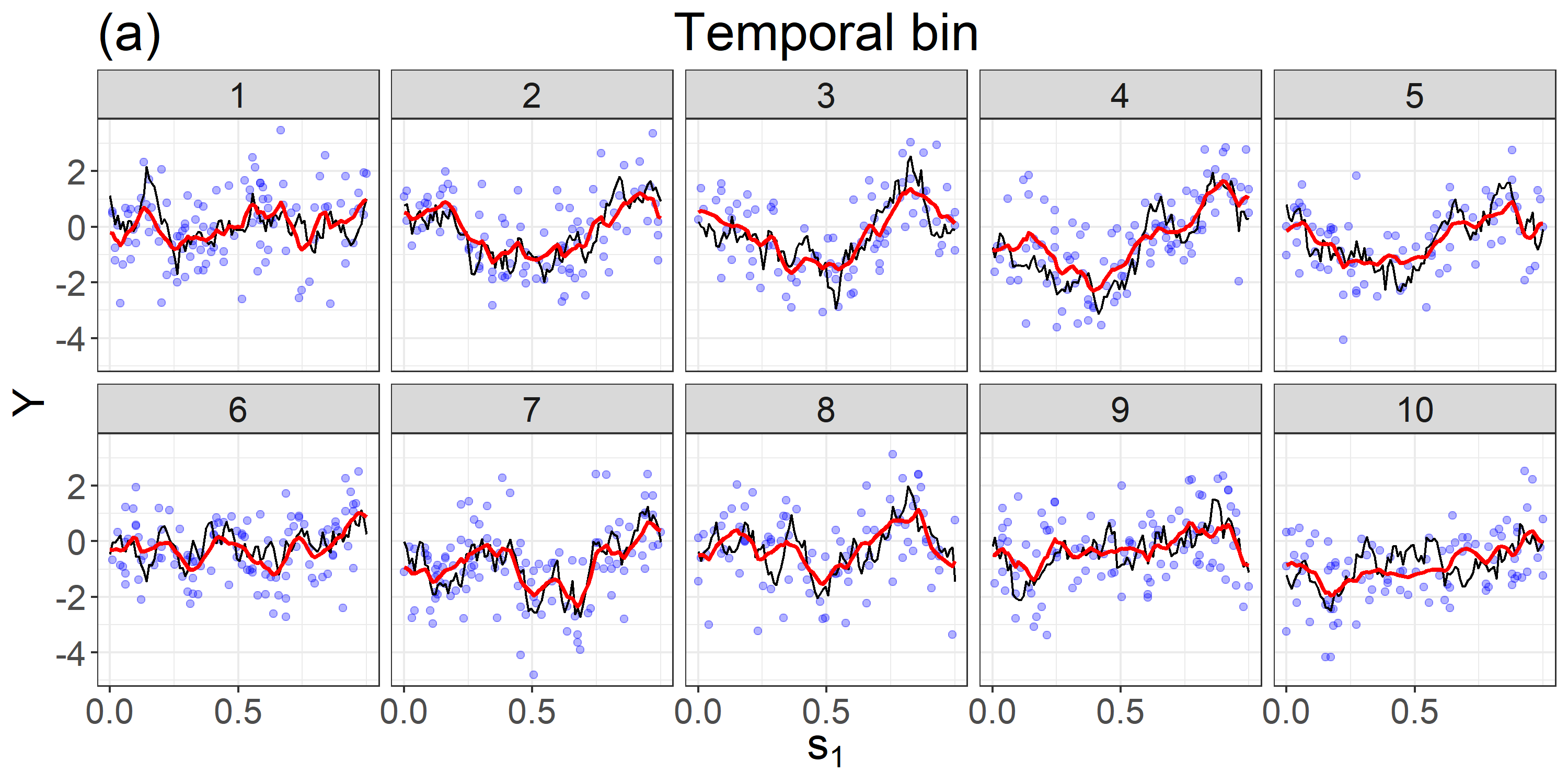}\\
	\includegraphics[width=2.1in]{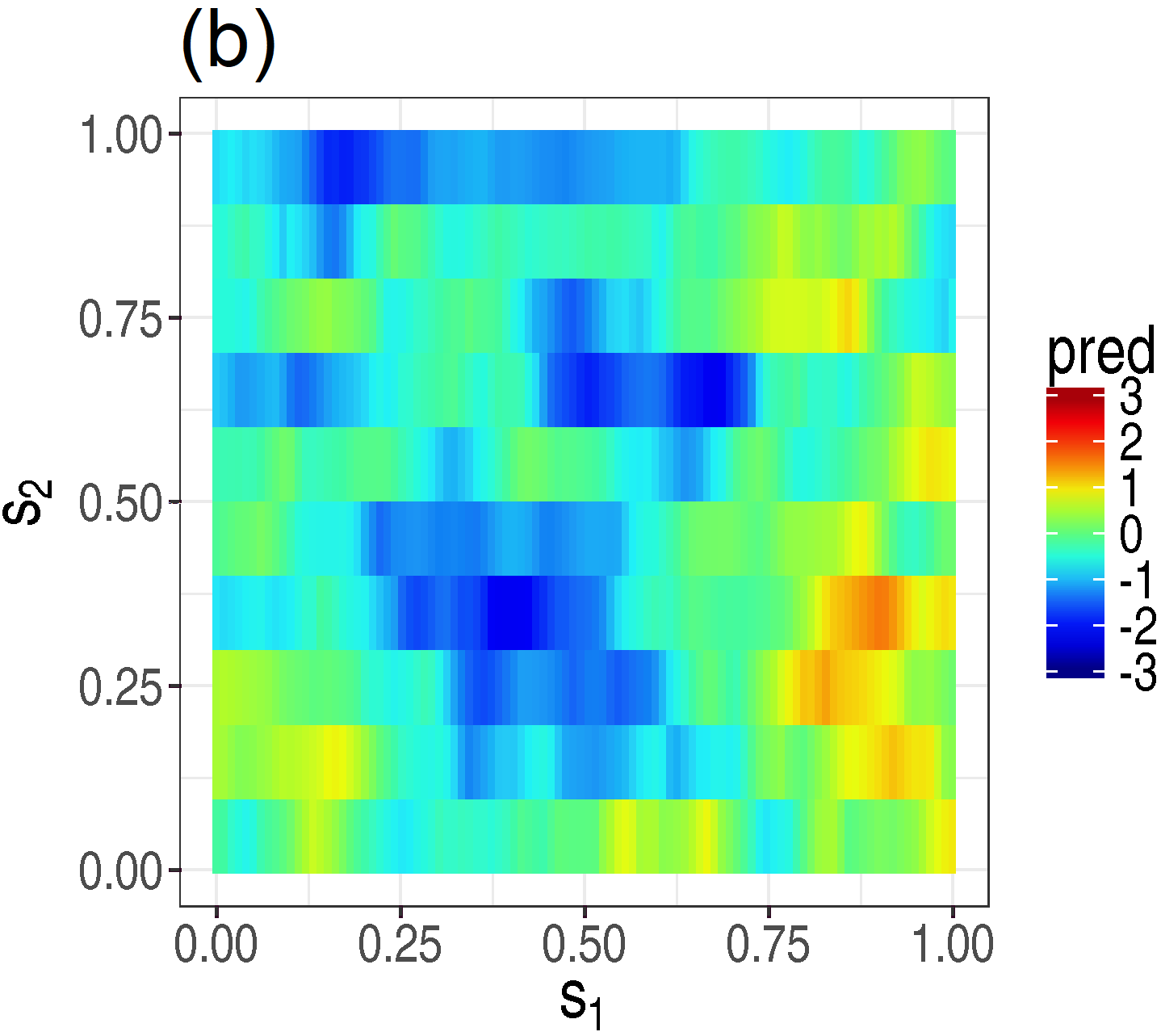}
	\includegraphics[width=2.1in]{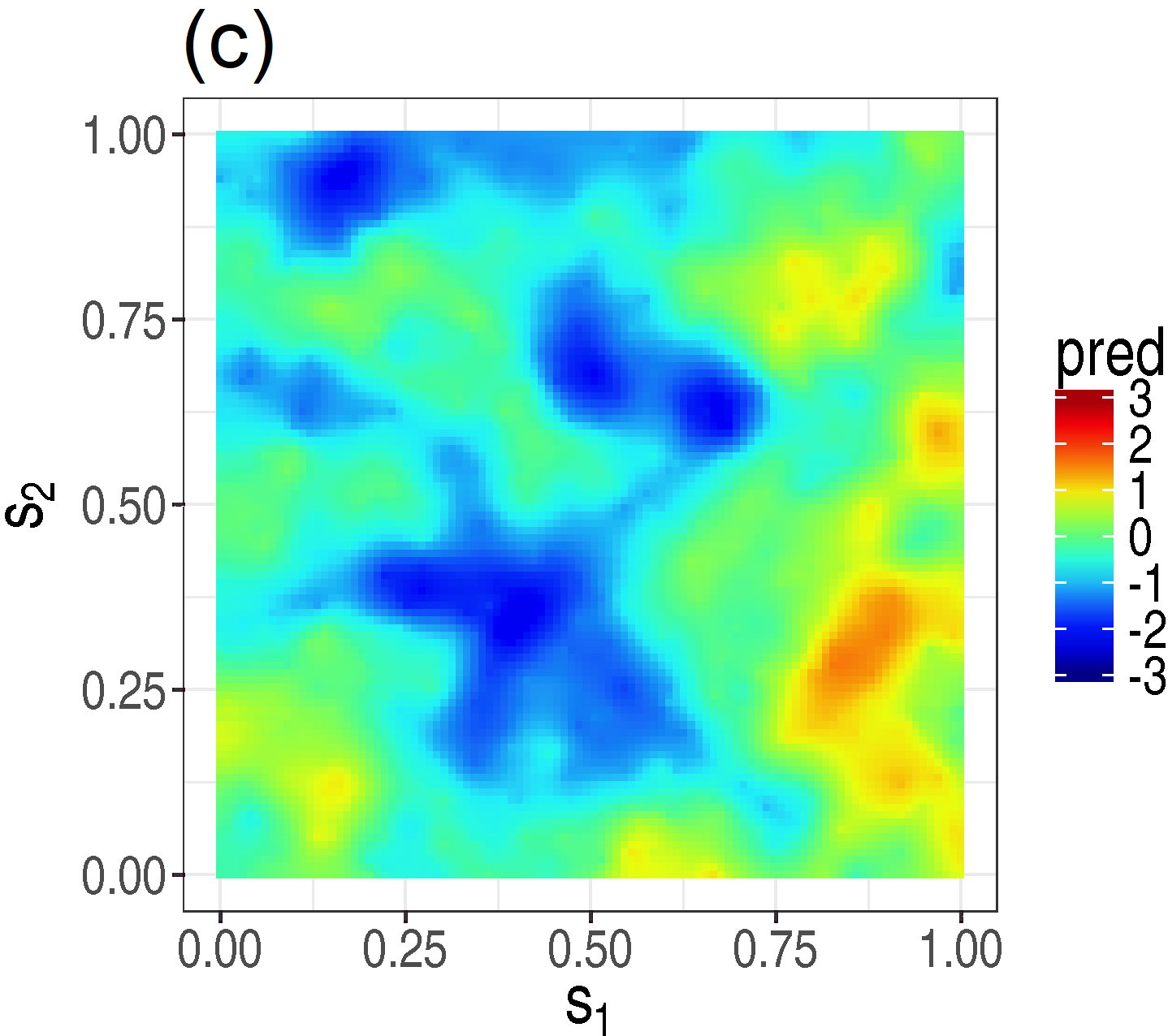}
\caption{Spatial prediction over $s_1 \in [0,1]$ using data in bins of $t (= s_2)$ of width 0.1. (\textbf{a})~One-dimensional spatial-only kriging of a two-dimensional field obtained by binning the data into fixed temporal bins (of $s_2$) of width 0.1. The data that fall into the bins are shown as blue dots, the~true (unobserved) field at the centre of the bin as a black solid line, and the spatial prediction as a red line. (\textbf{b}) The implied two-dimensional spatio-temporal prediction \blue{(pred)} surface (i.e., assuming temporal invariance within each bin), when carrying out prediction using fixed bins. (\textbf{c}) Predictions \blue{(pred)} obtained by using data in a moving window in $s_2$ and using optimal prediction for locations at the centre of the moving window. \label{fig:exp3}}
\end{figure}

{Example 4:} To elucidate further the deleterious impact of assuming temporal independence across bins, assume~that in the simulation experiment where $\svec = (s_1,s_2)'$, $s_1$ denotes one-dimensional space and $s_2$ denotes time. In this case, splitting the domain of $t (= s_2)$ into fixed bins of width 0.1~units, and~carrying out prediction in each of these bins (as a function of spatial location \mbox{$s_1 \in [0,1]$}), is~reasonable, since this process exhibits high temporal correlation within 0.1 units ($\corr(\svec,\svec + (0,0.1)') = 0.51$). Carrying out one-dimensional optimal spatial prediction in each bin using only the data in that bin yields the predictions shown in Figure \ref{fig:exp3}a, where we also show the process values for $t$ at the centre of each bin for comparison.

Validation can be carried out by allocating each datum used for diagnostics to its relevant temporal bin and computing the respective prediction standard errors. We found that the empirical version, RMSPE, at~the left-out diagnostic locations increases by approximately 15\% from 0.500 (in Example~1) to 0.574. Further, Figure \ref{fig:exp3}b shows that the implied space-time predictor exhibits `stripes' due to the fixed binning procedure (which assumes temporal invariance within each bin). In practice, deterioration is likely to be greater when spatial-covariance-function parameters are also estimated from the data, since only a fraction of the full dataset is available in each temporal bin.

\vspace{-6pt}

If dataset size or suspected temporal heterogeneity is the reason behind binning the data into temporal bins, and the main aim of the analysis is to obtain a good approximation to the optimal predictor $\E(Y(\cdot) \mid \Zvec)$, it is usually much better to consider a moving temporal window, where~the prediction at some space-time location $(s_1;t)$, for $t (= s_2)$, considers data in the temporal bin $[s_2 - \Delta/2, s_2 + \Delta/2]$, for temporal bin width $\Delta$. Then {spatio-temporal} prediction (and not just spatial prediction) is done based on the spatio-temporal data in this bin centred on the prediction location. This~is a type of local kriging \cite{Haas_1995} discussed further in Section \ref{sec:local} that, despite some theoretical flaws, in high SNR scenarios is well suited to carry out local (but not global) inference in space and time. In~Figure \ref{fig:exp3}b,c, we compare the prediction using fixed temporal bins with that using moving temporal bins. The RMSPE and the coverage computed from the latter predictions (Figure~\ref{fig:exp3}c) are virtually identical to those obtained from spatio-temporal kriging on $[0,1] \times [0,1]$ (Figure~\ref{fig:exp1}c). Results from \cite{Zeng_2017} also show the considerable improvement of moving-window spatio-temporal predictions compared to a sequence of spatial-only predictions. In Section \ref{sec:OCO-2} we use moving-window spatio-temporal kriging to obtain predictions and prediction standard errors of column-averaged CO$_2$ (XCO$_2$) from OCO-2.

\subsection{Local Prediction and Signal-To-Noise Ratio} \label{sec:local}


Local kriging reduces computational burden, since predictions are based on only a subset of the whole dataset. Local kriging has a place for answering local questions, since the prediction standard errors are appropriate for the local predictor. However, as was discussed in Section \ref{sec:intro}, local kriging yields predictions that are statistically incoherent over large scales. Continental-scale or oceanic-scale questions require a combination of local predictors, and the associated prediction standard error of aggregated local predictors requires a coherent global model. Consequently, predictors obtained from local kriging do not generally have correct coverage over large scales. 

If Level 3 products are primarily used to obtain local information, global incoherence may not be of prime concern in some applications. However, even when doing local-only predictions, one~must still be careful: The quality of the prediction when using only a subset of observations is highly determined by the SNR of the process. With low SNRs, that is when $\sigma^2_\epsilon$ is large with respect to $\sigma^2_Y$, the optimal predictor borrows more strength from observations that are far away from the prediction location, than with high SNRs. If the SNR is small, one may need to consider tens of thousands of observations to get a good approximation to the optimal predictor, which is impossible with the classic kriging methods. In this sense, local kriging, even when using moving windows, is not recommended in low SNR settings.

In contrast, a global-scale predictor like FRK is very useful where there is low SNR, since it can consider hundreds of thousands of data points with ease. Further, in Appendix \ref{app:A3}, we show that spatial-predictor smoothness (measured through second-order absolute difference quotients) increases with measurement-error variance (i.e., decreasing SNR). Thus, FRK is an ideal `big-data' approach for spatio-temporal prediction in low SNR settings. Furthermore, local modelling/prediction and FRK are not mutually exclusive. Indeed, it is often the case that a large amount of data is present and spatio-temporal domain subsets are needed, even when using state-of-the-art approaches that deal with a large number of basis functions. We demonstrate FRK in the context of a temporal moving window in Section \ref{sec:OCO-2}.

\begin{figure}[H]
	\centering
	\includegraphics[width=1.8in]{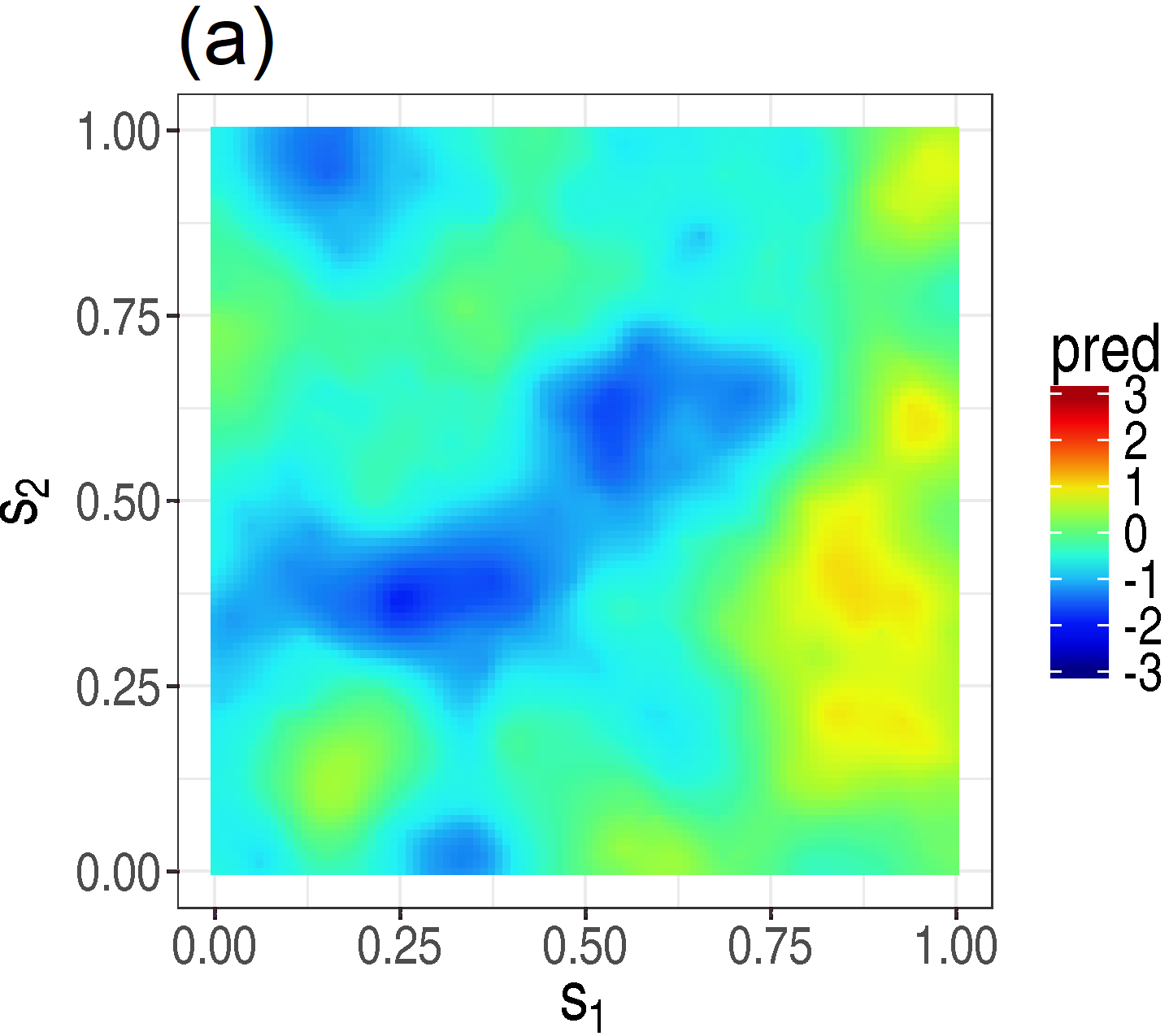}
	\includegraphics[width=1.8in]{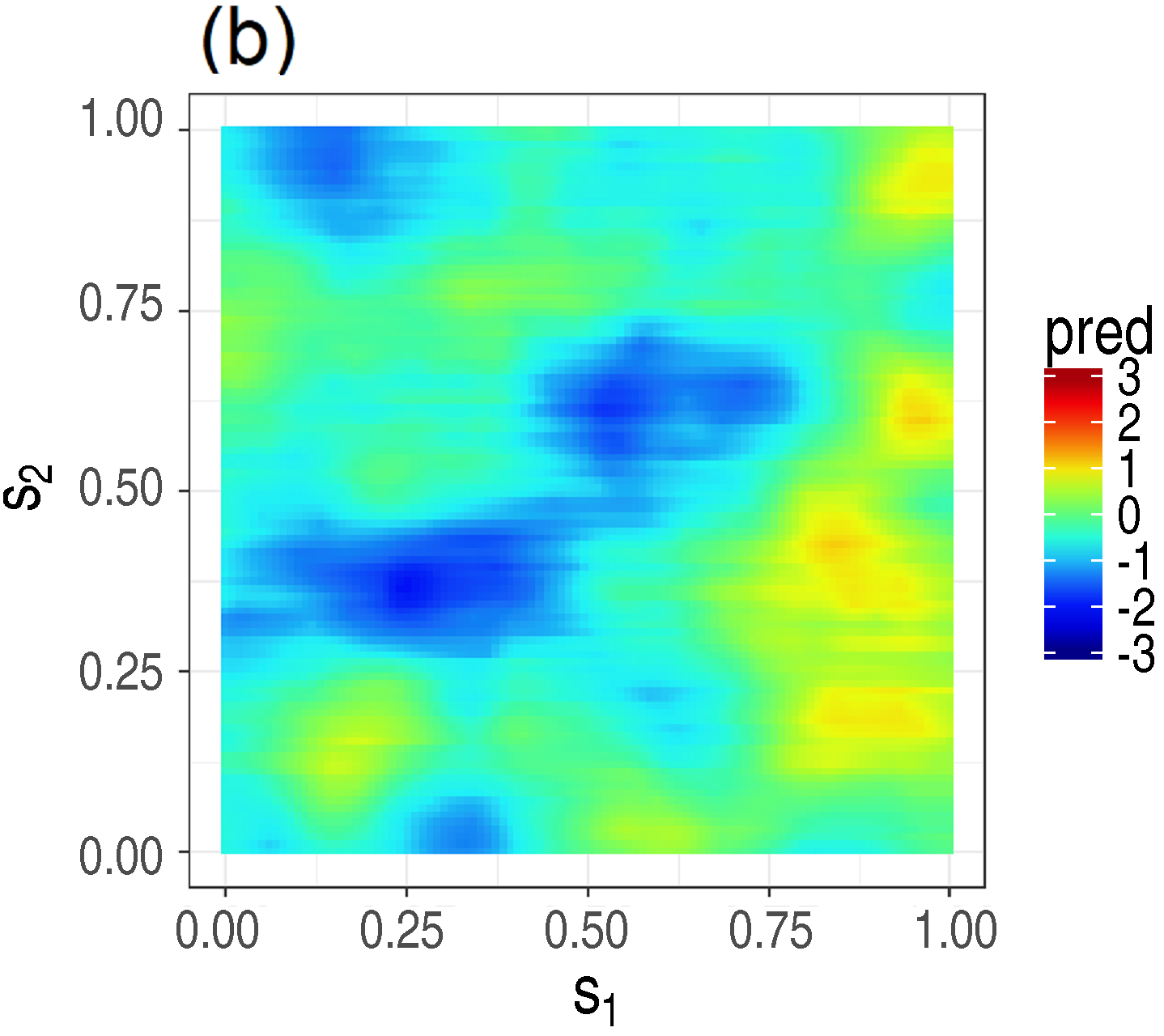}
	\includegraphics[width=1.8in]{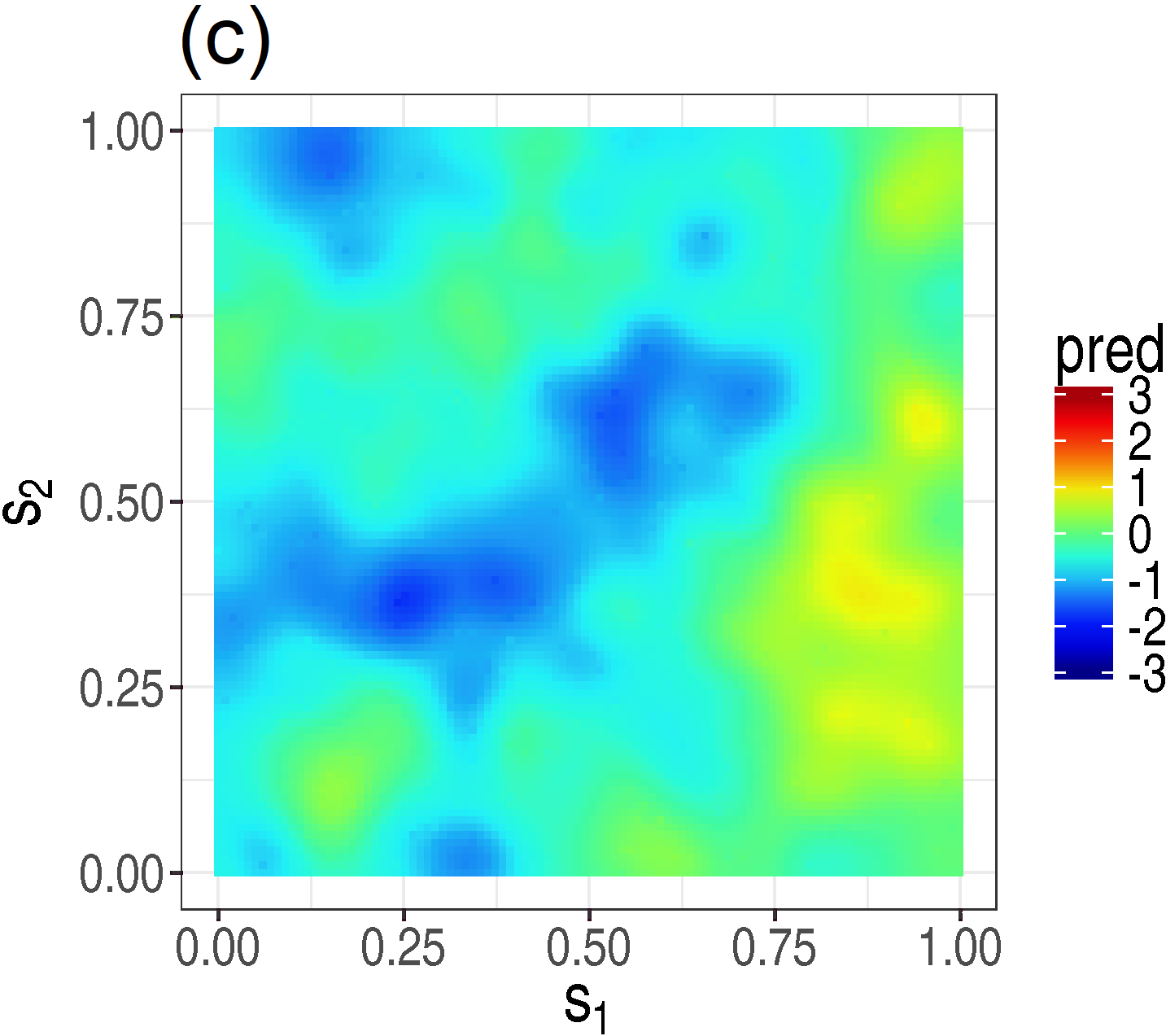}
	\caption{Spatial prediction of the true process (Figure~\ref{fig:exp1}b) using the experimental setup in Section~\ref{sec:spaces} and data with increased measurement-error variance, $\sigma^2_\epsilon = 10$. (\textbf{a}) Optimal prediction \blue{(pred)} using the simulated retrievals; (\textbf{b}) the prediction \blue{(pred)} from simple kriging on the prediction grid from the simulated retrievals using the exact model and a moving window in $s_2$ of width 0.1 units; (\textbf{c}) the prediction \blue{(pred)} from fixed rank kriging on the prediction grid using the simulated retrievals. \label{fig:exp5}}
\end{figure}

{Example 5: }Consider again the simulation experiment of Section \ref{sec:spaces}, but this time let $\sigma^2_\epsilon = 10$. The optimal prediction with this increased measurement-error variance is shown in Figure \ref{fig:exp5}a. Now the prediction obtained using the moving-window local kriging method described in Section \ref{sec:FixedWindow}, shown~in Figure \ref{fig:exp5}b, exhibits striping since predictions are based on relatively few, very noisy observations. The~RMSPE at the diagnostic locations when doing kriging with {all} the data is 0.685, but it increases to 0.707 when using a moving window of width 0.1 (despite use of the exact covariance function). The~prediction from FRK using the parameters estimated in Section~\ref{sec:FRK}, illustrated in Figure~\ref{fig:exp5}c, is clearly much closer to the optimal predictor, despite its being a misspecified model. The RMSPE using FRK in this case was 0.694. We can expect local kriging's relative RMSPE (relative to the optimal) to deteriorate further as the SNR decreases. In our experience, when the SNR is less than about $0.2$, local kriging begins to produce poor-quality pointwise predictions when compared to dimensionality-reduction methods (e.g., FRK) in models involving the exponential covariance function.

\section{OCO-2 Level 3 Products from V7r and V8r Lite Files}\label{sec:OCO-2}

In this section, we use FRK and a moving window in time to produce statistical Level 3 products from the OCO-2 Data Release 7r Lite File Version B7305Br and the OCO-2 Data Release 8r Lite File Version B8100r. The use of the bias-corrected Lite Files, as opposed to the raw retrievals, is~justified in Section \ref{sec:coverage}. The Level 3 products, which we name FRK Version 7r and FRK Version 8r products, are~made up of daily FRK predictions and prediction standard errors from 1 October 2014--28 February 2017. The FRK products, and animations of these, are provided at \url{https://niasra.uow.edu.au/cei/oco2level3}.  

\subsection{OCO-2 Data Preprocessing}\label{sec:OCO2_preproc}

The OCO-2 Release 7r Lite File \cite{OCO2v7r} includes bias corrections for footprints, parameters, and~scalings. All data in this File with some missing variables, with a warn \blue{level greater than} or equal to 15 and a quality flag of one, or with a reported retrieval standard error of more than 3~ppm, were~filtered out. 
The OCO-2 Release 8r Lite File \cite{OCO2v8r} accounts for stratospheric aerosols, thus~improving the data quality in the Southern Hemisphere, and it has larger global coverage due to improvements in pre-screeners. All data in this File with a quality flag of zero were kept for generating the product, irrespective of the warn level. For both versions, we did not make a distinction between the mode of operation (land nadir, land glint, ocean glint; see \cite{Wunch_2017} Section \ref{sec:pedag} for a concise summary of modes of operation) in which the retrieval was made when generating the products, and all retrievals obtained when the instrument was in target mode were filtered out.

The FRK Level 3 products were produced in process space using a moving window of 16 days' duration, which matches the repeat cycle of the satellite. We considered data in the Lite File between 24 September 2014 and 8 March 2017, and hence the FRK products were produced for days between 1 October 2014 and 28 February 2017. The standard errors of the retrievals returned by the Level 2 retrieval algorithm are known to be underestimated \cite{Hobbs_2017}. To cater for this, when the Lite File gave a standard error below 2 ppm, we raised it to 2 ppm, following a similar strategy in \cite{Nguyen_2017}. This may result in our FRK Level 3 products being under-confident; however, this is preferred to ones that are over-confident. We do not expect this adjustment to affect the relative quality of the Version 7r and the Version 8r FRK products: Indeed, when repeating the study discussed below with the standard error \blue{threshold set} to 0.5 ppm instead of 2 ppm, we obtained similar relative performance of the Version 7r and Version 8r FRK products but worse predictions (higher MAE and RMSPE) when assessing these products against TCCON.

All retrievals were subsequently aggregated into a 1 $\times$ 1 $\times$ 1 lon-lat-day grid. An aggregated retrieval \blue{and standard error} in any of these space-time cubes was found by averaging the retrievals and \blue{the retrieval standard errors} of all OCO-2 retrievals falling into the space-time cube. Specifically, denote the vector of $m_i$ retrievals falling into the $i$-th space-time cube as $\tilde \Zvec_i$ and the corresponding vector of retrieval standard errors as $\tilde \sigmab_i$. An aggregated retrieval $Z_i$ and retrieval standard error $\sigma_{\epsilon,i}$ for the $i$-th space-time cube were then found \blue{by averaging the vector elements},
$$
Z_i = \frac{1}{m_i}\sum_{l=1}^{m_i}\tilde{Z}_{i,l},\qquad \sigma_{\epsilon,i} = \frac{1}{m_i}\sum_{l=1}^{m_i}\tilde\sigma_{i,l}~~,
$$
for $i = 1,\dots,m$, where $m$ is the number of 1 $\times$ 1 $\times$ 1 lon-lat-day cubes containing one or more retrievals. The formula for $\sigma_{\epsilon,i}$ assumes (approximately) perfect correlation between the retrieval errors within the $i$-th space-time cube. These aggregated retrievals and retrieval standard errors were then used for generating the FRK Level 3 products. Notice that here, unlike in Section \ref{sec:pedag}, we allow for heteroscedasticity of the retrieval standard error.

\subsection{Implementation Details for FRK}

\begin{figure}[H]
	\centering
	\includegraphics[width=3in]{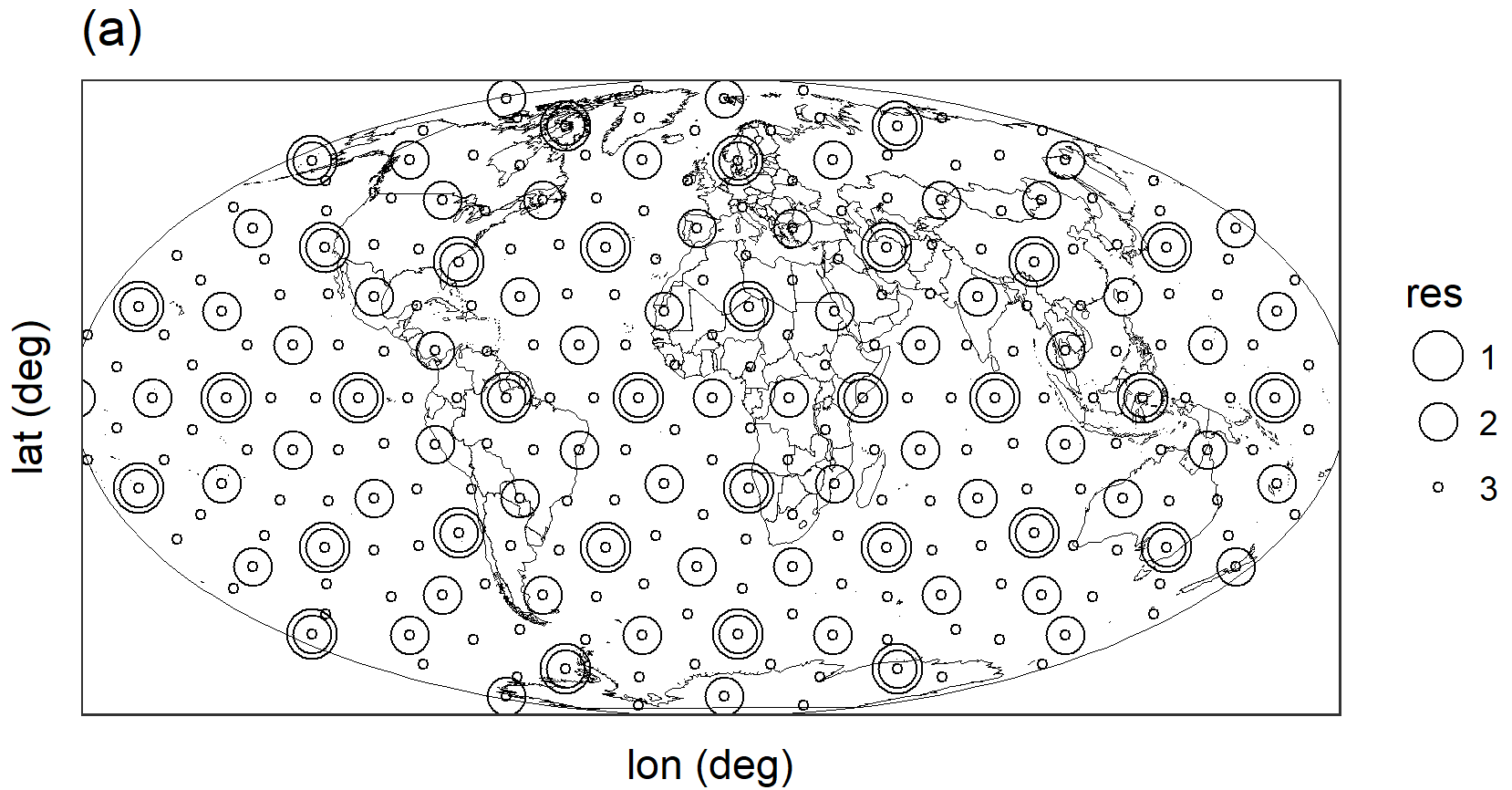}
	\includegraphics[width=3in]{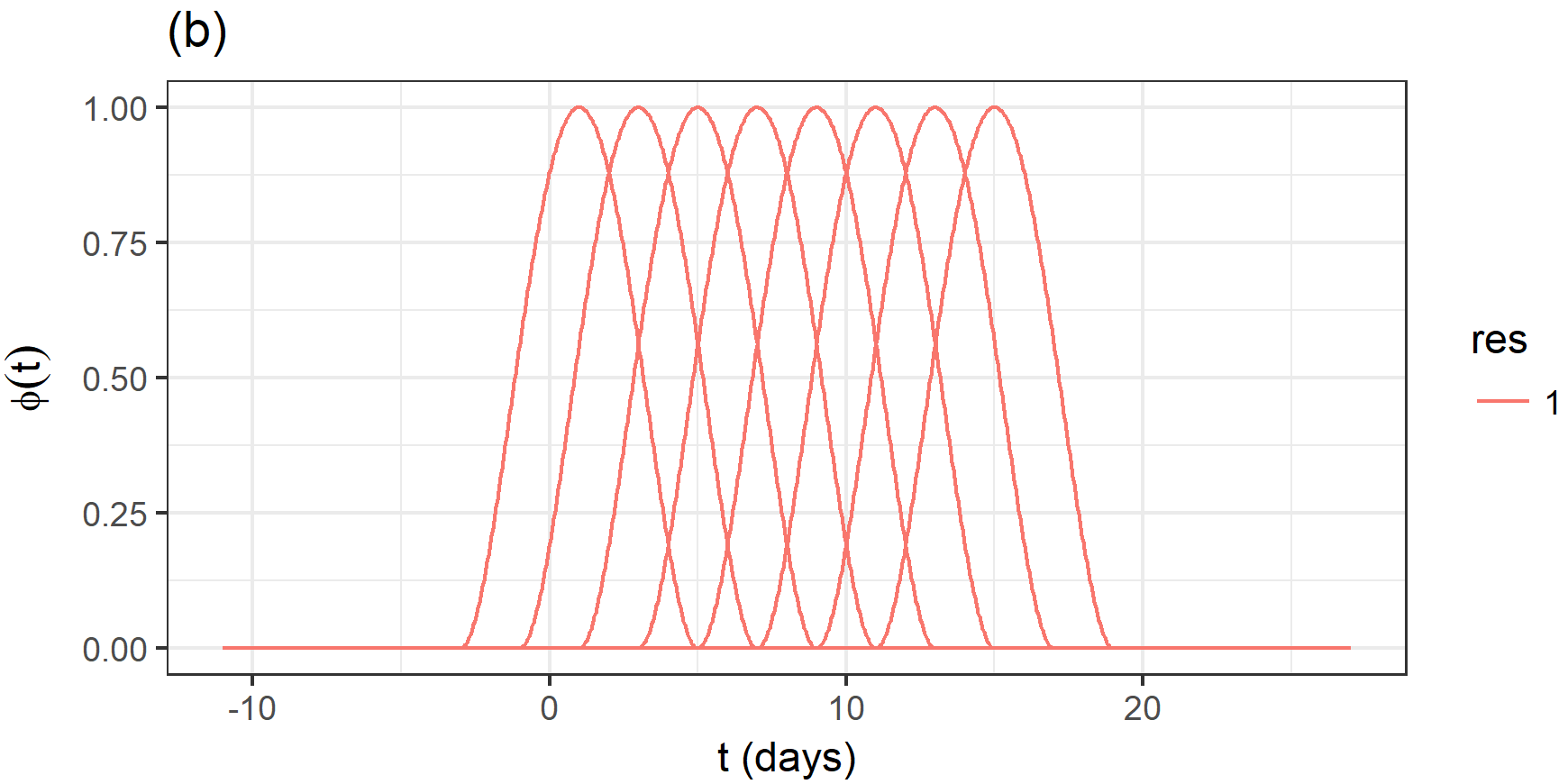}
	\caption{Spatio-temporal basis functions constructed by taking the tensor product of spatial and temporal basis functions. (\textbf{a}) The spatial basis functions used are bisquares over three different resolutions. The circle denotes the centroid of the bisquare while the circle size denotes the resolution (smaller indicates a finer resolution and hence a smaller aperture). (\textbf{b}) The temporal basis functions used were bisquares with centroids regularly placed between zero and 16 days in the moving window. \label{fig:STbasis}}
\end{figure}

A 16-day moving-window variant of FRK was used to construct the Level 3 products. Each~window was indexed by the eighth day in the window, at which a spatial global prediction was made, and the spatio-temporal domain of interest was spanned with 3168 spatio-temporal basis functions. These basis functions were constructed by finding the tensor product of 396 spatial bisquares arranged over three resolutions on the sphere, and eight temporal bisquares at a single temporal resolution regularly spaced in the 16-day window; see Figure \ref{fig:STbasis}. The eight temporal basis functions were chosen to allow the signal to considerably vary temporally within a 16-day window, while the number of resolutions for the spatial basis functions was capped so that the total number of basis functions was less than 4000, a point beyond which FRK begins to slow computationally. Spectral-based methods that may assist in choosing basis functions are outlined in \cite{Zammit_2012}. Note that a larger window is not considered for computational reasons; reduced rank methods that are able to deal with a larger number of basis functions are discussed in Section \ref{sec:discussion}.

In order to produce a prediction for the target day (i.e., the eighth day in the window), we require there to be data on the target day or, within its 16-day window, data on a day before and a day after the target day. Missing days in the Level 3 products are hence only present when there are gaps of eight~days or more in the respective Lite File. Since kriging methods are not well suited for extrapolation, predictions were only made for latitudes ranging between the lowest and highest data point on the target date. This resulted in Level 3 products with a latitude extent that changed slightly from day to day. Due to the increased data density in the Version 8r Lite File, the FRK Version 8r product has a considerably larger latitudinal span than the FRK Version 7r product.

To implement FRK, one needs to estimate parameters appearing in the matrix $\Kmat$, here notated as $\varthetab$, and to estimate the variance of the fine-scale process $\zeta(\cdot;\cdot)$. In our case, we let $\Kmat(\varthetab) = \textrm{bdiag}(\{\Kmat_q(\varthetab) : q = 1,\dots,n_{res}\})$, where $\textrm{bdiag}(\cdot)$ returns a block-diagonal matrix constructed from its arguments. The~matrices are: 
$$\Kmat_q(\varthetab) = (\vartheta_{1q}\exp(-d^{(s)}_{ijq}/\vartheta_{2q}-d^{(t)}_{ijq}/\vartheta_{3q}) : i,j = 1,\dots,r_q),$$ 
where $d^{(s)}_{ijq}$ and $d^{(t)}_{ijq}$ are the spatial and temporal distances between the centroids of the $i$-th and $j$-th basis functions at the $q$-th resolution, respectively; $r_q$ is the number of basis functions at the $q$-th resolution, $q = 1,\dots,n_{res}$; $n_{res} = 3$ is the number of resolutions; $\vartheta_{1q}$ is the marginal variance at the $q$-th resolution; $\vartheta_{2q}$ is the spatial e-folding length at the $q$-th spatial resolution; and $\vartheta_{3q}$ is the temporal e-folding length at the $q$-th spatial resolution. All parameters in each moving window were estimated using an expectation maximisation (EM) algorithm; see \cite{Zammit_2017} for details.

\begin{figure}[H]
	\centering
	\includegraphics[width=3in]{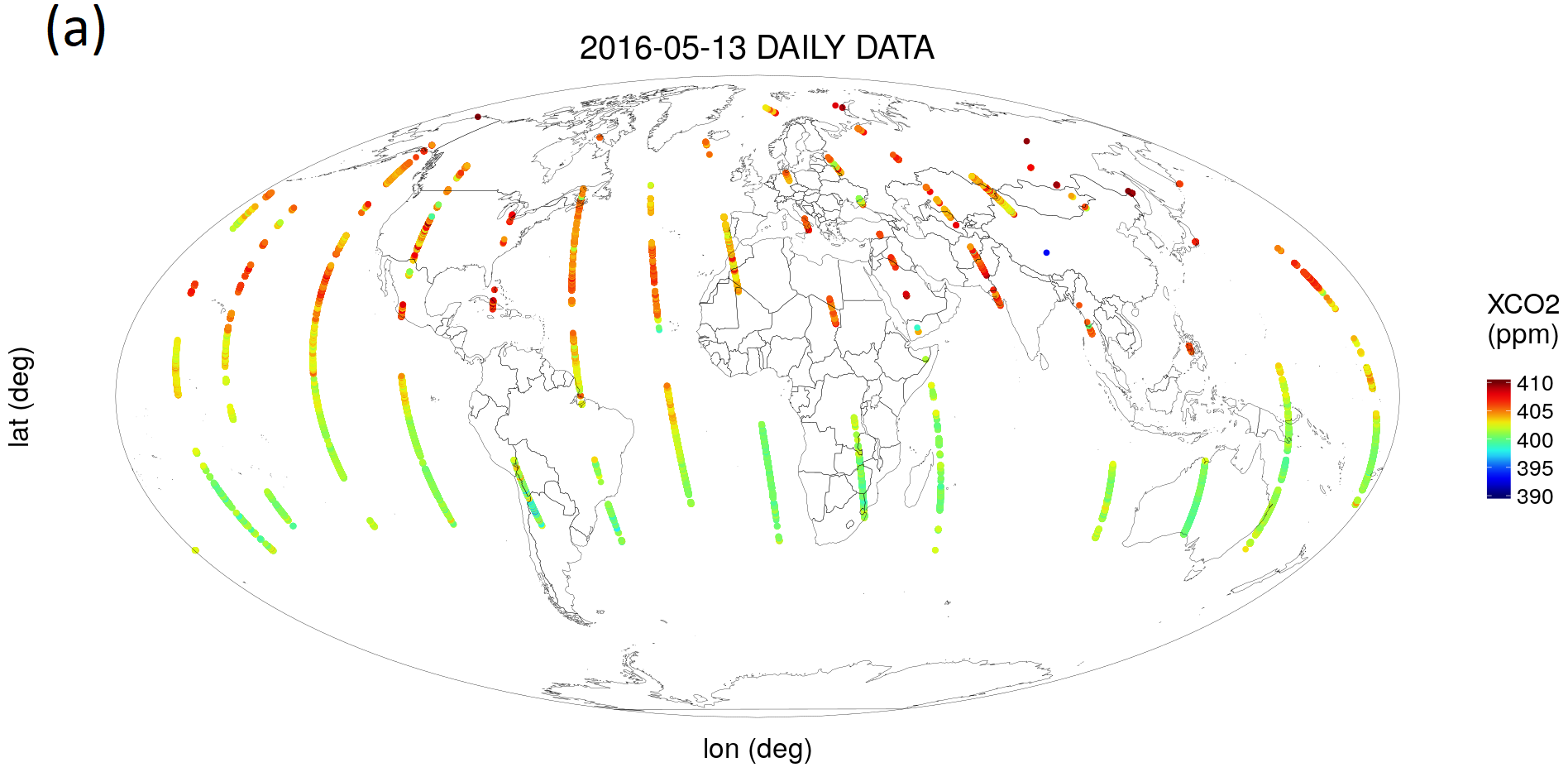}
	\includegraphics[width=3in]{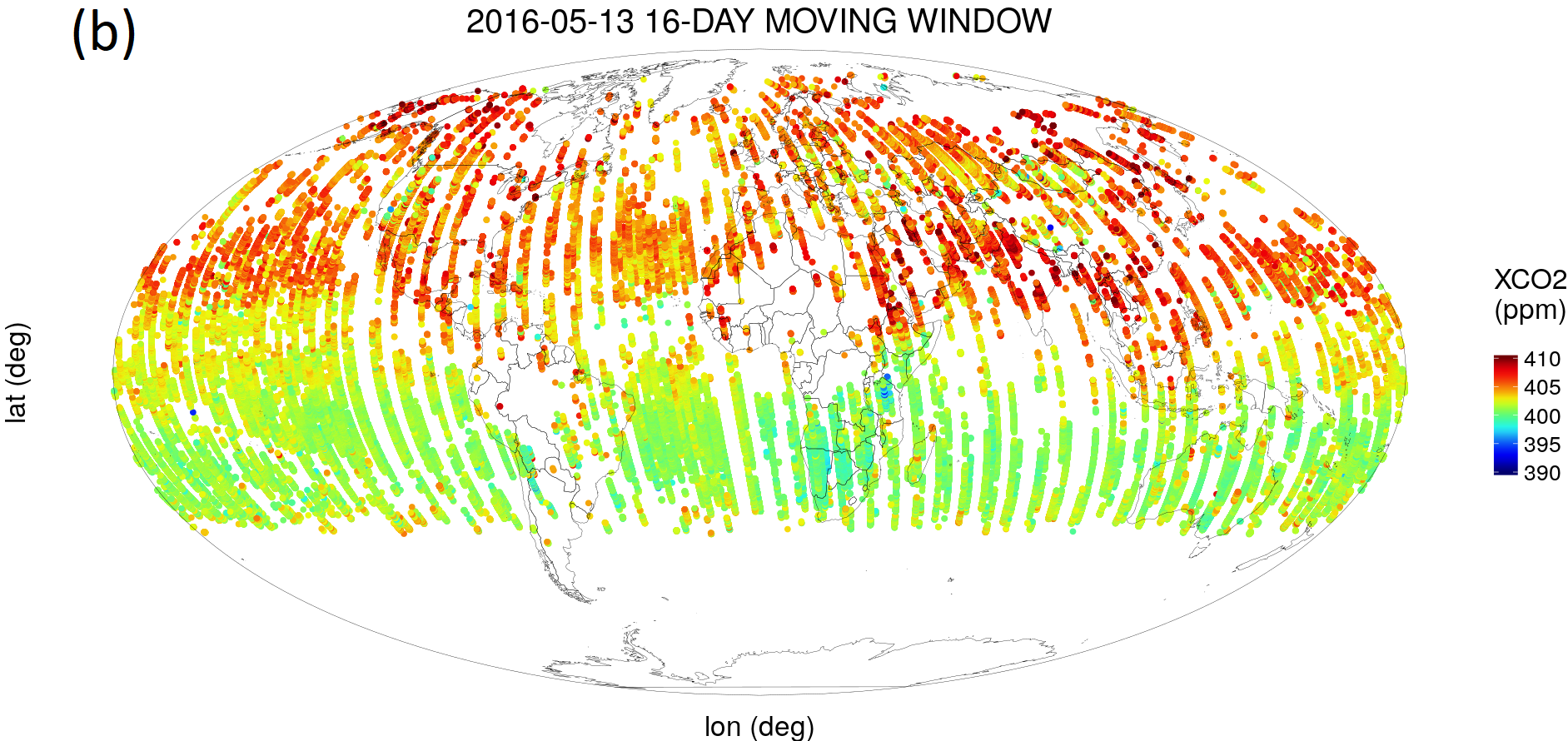}
	\includegraphics[width=3in]{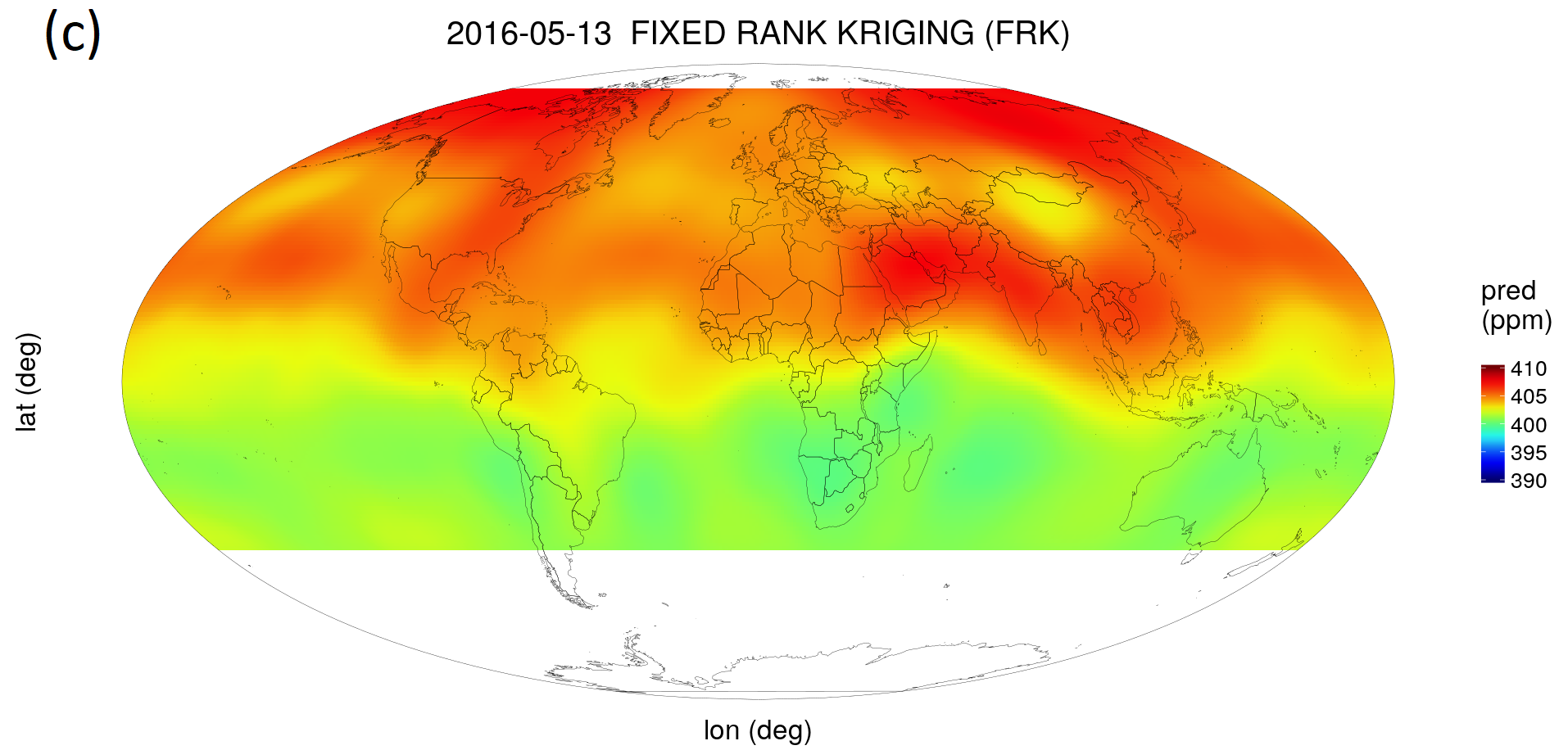}
   	\includegraphics[width=3in]{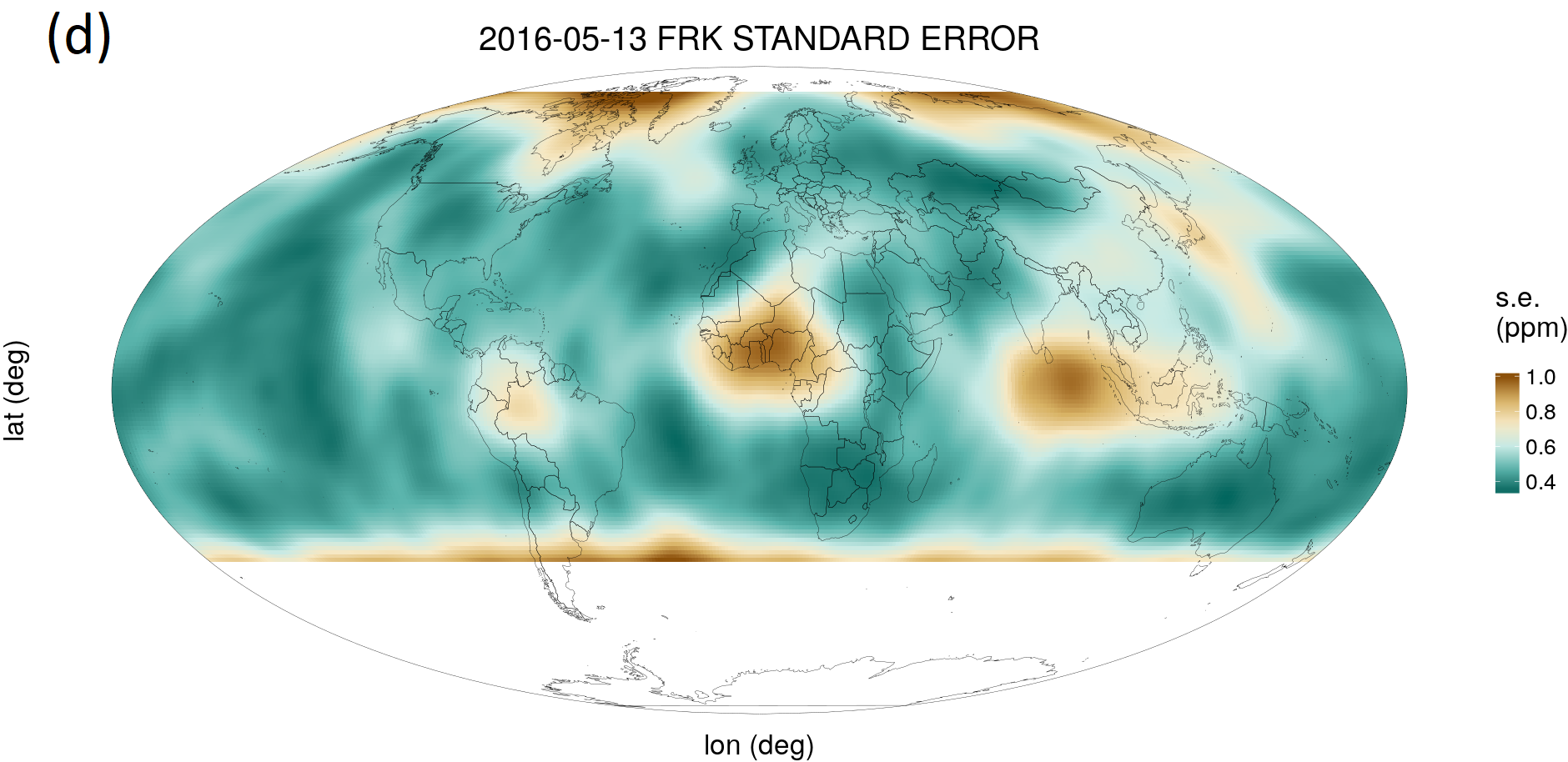}
   \includegraphics[width=3in]{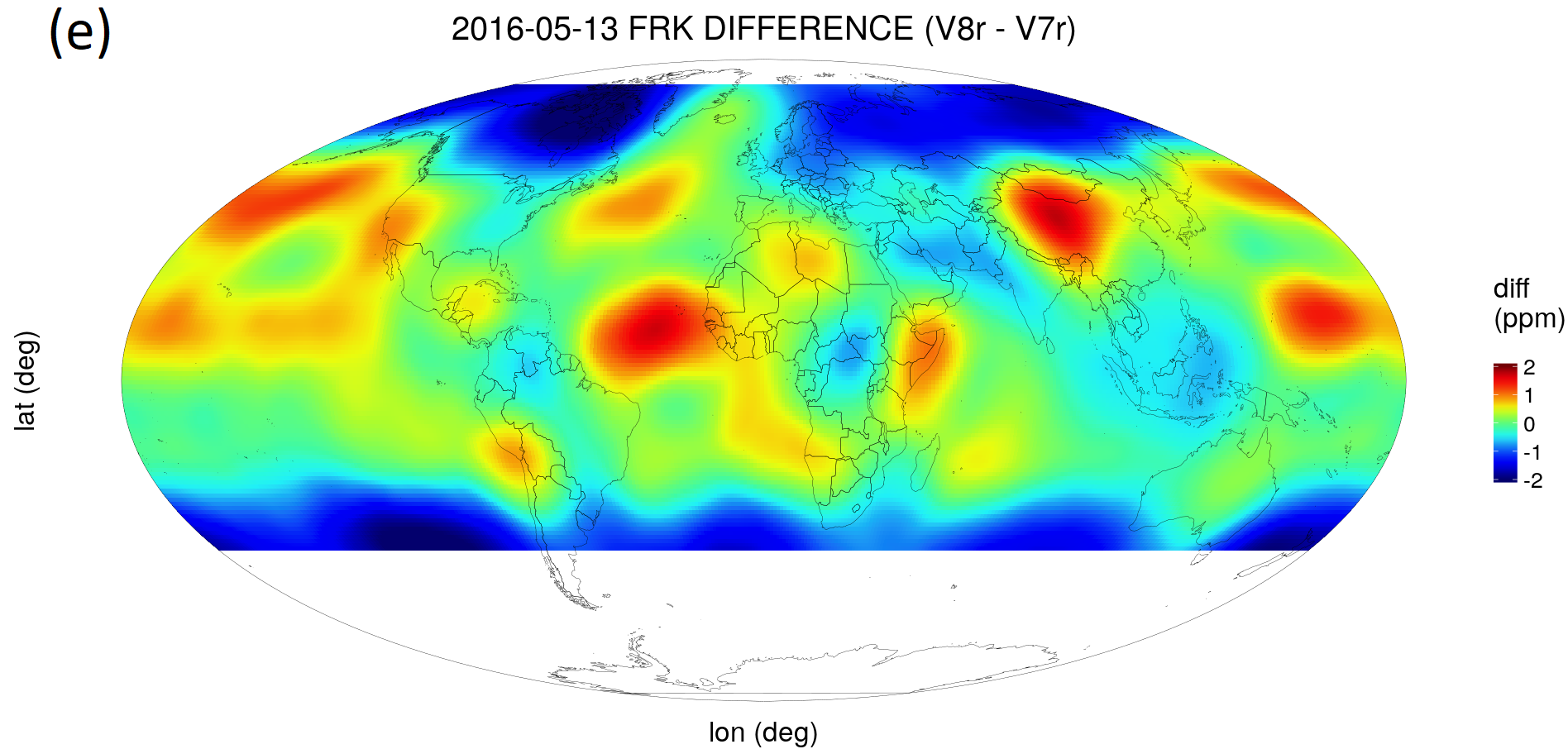}
   \includegraphics[width=3in]{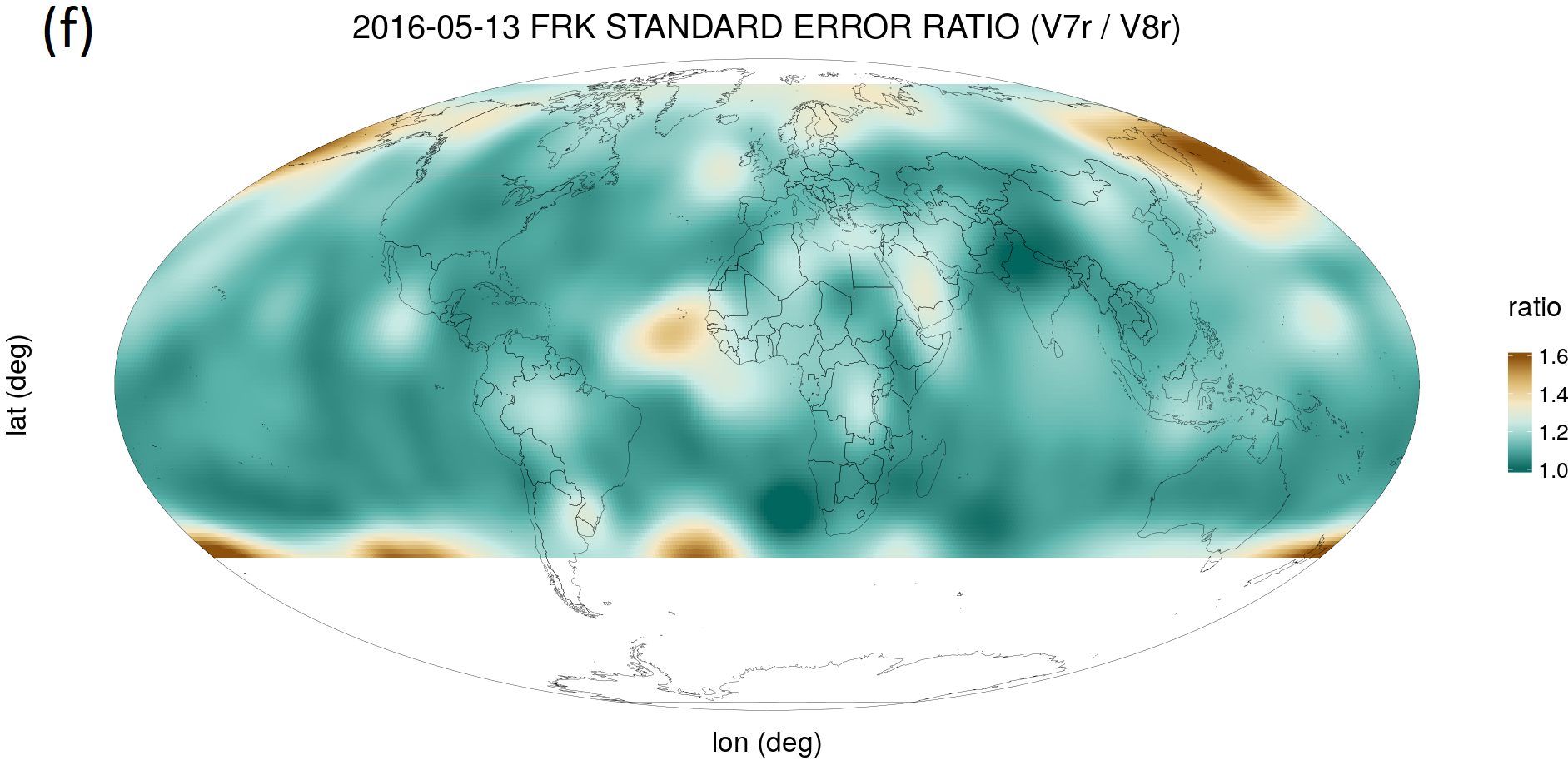}
	\caption{Fixed rank kriging (FRK) predictions of XCO$_2$ for 13 May 2016. (\textbf{a}) Orbiting Carbon Observatory-2 (OCO-2) Version 7r Lite File XCO$_2$ data with warn levels $\le 15$ on 13 May 2016; (\textbf{b})~same as ({a}), but for data between 6 May 2016 and 21 May 2016, inclusive; (\textbf{c}) global FRK Version 7r prediction \blue{(pred)} of XCO$_2$; (\textbf{d}) global FRK Version 7r prediction standard errors \blue{(s.e.)} ; (\textbf{e}) the difference \blue{(diff)} between the FRK Version 8r predictions and those of FRK Version 7r for 13 May 2016; (\textbf{f}) the ratio of the FRK Version 7r prediction standard errors to those of FRK Version 8r for 13 May 2016. \label{fig:Level3}}
\end{figure}

Predictions and prediction standard errors of the FRK Version 7r product for 13 May 2016, as~well as the retrievals obtained on that day and all retrievals in the 16-day window centred on that day (from~the Version 7r Lite File), are shown in Figure \ref{fig:Level3}a--d. The difference between the Version 8r and Version 7r predictions and the ratio of the prediction standard errors are shown in Figure~\ref{fig:Level3}e,f, respectively. From Figure \ref{fig:Level3}e we see that there are substantial differences in the predictions, with~regional variations on the order of 2 ppm. As expected, the prediction standard errors for the FRK Version 8r product are consistently lower than those from Version 7r. This is mostly due to there being more Version 8r retrievals with which to generate the Level 3 product.

Note that the FRK prediction in Figure \ref{fig:Level3}c appears smooth. This is partly due to our imposing a minimum retrieval standard error of 2 ppm (recall Appendix \ref{app:A3}) but also because the optimal prediction is unlikely to resemble what one would expect from a typical transport model (recall Section \ref{sec:smooth}). The quality of Level 3 products can only be properly assessed through validation, using~diagnostics. At least one of these diagnostics should take uncertainty into account. In Section \ref{sec:spaces} we introduced one such diagnostic, coverage. In the next section, we show why it is important to use data that has been corrected for bias when assessing coverage of the resulting product from those data.

\subsection{A Coverage Diagnostic in the Presence of Measurement Bias}\label{sec:coverage}

Recall that the true field (here, column-averaged CO$_2$) is denoted as $Y(\svec;t)$, at spatial location $\svec$ and time point $t$. 
A retrieval at $(\svec_i;t_i)$ results in the observation $Z(\svec_i;t_i)$ that is equal to the true value $Y(\svec_i;t_i)$ plus measurement error. In the case of raw OCO-2 retrievals, it is likely that the measurement error, $\epsilon_i$, has non-zero mean, resulting in a biased retrieval. The probabilistic considerations that follow can account for this by assuming that the bias at location $\svec_i$ and time $t_i$ is $\mu_\epsilon(\svec_i; t_i)$. 


Under an assumption of spatio-temporally varying measurement bias, the retrieval error, $$Z(\svec_i;t_i)-Y(\svec_i;t_i),$$
is (approximately) $N(\mu_\epsilon(\svec_i; t_i),\sigma_{\epsilon,i}^2)$. Hence, with probability 0.95, the error $Z(\svec_i;t_i)-Y(\svec_i;t_i)$ belongs to the interval, $(\mu_\epsilon(\svec_i; t_i) -1.96\sigma_{\epsilon,i}, \ \mu_\epsilon(\svec_i; t_i) + 1.96\sigma_{\epsilon,i})$. Equivalently, the true XCO$_2$ value, $Y(\svec;t)$, lies~in the random interval,
\begin {equation}\label {eq:1} (Z(\svec_i;t_i)- \mu_\epsilon(\svec_i; t_i)-1.96\sigma_{\epsilon,i},~ Z(\svec_i;t_i)-\mu_\epsilon(\svec_i; t_i) + 1.96\sigma_{\epsilon,i}), 
\end{equation} 
with probability 0.95. We refer to Equation \eqref{eq:1} as a 95\% prediction interval. This simple result is very important: It says that the measurement should \textit{first be corrected for bias}; then the 95\% prediction interval can be obtained by adding and subtracting $1.96\sigma_{\epsilon,i}$. In the context of OCO-2, this means that one should generate products based on the bias-corrected Lite Files, and not the raw retrievals, in~order to ensure reliable coverages.

Section \ref{sec:pedag} discusses how spatio-temporal kriging can be carried out to obtain a prediction $\hat{Y}(\svec^*;t^*)$ of $Y(\svec^*;t^*)$, even though there may be no data at $(\svec^*;t^*)$. There, we assumed that $\mu_\epsilon(\svec^*; t^*) = 0$; however, more generally, $\hat{Y}(\svec^*;t^*)$ is an unbiased predictor of $Y(\svec^*;t^*) + \mu_\epsilon(\svec^*; t^*)$. Further, the~kriging variance is:
\begin{equation*}
\sigma^2_k(\svec^*;t^*) = \E(\hat{Y}(\svec^*;t^*) - Y(\svec^*;t^*)-\mu_\epsilon(\svec^*; t^*))^2 = \var(\hat{Y}(\svec^*;t^*) -Y(\svec^*;t^*)),
\end{equation*}
since $\E(\hat{Y}(\svec^*;t^*)) = \E(Y(\svec^*;t^*)) + \mu_\epsilon(\svec^*; t^*)$. Hence, the prediction error, 
$$\hat{Y}(\svec^*;t^*)-Y(\svec^*;t^*),$$
has mean $\mu_\epsilon(\svec^*; t^*)$ and variance $\sigma^2_k(\svec^*;t^*)$. Provided this error is (approximately) normally distributed, we can see that what led to Equation \eqref{eq:1} also leads to the following (approximate) 95\% kriging prediction interval \textit{at any} location $(\svec^*;t^*)$, not just at retrieval locations. With probability 0.95, the true XCO$_2$ value at any given location $(\svec^*;t^*)$ lies in the random interval, 
\begin{equation*}
(\hat{Y}(\svec^*;t^*) - \mu_\epsilon(\svec^*; t^*) - 1.96\sigma_k(\svec^*;t^*),~~\hat{Y}(\svec^*;t^*) - \mu_\epsilon(\svec^*; t^*) + 1.96\sigma_k(\svec^*;t^*)).
\end{equation*} 

These same ideas can be used to compare two different XCO$_2$ values at $(\svec^*;t^*)$ that are each attempting to predict 
$Y(\svec^*;t^*)$. We denote the two predictors as $\hat{Y}_1(\svec^*;t^*)$ and $\hat{Y}_2(\svec^*;t^*)$ with prediction errors, $\hat{Y}_1(\svec^*;t^*) - Y(\svec^*;t^*)$ and $\hat{Y}_2(\svec^*;t^*) - Y(\svec^*;t^*)$, respectively, which have (approximately) normal distributions, $N(\mu_{\epsilon,1}(\svec^*; t^*),\sigma^2_{k,1})$ and $N(\mu_{\epsilon,2}(\svec^*; t^*),\sigma^2_{k,2})$, respectively. Then, a comparison of the two predictors is given by, 
\begin{equation}\label{eq:3}
\Delta(\svec^*;t^*) \equiv \hat{Y}_1(\svec^*;t^*) - \hat{Y}_2(\svec^*;t^*) = \left(\hat{Y}_1(\svec^*;t^*) - Y(\svec^*;t^*)\right) - \left(\hat{Y}_2(\svec^*;t^*) - Y(\svec^*;t^*)\right).
\end{equation}

From Equation \eqref{eq:3},
\begin{align*}
\mu_\Delta(\svec^*; t^*) &\equiv \E(\hat{Y}_1(\svec^*;t^*) - \hat{Y}_2(\svec^*;t^*)) = \mu_{\epsilon,1}(\svec^*; t^*)-\mu_{\epsilon,2}(\svec^*; t^*),\\
\sigma_\Delta^2 &\equiv \var (\hat{Y}_1(\svec^*;t^*) - \hat{Y}_2(\svec^*;t^*)) = \sigma_{k,1}^2 + \sigma^2_{k,2} - 2\rho\sigma_{k,1}\sigma_{k,2},
\end{align*}
where $\rho$ is the correlation between the two prediction errors. If $\hat{Y}_1$ and $\hat{Y}_2$ are from two different prediction methods based on the same measurements, then $\rho$ will be non-zero in general. 

However, in many important cases $\rho = 0$, which greatly simplifies the calculation of \mbox{$\sigma^2_\Delta = (\sigma_{k,1}^2 + \sigma^2_{k,2})$}. One such case is when $\hat{Y}_1(\svec^*;t^*)$ is a kriging predictor from OCO-2 data and $\hat{Y}_2(\svec^*;t^*)$ is a TCCON observation at location $(\svec^*;t^*)$. In this case, for known $\mu_{\epsilon,1}, \mu_{\epsilon,2}$, the~two prediction errors are statistically independent because the random component of the OCO-2 measurement error is independent of the TCCON measurement. Generally, the random interval, 
\begin{equation}\label{eq:4}
(\hat{Y}_1(\svec^*;t^*) - \hat{Y}_2(\svec^*;t^*)-\mu_\Delta(\svec^*; t^*) - 1.96 \sigma_\Delta,~~\hat{Y}_1(\svec^*;t^*) - \hat{Y}_2(\svec^*;t^*)-\mu_\Delta(\svec^*; t^*) + 1.96 \sigma_\Delta),
\end{equation}
contains zero with probability 0.95. If one consistently observes the interval not straddling zero, it~is an important diagnostic indicating that something is not right with $\mu_\Delta(\svec^*; t^*)$ or $\sigma_\Delta$. If initially $\mu_\Delta(\svec^*; t^*)$ is set equal to zero, and the problem is subsequently diagnosed as an undetected bias term, it can be added to the zero bias and the diagnosis based on Equation \eqref{eq:4} repeated. If the fraction of intervals containing zero (i.e., the coverage) is larger than 0.95, then the prediction interval is said to be conservative, which indicates that $\sigma_\Delta$ is too large. If the coverage is smaller than 0.95, the prediction interval is said to be liberal, which indicates that $\sigma_\Delta$ is too small (when it comes to prediction intervals, being conservative is preferable to being liberal). 


Different kriging predictors (i.e., different Level 3 products) can be compared via their coverage when compared to TCCON observations. A predictor with small RMSPE might actually have poor coverage, resulting in misleading inferences that declare a ``signal'' to be present when in fact it is not. Of course, some assumptions need to be made about the measurement biases, $\mu_{\epsilon,1}(\svec^*; t^*)$ and $\mu_{\epsilon,2}(\svec^*; t^*)$, in practice. When comparing to TCCON, it is reasonable to assume that $\mu_{\epsilon,2}(\svec^*; t^*) = 0$ at all TCCON locations (i.e., that TCCON measurements are unbiased). Then, $\mu_{\epsilon,1}(\svec^*; t^*)$ can be estimated by fitting a classical multivariate linear model to the differences between some simple predictions based on the raw OCO-2 retrievals, and other predictions that may be partially based on TCCON data if desired. This is what is done when constructing the Lite Files, where the regressors for $\mu_{\epsilon,1}(\svec^*; t^*)$ are based on physical attributes such as surface pressure and aerosol abundance. Once this estimate is made, it is then treated as fixed. Strictly, if TCCON data are used to estimate $\mu_{\epsilon,1}$ (as is done for constructing the Lite Files) then this estimate also depends on \blue{$\hat{Y}_2$}. However, allowing for this induced dependence is beyond the the scope of this paper. In what follows, we assume that the products generated from the Lite Files are unbiased in space and time and hence that coverages properly derived from this product are valid. 


\subsection{Comparison to TCCON Data}

TCCON is a ground-based network designed to provide observations of XCO$_2$ that can be directly compared to OCO-2 retrievals. In this study, we use TCCON data from the GGG2014 database \cite{Wunch_2017_TCCON}, specifically data collected from 24 stations (listed in Table \ref{tab:station_results1} in Appendix \ref{app:B1}), six in the Southern Hemisphere and 18 in the Northern Hemisphere \cite{Sherlock2014ll,Griffith2014wg,DeMaziere2014ra,Griffith2014db,Feist2014ae,Dubey2014ma,Blumenstock2014iz,Kawakami2014js,Wennberg2014ci,Iraci2014df,Morino2014tk,Goo2014ay,Wennberg2014oc,Morino2014rj,Wennberg2014pa,Sussmann2014gm,Warneke2014or,Te2014pr,Hase2014ka,Notholt2014br,Deutscher2014bi,Kivi2014so,Notholt2014ny,Strong2014eu}. We only consider the TCCON measurements at a station that fall within a 60-min time-window centred on the average local crossing time of the OCO-2 satellite over that station (usually in the early afternoon local time). The remaining TCCON measurements were then aggregated by site and by day in the same way OCO-2 was aggregated on the $1 \times 1 \times 1$ lon-lat-day grid; see Section \ref{sec:OCO2_preproc}. 

\begin{figure}[H]
	\centering
	\includegraphics[width=3in]{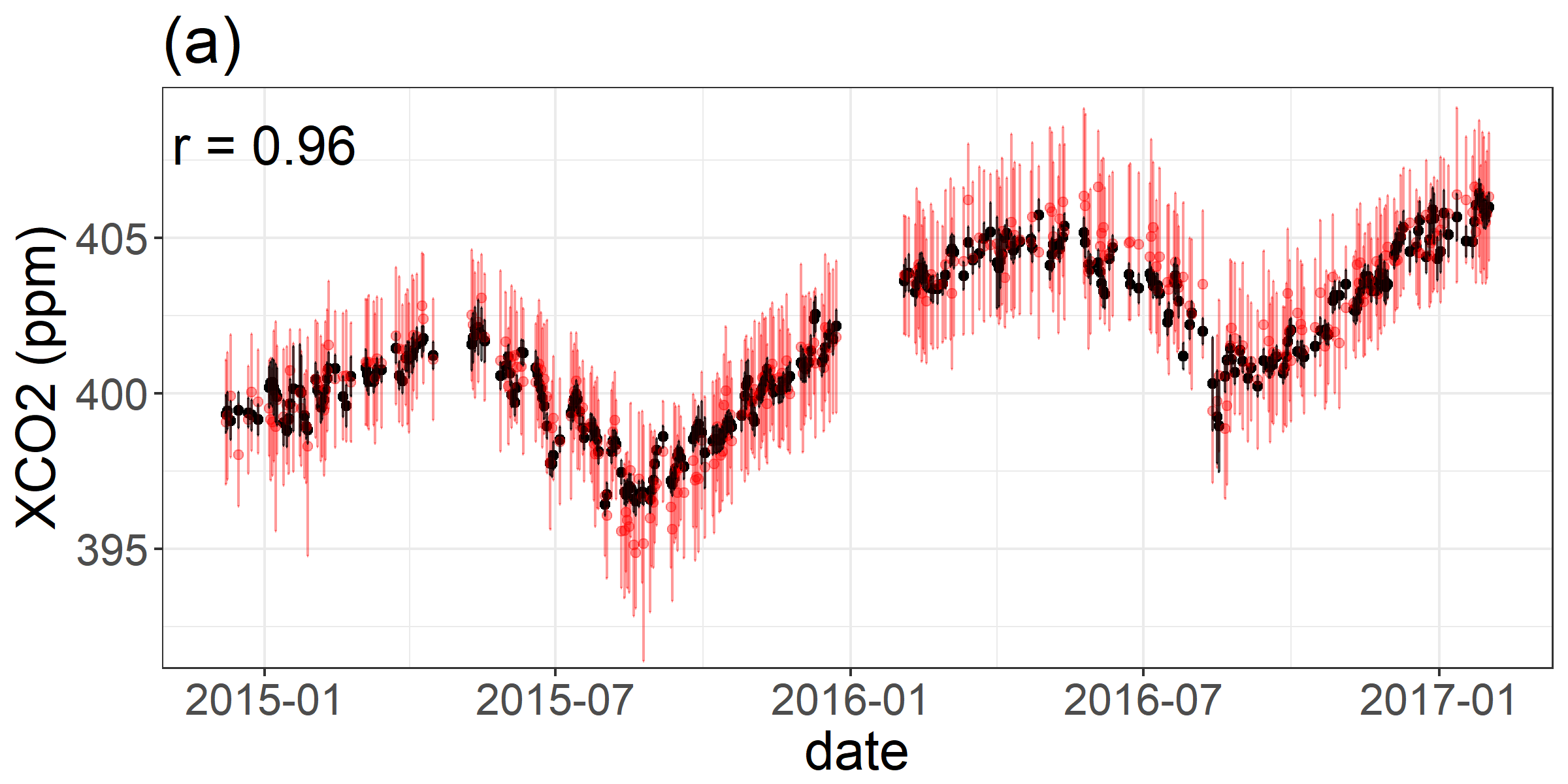}
	\includegraphics[width=3in]{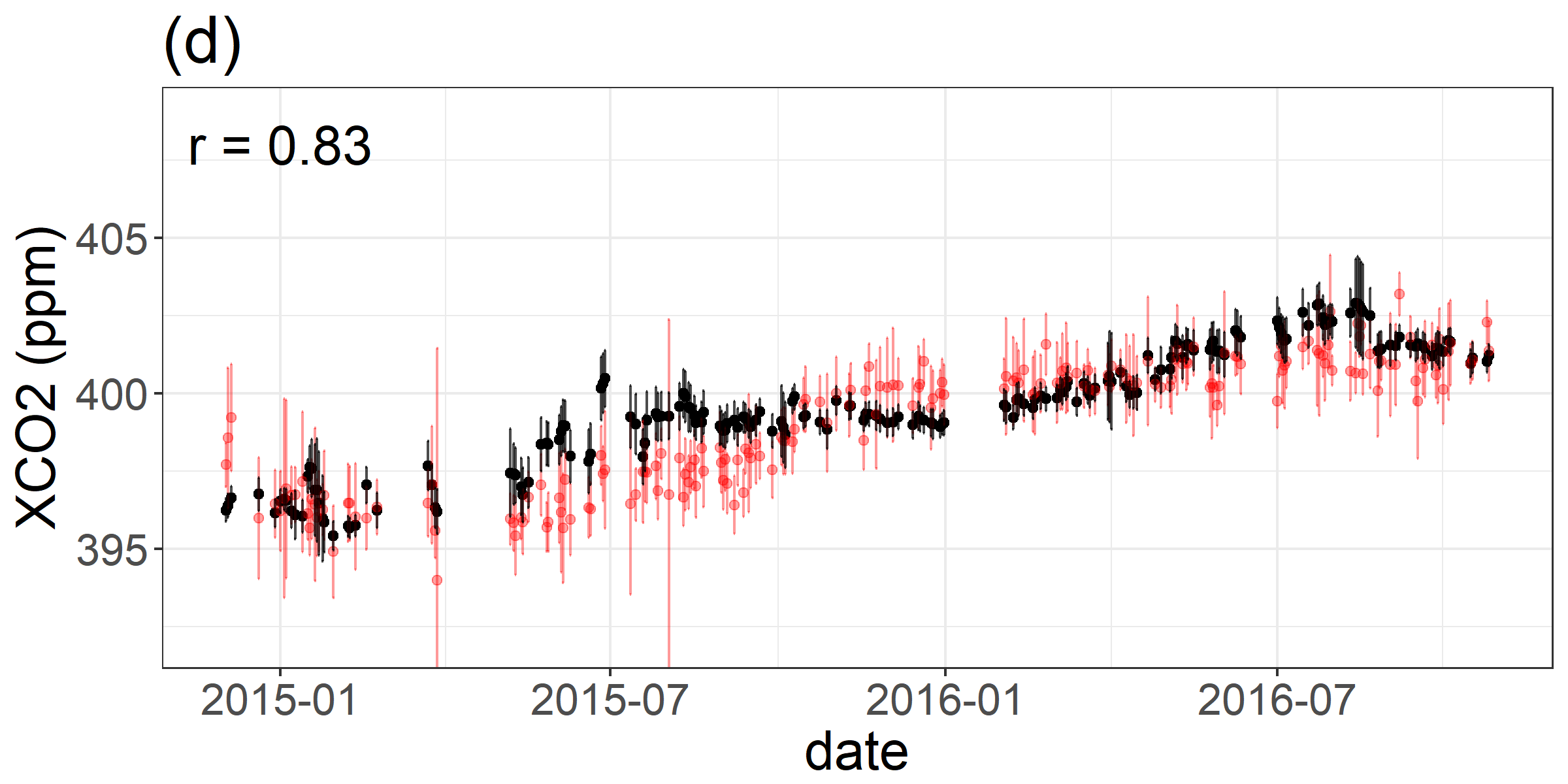}
	\includegraphics[width=3in]{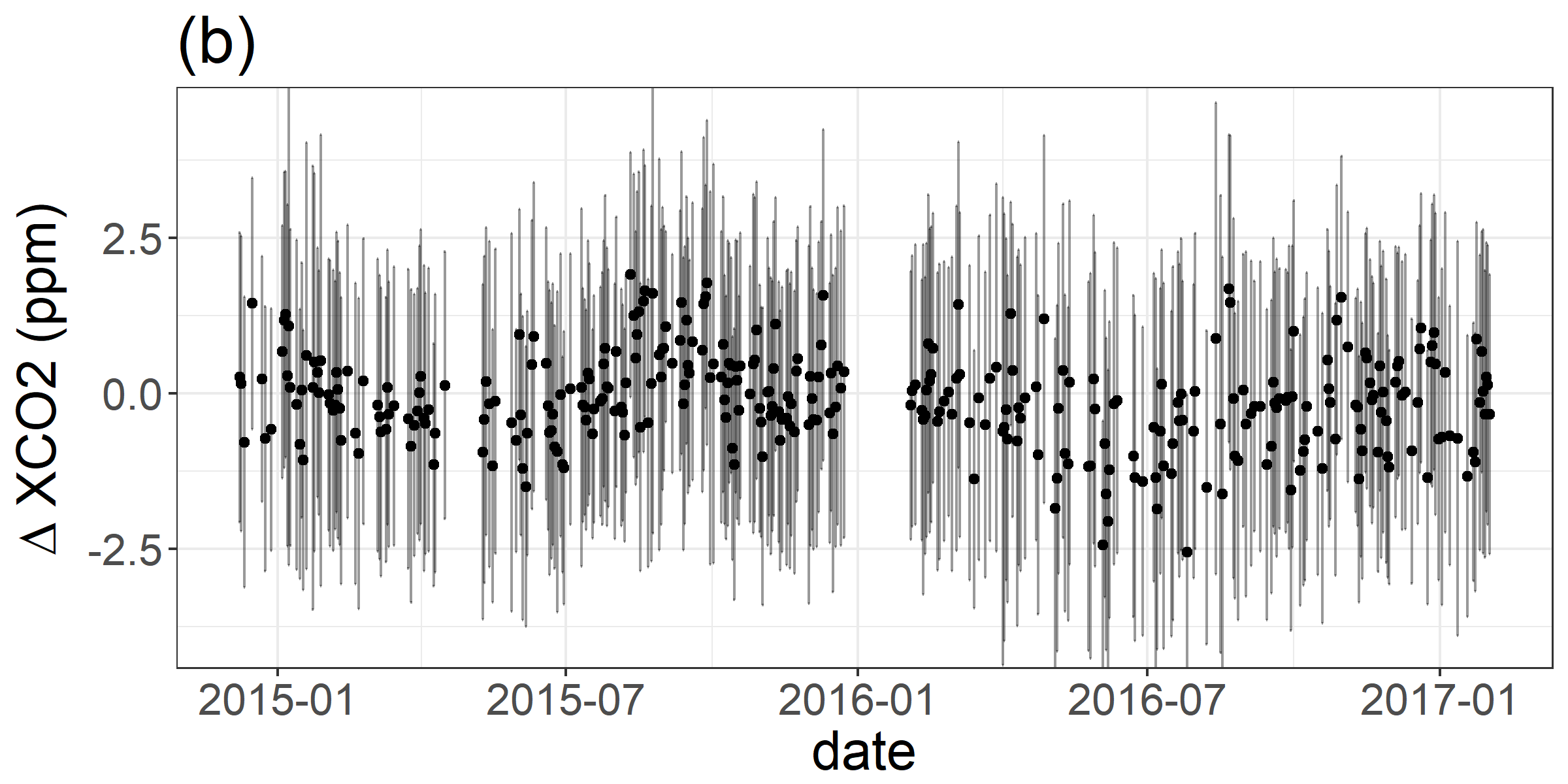}
   	\includegraphics[width=3in]{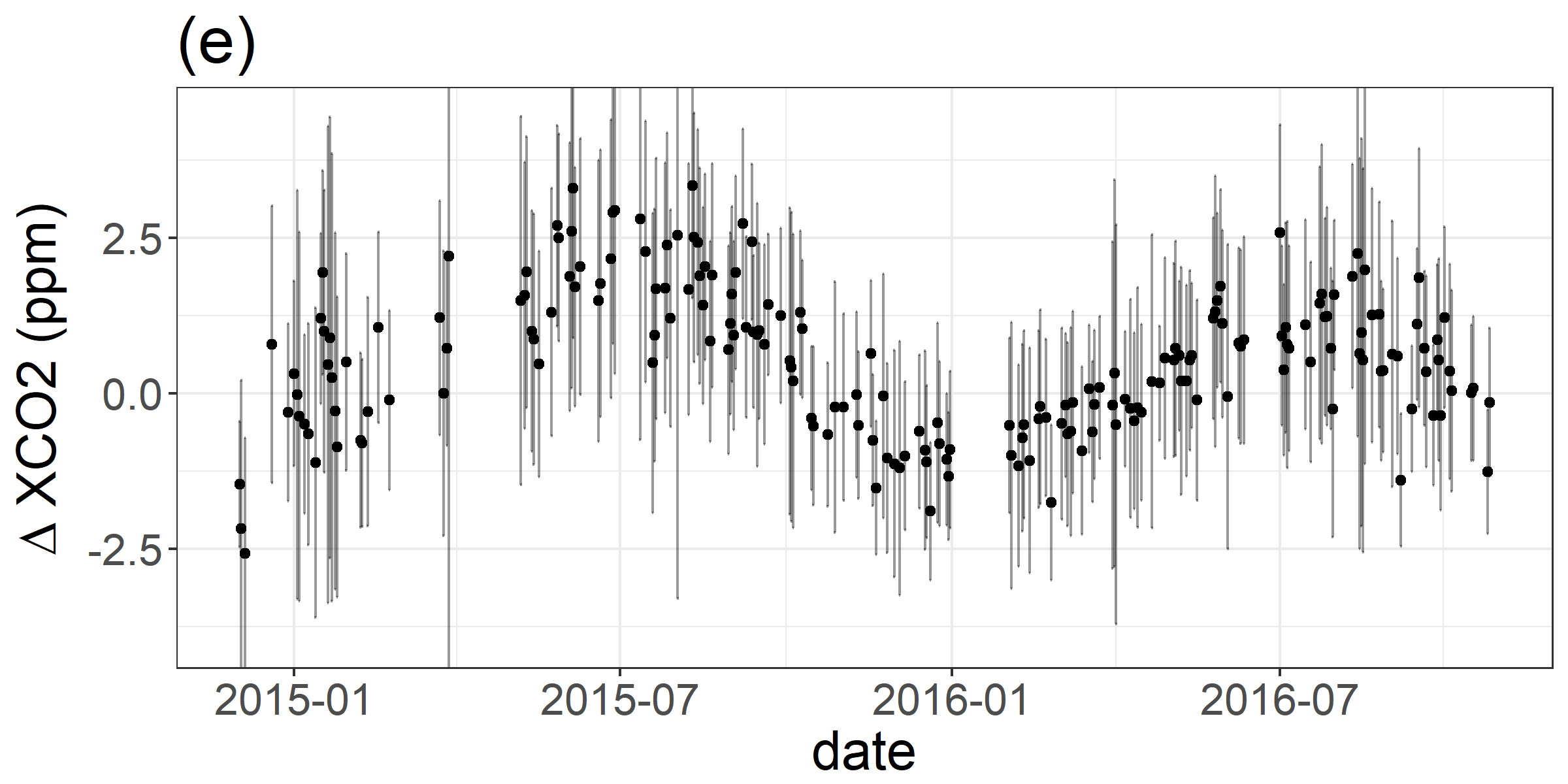}
   \includegraphics[width=3in]{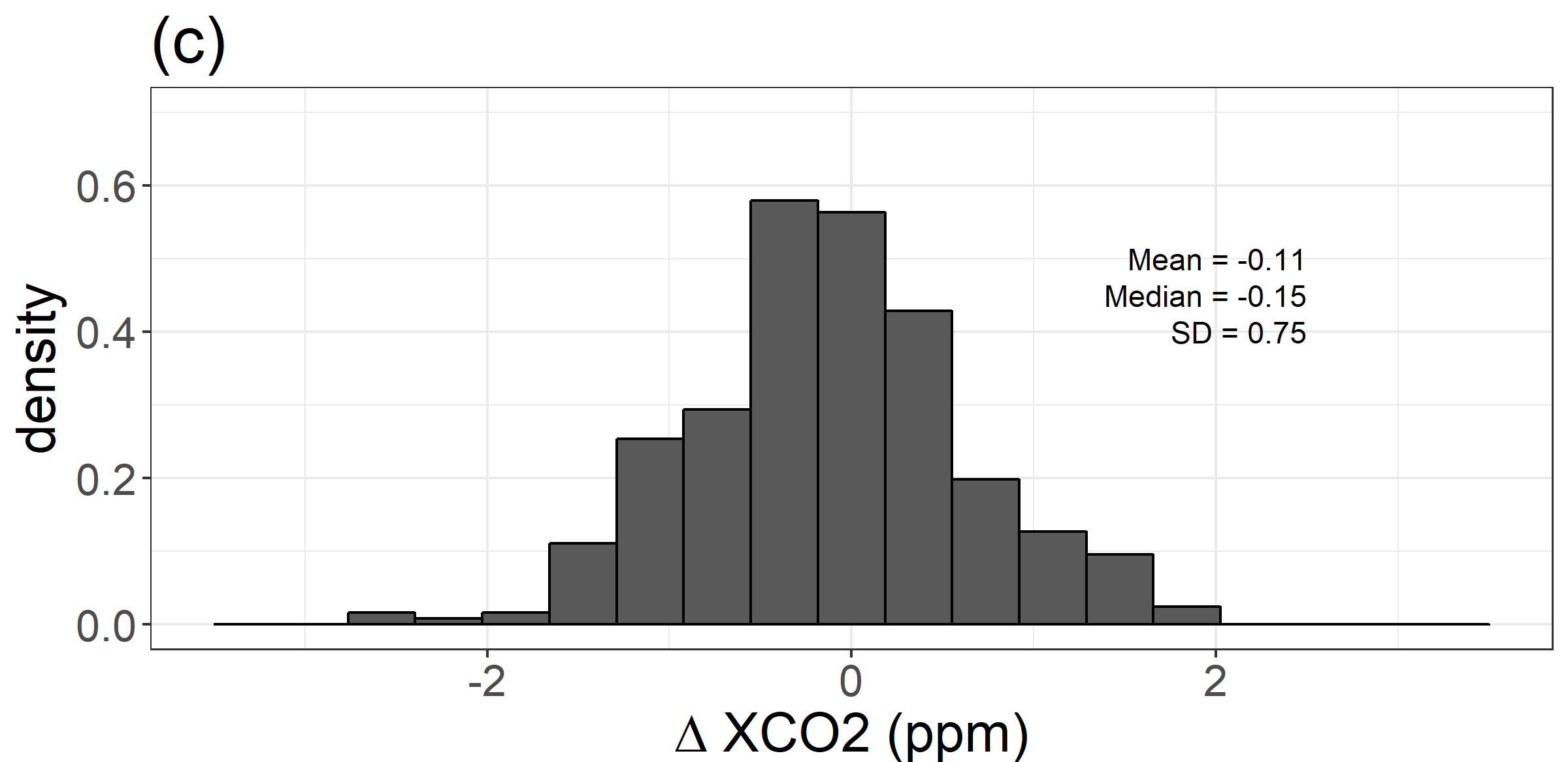}
   \includegraphics[width=3in]{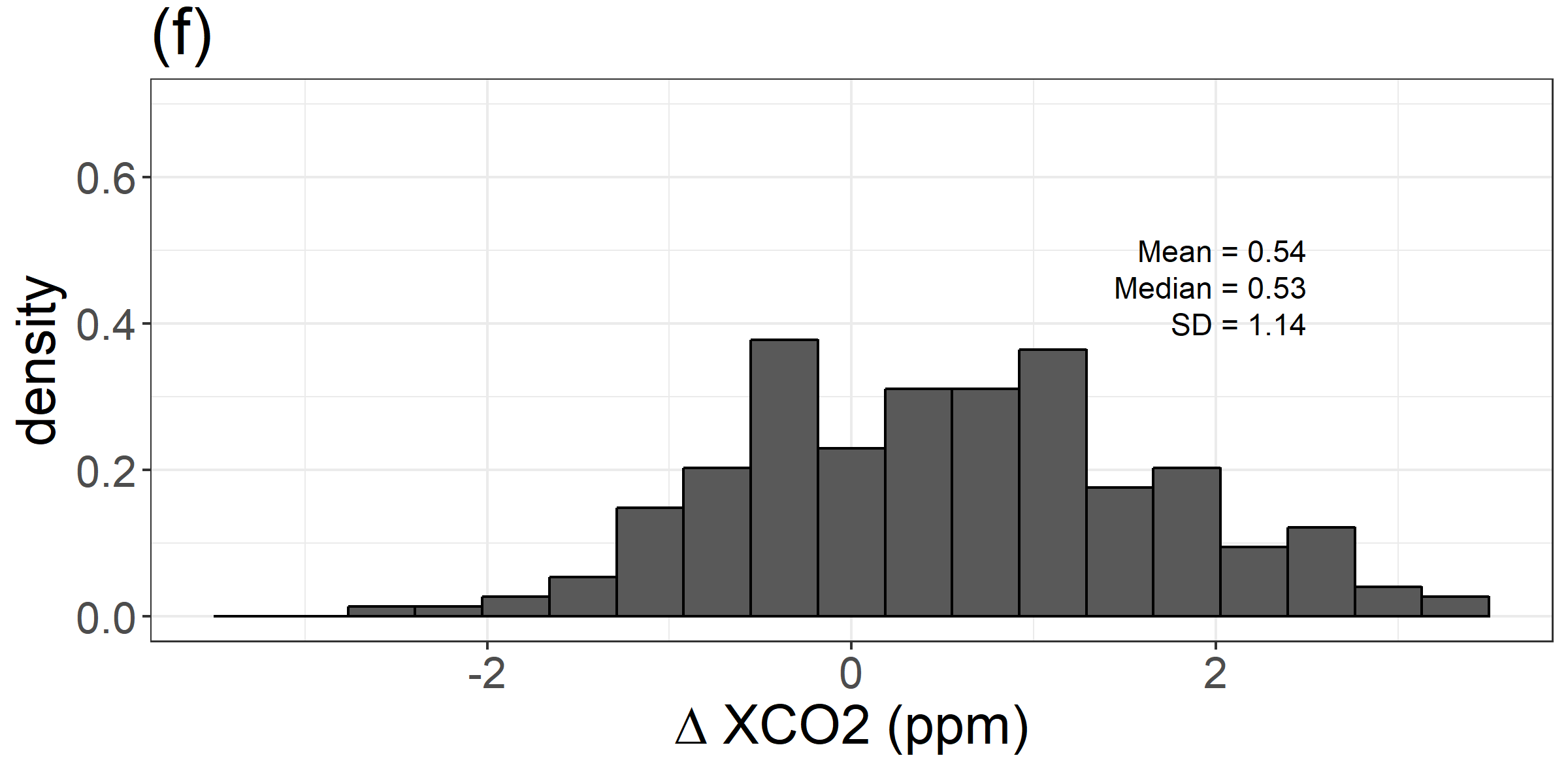}   
	\caption{Comparison between the Total Carbon Column Observing Network (TCCON) measurements of XCO$_2$ and the fixed rank kriging (FRK) Version 7r product. (\textbf{a}) Time series plots for the TCCON data (red) and the FRK product (black) at Lamont, Oklahoma. The black bars are the TCCON data $\pm 2$ times TCCON's measurement-error standard deviation, while the red error bars are the FRK predictions $\pm 2$ times the prediction standard error, respectively. (\textbf{b}) The differences between the TCCON measurements and the FRK predictions at Lamont, where the error bars are the difference $\pm 2$ times the square-root of the sum of the TCCON error variance and the FRK prediction variance. (\textbf{c})~Histogram of the differences between the TCCON measurements and the FRK predictions at Lamont. (\textbf{d}--\textbf{f}) Same as ({a}--{c}), but for the TCCON station at Wollongong, Australia. \label{fig:Lamont_Woll}}
\end{figure}

Unlike \cite{Wunch_2017, Liang_2017}, we do not rely on coincidence criteria to compare the FRK Level 3 products to TCCON measurements, since the Level 3 products are global and at a daily resolution. The availability of global daily Level 3 products increases the number of comparisons we can make, since FRK inferences on XCO$_2$ can be made for every $1 \times 1 \times 1$ lon-lat-day cube when TCCON data are available (except for days when OCO-2 is not making retrievals for eight consecutive days or longer, or when the TCCON station falls outside the latitude range of the retrievals in the 16-day window). In total, we have 3607 FRK Version 7r predictions and 4156 FRK Version 8r predictions that we can compare to TCCON values, globally, between October 2014 and February 2017. However, to make the comparisons fair, we~only consider Version 7r and Version 8r predictions that are associated with the same locations in space and time. This reduces the total number of comparisons to 3510. The FRK Version 7r predictions and the TCCON data at two stations (Lamont and Wollongong), together with their differences, are~shown in Figure \ref{fig:Lamont_Woll}a--f. Note how the `spread' of the differences at the two stations is similar, but~that the differences at the Wollongong station have a distinct seasonal cycle. This cycle is also apparent when using the coincidence criteria in (Figure~A1 (s) \cite{Wunch_2017}) and when using the Version 8r data (not~shown). 

\begin{figure}[H]
	\centering
	\includegraphics[height=2.3in]{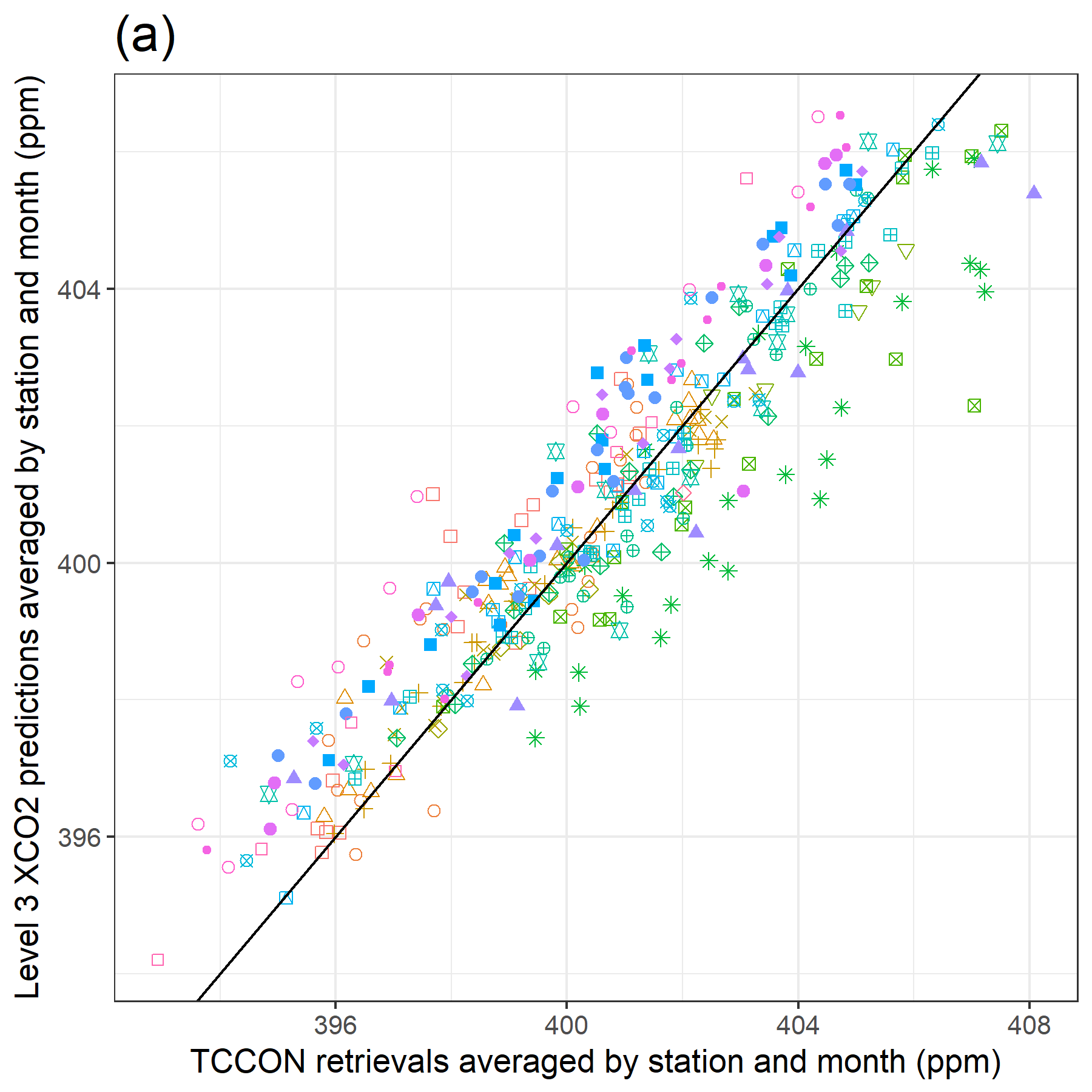}
   \includegraphics[height=2.3in]{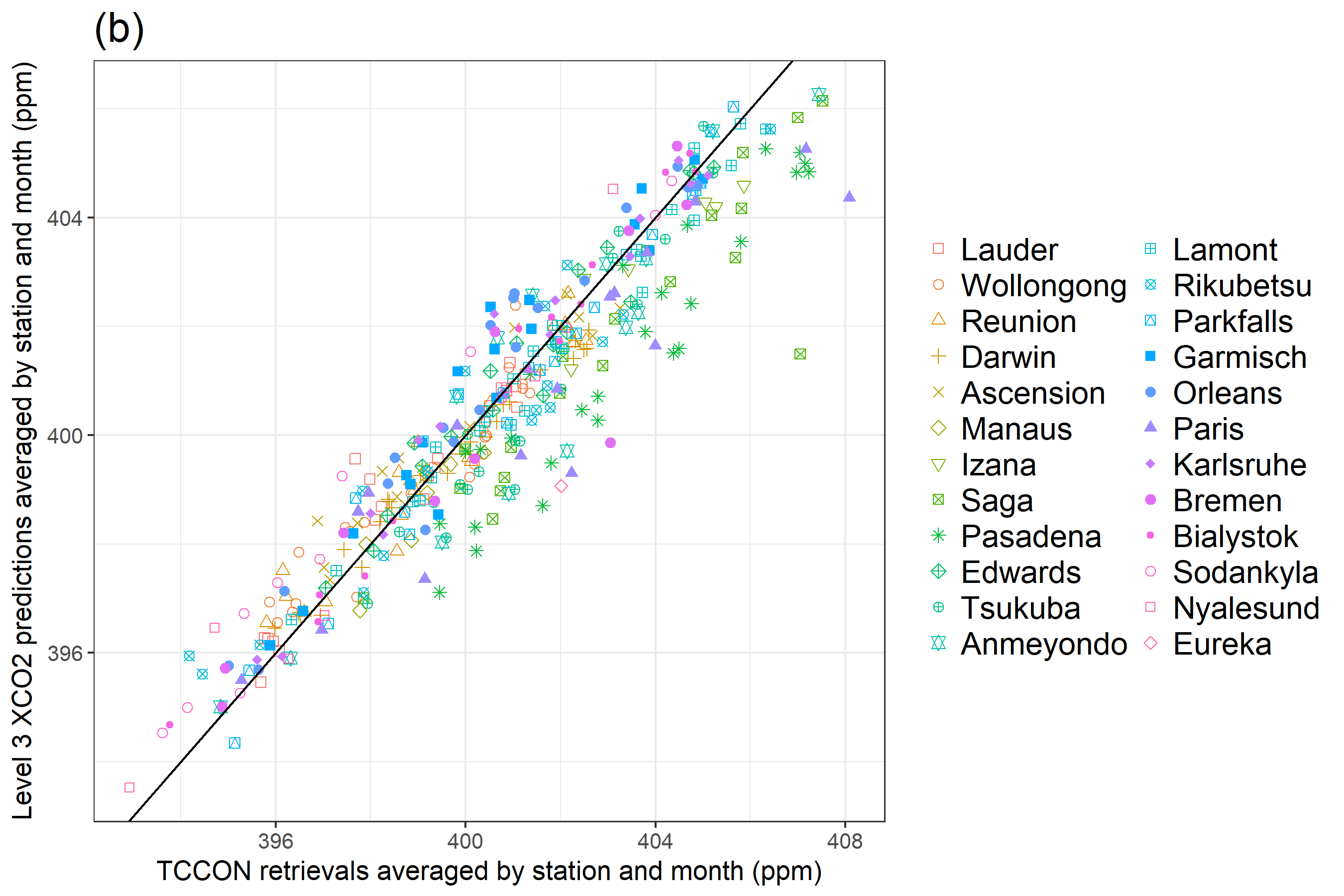}
	\caption{Scatter plots of monthly averages of the Total Carbon Column Observing Network (TCCON) measurements vs.~the fixed rank kriging (FRK) predictions, for 24 TCCON stations between October 2014 and February 2017. The line with unit slope constrained to pass through the origin is also shown (black solid line). (\textbf{a}) Comparison of TCCON data with the FRK Version 7r product; (\textbf{b}) comparison of TCCON data with the FRK Version 8r product. \label{fig:TCCON_differences}}
\end{figure}

Scatter plots showing the mean differences by station and month for FRK Versions 7r and 8r are given in Figure \ref{fig:TCCON_differences}a,b, respectively. The differences are randomly spread around the unit line, which~is what one would expect from Level 3 products generated from bias-corrected Level 2 retrievals. There~are two problematic stations in FRK Version 7r: those at Sodankyl{\"a} and Pasadena. The former is very far north and poses challenges to the OCO-2 retrieval algorithm due to high solar zenith angles and snowy scenes, while the TCCON station at Pasadena is situated in a megacity, namely greater Los Angeles \cite{Wunch_2017}. The city acts as a fine-scale strong source of carbon dioxide emissions that cannot be adequately captured in the Level 3 products that recall are on a $1 \times 1$ lon-lat grid. However, we remark that the Level 3 predictions corroborate well with the retrievals from the Edwards station, which is only about 100 km away from Pasadena. We observe that predictions at Sodankyl{\"a} from the FRK Version 8r product are much improved, and a comparison of Figure \ref{fig:TCCON_differences}a,b reveals a slightly smaller spread around the unit line for the FRK Version 8r product.

Detailed diagnostics by station for the FRK Version 7r product are given in Table \ref{tab:station_results1} in Appendix~\ref{app:B1}. These include the MPE, the MAPE, the RMSPE, the coefficient of determination ($R^2$), the slope of the regression line constrained to pass through zero (Slope) and the empirical 95\% coverage (95\% Cov.).

The Version 7r RMSPEs given in Table \ref{tab:station_results1} are close to, but overall an improvement over, those~provided by \cite{Wunch_2017} (who also considered the Version 7r Lite File with warn levels $\le 15$), where~only OCO-2 retrievals coincident with TCCON measurements were considered. Unfortunately, detailed~comparisons to \cite{Wunch_2017} cannot be made due to the intrinsically different methodologies and the different number of observations on which the diagnostics are based. However, the agreement is reassuring, since our comparisons are made for days when OCO-2 retrievals are not necessarily coincident with TCCON measurements. The empirical coverage is also reasonable in most cases.

Similar diagnostics are available for the FRK Version 8r product in Table \ref{tab:station_results2} in Appendix \ref{app:B1}. A~comparison of the two tables reveals an overall improvement of the FRK Version 8r product over the Version 7r product at the TCCON sites, with MPEs at Wollongong and Lauder in the Southern Hemisphere \blue{substantially reduced}. This corroborates the anticipated improvements of Version 8r over Version 7r that include, amongst other things, accounting for stratospheric aerosols in the Level~2~algorithm. 

Summary diagnostics across all TCCON stations for both FRK products are given in Table \ref{tab:station_results_all}, with Pasadena included or excluded in the summaries. For the FRK Version 7r product, the overall agreement between the FRK prediction and TCCON is quite good, with a global MPE (bias) of 0.08~ppm, an RMSPE of 1.36 ppm, and coefficient of determination $R^2 = 0.80$. The slope of the regression line (treating TCCON as a perfect covariate) forced to pass through zero is 1.00019. If Pasadena is omitted from the statistics obtained based on all stations, then the global MPE (bias) increases slightly to 0.35~ppm, but the RMSPE decreases to 1.15 ppm. This compares well with \cite{Chevallier_2017} who also omitted Pasadena from the study, but considered OCO-2 retrievals over a smaller time horizon (comparisons there found a similar global bias as we found, and an RMSPE of 1.5 ppm). For the FRK Version 8r product, all diagnostics fared better than those for the FRK Version 7r product. Uncertainty was also better captured: an 86\% empirical coverage was achieved when Pasadena was included, and 92\% was achieved when excluded.


\begin{table}[H]
 \caption{Summary statistics of the differences between the fixed rank kriging predictions and the Total Carbon Column Observing Network (TCCON) measurements between 1 October 2014 and 28 February 2017 and across 24 TCCON stations \blue{(or 23 without Pasadena -- w/o Pas.)}. Summary statistics include the mean prediction error (MPE), the mean absolute error (MAPE), the root-mean-squared prediction error (RMSPE), the coefficient of determination ($R^2$), the slope (Slope) of the regression line constrained to pass through $(0,0)$ and the empirical 95\% coverage (95\% Cov.). The number of observations considered in each row is denoted as~$N$.\label{tab:station_results_all}}  
	\centering
	
\begin{tabular}{cccccccc}
 \toprule
 & \textbf{N} & \textbf{MPE (Bias)} & \textbf{MAPE} & \textbf{RMSPE} & \textbf{R$^2$} & \textbf{Slope} & \textbf{95\% Cov.} \\
 \midrule
 Total v7r & 3510 & 0.08 & 1.02 & 1.36 & 0.80 & 1.000 & 0.80 \\ 
Total v7r (w/o Pas.
) & 3067 & 0.35 & 0.88 & 1.15 & 0.85 & 1.001 & 0.85 \\ 
 \midrule
 Total v8r & 3510 & $-$0.22 & 0.85 & 1.16 & 0.86 & 0.999 & 0.86 \\ 
Total v8r (w/o Pas.) & 3067 & 0.01 & 0.71 & 0.94 & 0.89 & 1.000 & 0.92 \\ 
 \bottomrule
\end{tabular}
\end{table}

\section{Conclusions}\label{sec:discussion}

In this article, we have presented statistical approaches to generating Level 3 products that contain maps of predictions and prediction standard errors, from satellite remote sensing retrievals. We showed that these products are likely to appear `smooth', but that this is necessary to achieve optimality. We also showed that various local approaches to generating the Level 3 products each have their own strengths and weaknesses. In particular, we showed that moving-window methods can be expected to produce better predictions than `blocking' methods, despite their theoretical limitations, but that their performance can suffer in the presence of low SNR. When both data size and SNR are an issue, we showed that reduced-rank methods such as FRK are a viable and attractive way forward.

We used FRK to compare OCO-2 retrievals in the Version 7r and Version 8r Lite Files to TCCON data. The advantage of FRK over `coincident methods' \cite{Wunch_2017,Liang_2017} is that it increases the number of comparisons one can make, and it obviates the need to `extend' the coincident region to ensure that there are sufficient retrievals to make a comparison possible. We found that the Level 3 FRK maps from OCO-2 retrievals have favourable diagnostics at validation (in this case, TCCON) locations, both~in terms of prediction accuracy and in terms of uncertainty quantification. We also found that FRK Version 8r fared much better than FRK Version 7r on all diagnostics we considered. This improvement was expected since the bias correction in the Lite Files for Version 8r is based on more data than that for Version 7r. Furthermore, one should refrain from concluding that the FRK Version 8r product is superior to the Version 7r product globally, since estimates of the bias corrections in the Lite Files do use TCCON data; \blue{out-of-sample validation data} would be required to make this claim.

The FRK product we generated used a 16-day moving window to predict \blue{XCO$_2$, the column-averaged CO$_2$,} on the eighth day. Hence, with this product, one is not able to obtain prediction standard errors over temporal aggregates (say monthly averages of XCO$_2$). To remedy this, one would need to retain spatio-temporal predictions, prediction variances and covariances, over a time span that is at least as large as that of the desired temporal aggregation level (which might necessitate the use of a larger window). This would require considerable computational effort and possibly a different modelling framework. 

In this article, we used the statistical technique of FRK to generate the Level 3 products, in~part because it is fast and has been seen to work well with satellite remote sensing data elsewhere (e.g.,~\cite{Nguyen_2014b}), and it extends elegantly to predictions of aggregations of the process (i.e., change-of-support). Recent~years have seen the development of other methods built on the concept of dimension-reduction that may find use in the generation of statistical Level 3 products. These include fitting a stochastic partial differential equation model through finite elements \cite{Lindgren_2011} and multi-resolution approximations to Gaussian processes \cite{Katzfuss_2017}. There are variants of FRK that impose a sparse structure on $\Kmat^{-1}$ \cite{Nychka_2015} to speed up computations and model finer scales; these have also been shown to work well in practice. Irrespective of the statistical method used for generating a Level 3 product, diagnostics such as those presented here should be used for validation. In Section \ref{sec:coverage}, we show how the important notion of coverage can be used to assess uncertainty quantification, specifically the prediction standard errors, of a Level 3 product.

{R Software} \cite{R} code and instructions for reproducing the results in this paper can be found at \url{https://github.com/andrewzm/oco2-frk}.

\vspace{6pt} 

\acknowledgments{N.C.'s
 research was supported by a 2015--2017 Australian Research Council Discovery Grant, Number DP150104576. The OCO-2 data were produced by the OCO-2 project at the Jet Propulsion Laboratory, California Institute of Technology, and obtained from the OCO-2 data archive maintained at the NASA Goddard Earth Science Data and Information Services Center: \url{http://disc.gsfc.nasa.gov/}. TCCON data were obtained from the TCCON data archive hosted by the Carbon Dioxide Information Analysis Center, Oak Ridge National Laboratory: \url{http://tccondata.org/}.}

\authorcontributions{A.Z.-M. and N.C. wrote the paper. A.Z.-M. performed the experiments for Section \ref{sec:pedag}. C.S.~did the data pre-processing and generated the Level 3 products for Section \ref{sec:OCO-2}.}

\conflictsofinterest{The authors declare no conflict of interest.} 

\abbreviations{The following abbreviations are used in this manuscript:\\

\noindent 
\begin{tabular}{@{}ll}
 CO$_2$ & carbon dioxide \\
 FRK & fixed rank kriging\\
 GOSAT & Greenhouse gases Observing SATellite \\
 MAPE & mean absolute prediction error \\
 MPE & mean prediction error \\
 OCO-2 & Orbiting Carbon Observatory-2\\
 RMSPE & root-mean-squared prediction error\\
 SNR & signal-to-noise ratio\\
 TCCON & Total Carbon Column Observing Network \\
 XCO$_2$ & column-averaged carbon dioxide
\end{tabular}}

\appendixtitles{no} 
\appendixsections{multiple} 
\appendix
\section{}
\appendixtitles{yes}
\subsection{Variance Reduction of the Smoother} \label{app:A1}

In this section, we prove that $\var(D_{E(Y|Z)}(\svec_1^*,\svec_2^*,\svec_3^*)) = \var(D_{Y}(\svec_1^*,\svec_2^*,\svec_3^*)) - c$, where $c \blue{~\ge~} 0$. Let~$\Yvec^* \equiv (Y(\svec_1^*),Y(\svec_2^*),Y(\svec_3^*))'$. The law of total variance states that for any two random vectors $\Yvec^*$ and $\Zvec$,
\begin{equation}\label{eq:TotalVar}
\E(\var(\Yvec^* \mid \Zvec)) = \var(\Yvec^*) - \var(\E(\Yvec^* \mid \Zvec)).
\end{equation}

Therefore, for any vector $\avec \in \mathbb{R}^3$,
$$
\avec'\E(\var(\Yvec^* \mid \Zvec))\avec = \avec'\var(\Yvec^*)\avec - \avec'\var(\E(\Yvec^* \mid \Zvec))\avec \ge 0.
$$

In our simulations, the processes are Gaussian, and hence, the conditional variance does not depend on the data. Therefore, $\E(\var(\Yvec^* \mid \Zvec)) = \var(\Yvec^* \mid \Zvec)$, and:
$$
\avec'\var(\E(\Yvec^* \mid \Zvec))\avec = \avec'\var(\Yvec^*)\avec - \avec'\var(\Yvec^* \mid \Zvec)\avec \ge 0.
$$

Setting $\avec = \frac{1}{h^2}(1,-2,1)'$ and defining $\sigma^2_{D_Y} \equiv \avec'\var(\Yvec^*)\avec$ and $c \equiv \avec'\var(\Yvec^* \mid \Zvec)\avec$ completes the~proof.

\subsection{Recovering the Optimal Predictor with FRK} \label{app:A2}

In this section, we show that FRK can be set up in such a way to yield the optimal predictor exactly. Suppose we have a process, $\{Y(\svec): \svec \in D\}$, with known covariance function $C_Y(\svec,\uvec)$, for $\svec,\uvec \in D$. Let $\svec_1,\dots,\svec_m$ be the $m$ observation locations and define,
\begin{align*}
\cvec(\svec^*)' \equiv (C_Y(\svec^*,\svec_i): i = 1,\dots, m),\\
\Cmat \equiv (C_Y(\svec_i, \svec_j): i,j = 1,\dots, m).
\end{align*}

Recall the measurement model,
$$
Z(\svec_i) = Y(\svec_i) + \epsilon_i,\quad i = 1,\dots,m,
$$
where $Y(\cdot)$ and $\{\epsilon_i\}$ are independent, $\{\epsilon_i\}$ are mutually independent and where $\epsilon_i \sim \Gau(0,\sigma^2_\epsilon)$ has known, constant variance $\sigma^2_\epsilon$. The aim here is to define a process $\{\tilde{Y}(\svec):\svec \in D\},$ decomposed~as $\tilde{Y}(\svec) = \sum_{i=1}^r\phi_i(\svec)\eta_i$, that has the same optimal predictor as that of the original process. That is, we~require that: 
$$
\E(Y(\svec^*) \mid \Zvec) = \E(\tilde{Y}(\svec^*) \mid \Zvec).
$$

The decomposition that achieves this is similar to that used in the {predictive process} \cite{Banerjee_2008}. In~particular, let $\phib(\svec^*)' = \cvec(\svec^*)'(\Cmat + \sigma^2_\epsilon\Imat)^{-1}$, and let $\etab \mid \Zvec \sim \Gau(\Zvec,\Kmat)$, with $\Kmat = (\Cmat + \sigma^2_\epsilon\Imat)$. Then, 
\begin{equation}
\tilde{Y}(\svec^*) = \cvec(\svec^*)'(\Cmat + \sigma^2_\epsilon\Imat)^{-1}\etab.\label{eq:Ystar}
\end{equation}

The conditional expectation of Equation \eqref{eq:Ystar} with respect to $\Zvec$ is:
$$
\E(\tilde{Y}(\svec^*) \mid \Zvec) = \cvec(\svec^*)'(\Cmat + \sigma^2_\epsilon\Imat)^{-1}\Zvec = \E(Y(\svec^*) \mid \Zvec),
$$
which is the optimal predictor; see \cite{Rasmussen_2006} (Chapter~2). Interestingly, this setup of FRK will yield the following prediction variance: 
$$
\var(\tilde{Y}(\svec^*) \mid \Zvec) = \cvec(\svec^*)'(\Cmat + \sigma^2_\epsilon\Imat)^{-1} \cvec(\svec^*),
$$
which can be shown to be equal to the variance of the optimal predictor $ \E(Y(\svec^*) \mid \Zvec)$, rather than the conditional variance (conditional on $\Zvec$) of $Y(\svec^*)$, and in this case, it is important that the variances are interpreted as such. 

\subsection{Predictor `Smoothness' Increases with Measurement-Error Variance} \label{app:A3}

In this section, we prove that the smoothness of the optimal predictor (the conditional expectation) increases with measurement-error variance.

Recall from Appendix \ref{app:A1} that the relative smoothness (expected magnitude of second-order derivatives) of the optimal predictor with respect to that of the process decreases as $c$ increases. Moreover, in Appendix \ref{app:A1}, we noted that $c \equiv \avec'\var(\Yvec^* \mid \Zvec)\avec$ and $\avec = \frac{1}{h^2}(1,-2,1)'$. Now, from the discussion in Appendix \ref{app:A2}, we see that: 
$$\var(\E(\Yvec^* \mid \Zvec)) = (\Cmat^{*})'(\Cmat + \sigma^2_\epsilon\Imat)^{-1} \Cmat^*,$$
where $\Cmat^* \equiv (\cvec(\svec_1^*),\cvec(\svec_2^*),\cvec(\svec_3^*))$ is assumed to have full rank. Therefore:
$$
c = \avec'(\var(\Yvec^*) - (\Cmat^{*})'(\Cmat + \sigma^2_\epsilon\Imat)^{-1} \Cmat^*) \avec.
$$ 

Consider two measurement-error variances $\sigma^2_{\epsilon,1}$ and $\sigma^2_{\epsilon,2}$, with $\sigma^2_{\epsilon,2} > \sigma^2_{\epsilon,1}$. Define:
\begin{align*}
c_1 \equiv \avec'(\var(\Yvec^*) - (\Cmat^{*})'(\Cmat + \sigma^2_{\epsilon,1}\Imat)^{-1} \Cmat^*) \avec,\\
c_2 \equiv \avec'(\var(\Yvec^*) - (\Cmat^{*})'(\Cmat + \sigma^2_{\epsilon,2}\Imat)^{-1} \Cmat^*) \avec.
\end{align*}

We can use simple linear algebra to show that $\sigma^2_{\epsilon,2} > \sigma^2_{\epsilon,1}$ implies $c_2 > c_1$, as follows. The~matrices $(\Cmat + \sigma^2_{\epsilon,2}\Imat) > (\Cmat + \sigma^2_{\epsilon,1}\Imat)$, in the sense that their difference is a positive-definite matrix. Hence,~$(\Cmat + \sigma^2_{\epsilon_B}\Imat)^{-1} < (\Cmat + \sigma^2_{\epsilon_A}\Imat)^{-1}$, in the sense that their difference is a negative-definite matrix, and consequently:
\begin{align*}
c_1 - c_2 &= \avec'(\Cmat^{*})'[(\Cmat + \sigma^2_{\epsilon,2}\Imat)^{-1} - (\Cmat + \sigma^2_{\epsilon,1}\Imat)^{-1}  ]\Cmat^*\avec < 0.
\end{align*}

Hence, $c_2 > c_1$, as required. 
\appendixtitles{no}
\section{} \label{app:B1}


\begin{table}[H]
\renewcommand\thetable{A1}
	\caption{Summary statistics of the differences between the fixed rank kriging Version 7r product predictions and the Total Carbon Column Observing Network measurements between 1 October 2014 and 28 February 2017. Summary statistics include the mean prediction error (MPE), the mean absolute prediction error (MAPE), the root-mean-squared prediction error (RMSPE), the coefficient of determination ($R^2$), the slope (Slope) of the regression line constrained to pass \textls[-10]{through (0,0) and the empirical 95\% coverage (95\% Cov.). The number of observations considered in each row is denoted as}~$N$.\label{tab:station_results1}}
	\centering
\begin{tabular}{cccccccc}
 \toprule
\textbf{Station Name} & \textbf{N} & \textbf{MPE (Bias)} & \textbf{MAPE} & \textbf{RMSPE} & \textbf{R$^2$} & \textbf{Slope} & \textbf{95\% Cov.} \\
 \midrule
 Eureka & 5 & $-$1.49 & 1.49 & 1.72 & 0.98 & 0.996 & 0.60 \\ 
 \blue{Ny {\AA}lesund} & 31 & 1.15 & 1.30 & 1.57 & 0.92 & 1.003 & 0.81 \\ 
 \blue{Sodankyl{\"a}} & 112 & 2.08 & 2.08 & 2.28 & 0.94 & 1.005 & 0.29 \\ 
 Bialystok & 106 & 1.22 & 1.25 & 1.46 & 0.94 & 1.003 & 0.62 \\ 
 Bremen & 23 & 1.28 & 1.45 & 1.61 & 0.93 & 1.003 & 0.70 \\ 
 Karlsruhe & 103 & 1.05 & 1.17 & 1.42 & 0.92 & 1.003 & 0.66 \\ 
 Paris & 81 & 0.06 & 1.14 & 1.40 & 0.84 & 1.000 & 0.84 \\ 
 Orleans & 146 & 0.96 & 1.09 & 1.31 & 0.90 & 1.002 & 0.71 \\ 
 Garmisch & 101 & 1.14 & 1.25 & 1.47 & 0.89 & 1.003 & 0.59 \\ 
 Parkfalls & 190 & 0.51 & 0.82 & 1.08 & 0.91 & 1.001 & 0.87 \\ 
 Rikubetsu & 56 & 0.29 & 0.86 & 1.11 & 0.91 & 1.001 & 0.86 \\ 
 Lamont & 342 & $-$0.11 & 0.59 & 0.76 & 0.93 & 1.000 & 0.99 \\ 
 Anmeyondo & 48 & 0.33 & 1.22 & 1.43 & 0.84 & 1.001 & 0.73 \\ 
 Tsukuba & 137 & $-$0.25 & 0.96 & 1.29 & 0.72 & 0.999 & 0.87 \\ 
 Edwards & 337 &$ -$0.03 & 0.79 & 0.97 & 0.85 & 1.000 & 0.92 \\ 
 Pasadena & 443 & $-$1.82 & 1.98 & 2.32 & 0.71 & 0.995 & 0.44 \\ 
 Saga & 76 & $-$0.86 & 1.09 & 1.36 & 0.89 & 0.998 & 0.86 \\ 
 Izana & 17 & $-$1.08 & 1.08 & 1.23 & 0.90 & 0.997 & 0.71 \\ 
 Manaus & 38 & $-$0.05 & 0.66 & 0.85 & 0.47 & 1.000 & 0.97 \\ 
 Ascension & 210 & 0.32 & 0.68 & 0.91 & 0.79 & 1.001 & 0.99 \\ 
 Darwin & 284 & $-$0.06 & 0.49 & 0.63 & 0.92 & 1.000 & 1.00 \\ 
 Reunion & 243 & 0.23 & 0.58 & 0.71 & 0.91 & 1.001 & 0.94 \\ 
 Wollongong & 201 & 0.54 & 1.02 & 1.26 & 0.69 & 1.001 & 0.82 \\ 
 Lauder & 180 & 0.71 & 0.84 & 1.15 & 0.83 & 1.002 & 0.82 \\ 
 \bottomrule
\end{tabular}
\end{table}

\begin{table}[H]
\renewcommand\thetable{A2}
\caption{Same as Table \ref{tab:station_results1}, but for the fixed rank kriging Version 8r product. \label{tab:station_results2}}
	\centering
\begin{tabular}{cccccccc}
 \toprule
\textbf{Station Name} & \textbf{N} & \textbf{MPE (Bias)} & \textbf{MAPE} & \textbf{RMSPE} & \textbf{R$^2$} & \textbf{Slope} & \textbf{95\% Cov.} \\
 \midrule
 Eureka & 5 & $-$1.48 & 1.59 & 2.00 & 0.73 & 0.996 & 0.40 \\ 
 \blue{Ny {\AA}lesund} & 31 & 0.60 & 1.22 & 1.64 & 0.83 & 1.002 & 0.71 \\ 
 \blue{Sodankyl{\"a}} & 112 & 0.77 & 0.96 & 1.23 & 0.94 & 1.002 & 0.70 \\ 
 Bialystok & 106 & 0.18 & 0.60 & 0.76 & 0.95 & 1.000 & 0.94 \\ 
 Bremen & 23 & 0.18 & 0.86 & 1.12 & 0.91 & 1.000 & 0.87 \\ 
 Karlsruhe & 103 & 0.33 & 0.76 & 1.01 & 0.92 & 1.001 & 0.82 \\ 
 Paris & 81 & $-$0.63 & 1.22 & 1.55 & 0.82 & 0.998 & 0.72 \\ 
 Orleans & 146 & 0.31 & 0.79 & 0.97 & 0.89 & 1.001 & 0.79 \\ 
 Garmisch & 101 & 0.50 & 0.85 & 1.06 & 0.89 & 1.001 & 0.80 \\ 
 Parkfalls & 190 & $-$0.13 & 0.71 & 0.93 & 0.91 & 1.000 & 0.94 \\ 
 Rikubetsu & 56 & 0.00 & 0.90 & 1.07 & 0.91 & 1.000 & 0.91 \\ 
 Lamont & 342 & $-$0.22 & 0.59 & 0.75 & 0.93 & 0.999 & 0.99 \\ 
 Anmeyondo & 48 & $-$0.30 & 1.10 & 1.42 & 0.85 & 0.999 & 0.85 \\ 
 Tsukuba & 137 & $-$0.56 & 1.08 & 1.40 & 0.72 & 0.999 & 0.84 \\ 
 Edwards & 337 & 0.15 & 0.60 & 0.77 & 0.91 & 1.000 & 0.97 \\ 
 Pasadena & 443 & $-$1.77 & 1.86 & 2.15 & 0.79 & 0.996 & 0.44 \\ 
 Saga & 76 & $-$1.20 & 1.25 & 1.52 & 0.91 & 0.997 & 0.78 \\ 
 Izana & 17 & $-$0.92 & 0.96 & 1.08 & 0.88 & 0.998 & 0.65 \\ 
 Manaus & 38 & $-$0.35 & 0.62 & 0.77 & 0.69 & 0.999 & 1.00 \\ 
 Ascension & 210 & 0.36 & 0.69 & 0.89 & 0.82 & 1.001 & 1.00 \\ 
 Darwin & 284 & $-$0.17 & 0.51 & 0.61 & 0.94 & 1.000 & 1.00 \\ 
 Reunion & 243 & 0.00 & 0.51 & 0.62 & 0.93 & 1.000 & 0.99 \\ 
 Wollongong & 201 & 0.12 & 0.70 & 0.86 & 0.84 & 1.000 & 0.94 \\ 
 Lauder & 180 & 0.16 & 0.43 & 0.61 & 0.92 & 1.000 & 1.00 \\ 
 \bottomrule
\end{tabular}
\end{table}

\reftitle{References}


\begin{thebibliography}{999}

\bibitem[Chevallier \em{et~al.}(2017)Chevallier, Broquet, Pierangelo, and
 Crisp]{Chevallier_2017}
Chevallier, F.; Broquet, G.; Pierangelo, C.; Crisp, D.
\newblock Probabilistic global maps of the {CO$_2$} column at daily and monthly
 scales from sparse satellite measurements.
\newblock {\em J. Geophys. Res. Atmos.} {\bf 2017}, {\em
 122},~7614--7629.

\bibitem[Tiwari \em{et~al.}(2006)Tiwari, Gloor, Engelen, Chevallier,
 R{\"o}denbeck, K{\"o}rner, Peylin, Braswell, and Heimann]{Tiwari_2006}
Tiwari, Y.K.; Gloor, M.; Engelen, R.J.; Chevallier, F.; R{\"o}denbeck, C.;
 K{\"o}rner, S.; Peylin, P.; Braswell, B.H.; Heimann, M.
\newblock {Comparing CO$_2$ retrieved from Atmospheric Infrared Sounder with
 model predictions: Implications for constraining surface fluxes and
 lower-to-upper troposphere transport}.
\newblock {\em J. Geophys. Res.~Atmos.} {\bf 2006}, {\em
 111}, doi:10.1029/2005JD006681.

\bibitem[Hammerling \em{et~al.}(2012)Hammerling, Michalak, O'Dell, and
 Kawa]{Hammerling_2012}
Hammerling, D.M.; Michalak, A.M.; O'Dell, C.; Kawa, S.R.
\newblock Global CO$_2$ distributions over land from the Greenhouse Gases
 Observing Satellite (GOSAT).
\newblock {\em Geophys. Res. Lett.} {\bf 2012}, {\em 39}, doi:10.1029/2012GL051203.

\bibitem[Inoue \em{et~al.}(2013)Inoue, Morino, Uchino, Miyamoto, Yoshida,
 Yokota, Machida, Sawa, Matsueda, Sweeney, Tans, Andrews, Biraud, Tanaka,
 Kawakami, and Patra]{Inoue_2013}
Inoue, M.; Morino, I.; Uchino, O.; Miyamoto, Y.; Yoshida, Y.; Yokota, T.;
 Machida, T.; Sawa, Y.; Matsueda,~H.; Sweeney, C.; et al.
\newblock {Validation of XCO$_2$ derived from SWIR spectra of GOSAT TANSO-FTS
 with aircraft measurement data}.
\newblock {\em Atmos. Chem. Phys.} {\bf 2013}, {\em
 13},~9771--9788.

\bibitem[Butz \em{et~al.}(2011)Butz, Guerlet, Hasekamp, Schepers, Galli, Aben,
 Frankenberg, Hartmann, Tran, Kuze, Keppel-Aleks, Toon, Wunch, Wennberg,
 Deutscher, Griffith, Macatangay, Messerschmidt, Notholt, and
 Warneke]{Butz_2011}
Butz, A.; Guerlet, S.; Hasekamp, O.; Schepers, D.; Galli, A.; Aben, I.;
 Frankenberg, C.; Hartmann, J.M.; Tran,~H.; Kuze, A.; et al.
\newblock Toward accurate CO$_2$ and CH$_4$ observations from GOSAT.
\newblock {\em Geophys. Res. Lett.} {\bf 2011}, {\em 38}, doi:10.1029/2011GL047888.

\bibitem[Katzfuss and Cressie(2011)]{Katzfuss_2011}
Katzfuss, M.; Cressie, N.
\newblock {Spatio-temporal smoothing and EM estimation for massive
 remote-sensing data sets}.
\newblock {\em J. Time Ser. Anal.} {\bf 2011}, {\em 32},~430--446.

\bibitem[Zeng \em{et~al.}(2017)Zeng, Lei, Strong, Jones, Guo, Liu, Deng,
 Deutscher, Dubey, Griffith, Hase, Henderson, Kivi, Lindenmaier, Morino,
 Notholt, Ohyama, Petri, Sussmann, Velazco, and Wennberg]{Zeng_2017}
Zeng, Z.C.; Lei, L.; Strong, K.; Jones, D.B.A.; Guo, L.; Liu, M.; Deng, F.;
 Deutscher, N.M.; Dubey, M.K.; Griffith, D.W.T.; et al.
\newblock Global land mapping of satellite-observed CO$_2$ total columns using
 spatio-temporal geostatistics.
\newblock {\em Int. J. Digit. Earth} {\bf 2017}, {\em
 10},~426--456.

\bibitem[Nguyen \em{et~al.}(2014)Nguyen, Osterman, Wunch, O'Dell, Mandrake,
 Wennberg, Fisher, and Castano]{Nguyen_2014b}
Nguyen, H.; Osterman, G.; Wunch, D.; O'Dell, C.; Mandrake, L.; Wennberg, P.;
 Fisher, B.; Castano, R.
\newblock A~method for colocating satellite X$_{CO_2}$ data to ground-based
 data and its application to ACOS-GOSAT and TCCON.
\newblock {\em Atmos. Meas. Tech.} {\bf 2014}, {\em
 7},~2631--2644.

\bibitem[Jing \em{et~al.}(2014)Jing, Shi, Wang, and Sussmann]{Jing_2014}
Jing, Y.; Shi, J.; Wang, T.; Sussmann, R.
\newblock Mapping global atmospheric CO$_2$ concentration at high
 spatiotemporal resolution.
\newblock {\em Atmosphere} {\bf 2014}, {\em 5},~870--888.

\bibitem[Tadi{\'c} \em{et~al.}(2015)Tadi{\'c}, Qiu, Yadav, and
 Michalak]{Tadic_2015}
Tadi{\'c}, J.; Qiu, X.; Yadav, V.; Michalak, A.
\newblock Mapping of satellite Earth observations using moving window block
 kriging.
\newblock {\em Geosci. Model Dev.} {\bf 2015}, {\em
 8},~3311--3319.

\bibitem[Haas(1995)]{Haas_1995}
Haas, T.C.
\newblock Local prediction of a spatio-temporal process with an application to
 wet sulfate deposition.
\newblock {\em J. Am. Stat. Assoc.} {\bf 1995},
 {\em 90},~1189--1199.

\bibitem[Cressie and Johannesson(2008)]{Cressie_2008}
Cressie, N.; Johannesson, G.
\newblock {Fixed Rank Kriging} for very large spatial data sets.
\newblock {\em J. R. Stat. Soc. Ser. B} {\bf 2008},
 {\em 70},~209--226.

\bibitem[Nguyen \em{et~al.}(2012)Nguyen, Cressie, and Braverman]{Nguyen_2012}
Nguyen, H.; Cressie, N.; Braverman, A.
\newblock Spatial statistical data fusion for remote sensing applications.
\newblock {\em J. Am. Stat. Assoc.} {\bf 2012},
 {\em 107},~1004--1018.

\bibitem[Nguyen \em{et~al.}(2014)Nguyen, Katzfuss, Cressie, and
 Braverman]{Nguyen_2014a}
Nguyen, H.; Katzfuss, M.; Cressie, N.; Braverman, A.
\newblock Spatio-temporal data fusion for very large remote sensing datasets.
\newblock {\em Technometrics} {\bf 2014}, {\em 56},~174--185.

\bibitem[Alkhaled \em{et~al.}(2008)Alkhaled, Michalak, Kawa, Olsen, and
 Wang]{Alkhaled_2008}
Alkhaled, A.A.; Michalak, A.M.; Kawa, S.R.; Olsen, S.C.; Wang, J.W.
\newblock A global evaluation of the regional spatial variability of column
 integrated CO$_2$ distributions.
\newblock {\em J. Geophys. Res. Atmos.} {\bf 2008}, {\em
 113}, doi:10.1029/2007JD009693.

\bibitem[Engelen \em{et~al.}(2009)Engelen, Serrar, and
 Chevallier]{Engelen_2009}
Engelen, R.J.; Serrar, S.; Chevallier, F.
\newblock Four-dimensional data assimilation of atmospheric CO$_2$ using AIRS
 observations.
\newblock {\em J. Geophys. Res. Atmos.} {\bf 2009}, {\em
 114}, doi:10.1029/2008JD010739.

\bibitem[Cressie(1993)]{Cressie_1993}
Cressie, N.
\newblock {\em Statistics for Spatial Data}; John Wiley
 \& Sons: New York, NY, USA, 1993.

\bibitem[Eldering \em{et~al.}(2017)Eldering, O'Dell, Wennberg, Crisp, Gunson,
 Viatte, Avis, Braverman, Castano, Chang, Chapsky, Cheng, Connor, Dang, Doran,
 Fisher, Frankenberg, Fu, Granat, Hobbs, Lee, Mandrake, McDuffie, Miller,
 Myers, Natraj, O'Brien, Osterman, Oyafuso, Payne, Pollock, Polonsky, Roehl,
 Rosenberg, Schwandner, Smyth, Tang, Taylor, To, Wunch, and
 Yoshimizu]{Eldering_2017}
Eldering, A.; O'Dell, C.W.; Wennberg, P.O.; Crisp, D.; Gunson, M.R.; Viatte,
 C.; Avis, C.; Braverman, A.; Castano, R.; Chang, A.; et al.
\newblock The Orbiting Carbon Observatory-2: First 18~months of science data
 products.
\newblock {\em Atmos. Meas. Tech.} {\bf 2017}, {\em
 10},~549--563.

\bibitem[{OCO-2 Science Team} \em{et~al.}(2015){OCO-2 Science Team}, Gunson,
 and Eldering]{OCO2v7r}
{OCO-2 Science Team}; Gunson, M.; Eldering, A.
\newblock {{OCO-2 Level 2 Bias-Corrected XCO2 and Other Select Fields from the
 Full-Physics Retrieval Aggregated as Daily Files, Retrospective Processing V7r}}; Goddard Earth
 Sciences Data and Information Services Center (GES DISC): Greenbelt, MD, USA, 2015.
\newblock Available online: \mbox{\url{https://disc.gsfc.nasa.gov/datasets/OCO2\_L2\_Lite\_FP\_V7r/summary}} (accessed on \blue{20 January 2018}). 

\bibitem[{OCO-2 Science Team} \em{et~al.}(2017){OCO-2 Science Team}, Gunson,
 and Eldering]{OCO2v8r}
{OCO-2 Science Team}; Gunson, M.; Eldering, A.
\newblock {{OCO-2 Level 2 Bias-Corrected XCO2 and Other Select Fields from the
 Full-Physics Retrieval Aggregated as Daily Files, Retrospective Processing V8r}}; Goddard Earth
 Sciences Data and Information Services Center (GES DISC): Greenbelt, MD, USA, 2017.
\newblock Available online: \mbox{\url{https://disc.gsfc.nasa.gov/datasets/OCO2\_L2\_Lite\_FP\_V8r/summary}} (accessed on \blue{20 January 2018}). 


\bibitem[Wunch \em{et~al.}(2011)Wunch, Toon, Blavier, Washenfelder, Notholt,
 Connor, Griffith, Sherlock, and Wennberg]{Wunch_2011_TCCON}
Wunch, D.; Toon, G.C.; Blavier, J.F.L.; Washenfelder, R.A.; Notholt, J.;
 Connor, B.J.; Griffith, D.W.; Sherlock,~V.; Wennberg, P.O.
\newblock The Total Carbon Column Observing Network.
\newblock {\em Philos. Trans. R. Soc. Lond. A
 Math. Phys. Eng. Sci.} {\bf 2011}, {\em
 369},~2087--2112.

\bibitem[Wunch \em{et~al.}(2017)Wunch, Toon, Sherlock, Deutscher, Liu, Feist,
 and Wennberg]{Wunch_2017_TCCON}
Wunch, D.; Toon, G.C.; Sherlock, V.; Deutscher, N.M.; Liu, C.; Feist, D.G.;
 Wennberg, P.O.
\newblock Documentation for the 2014 TCCON Data Release. 2017.
\newblock \url{http://dx.doi.org/10.14291/tccon.ggg2014.documentation.r0/1221662} (accessed on \blue{20 January 2018}). 

\bibitem[Cressie and Wikle(2011)]{Cressie_2011}
Cressie, N.; Wikle, C.K.
\newblock {\em Statistics for Spatio-Temporal Data}; John Wiley and Sons:
 Hoboken, NJ, USA, 2011.

\bibitem[Zhang \em{et~al.}(2008)Zhang, Craigmile, and Cressie]{Zhang_2008}
Zhang, J.; Craigmile, P.F.; Cressie, N.
\newblock Loss function approaches to predict a spatial quantile and its
 exceedance region.
\newblock {\em Technometrics} {\bf 2008}, {\em 50},~216--227.

\bibitem[Gneiting and Katzfuss(2014)]{Gneiting_2014}
Gneiting, T.; Katzfuss, M.
\newblock Probabilistic forecasting.
\newblock {\em Annu. Rev. Stat. Its Appl.} {\bf 2014},
 {\em 1},~125--151.

\bibitem[Aldworth and Cressie(2003)]{Aldworth_2003}
Aldworth, J.; Cressie, N.
\newblock Prediction of nonlinear spatial functionals.
\newblock {\em J. Stat. Plan. Inference} {\bf 2003}, {\em
 112},~3--41.

\bibitem[Johnson \em{et~al.}(1994)Johnson, Kotz, and
 Balakrishnan]{Johnson_1994}
Johnson, N.L.; Kotz, S.; Balakrishnan, N.
\newblock {\em Continuous Univariate Distributions, Volume 1}, 2nd ed.; Wiley:~New~York, NY, USA, 1994.

\bibitem[Wikle(2010)]{Wikle_2010}
Wikle, C.K.
\newblock Low-rank representations for spatial processes. In {\em Handbook of
 Spatial Statistics}; Gelfand, A.E., Diggle, P., Guttorp, P., Fuentes, M.,
 Eds.; Chapman and Hall/CRC: Boca Raton, FL, USA, 2010; pp. 107--118.

\bibitem[Zammit-Mangion and Cressie(2017)]{Zammit_2017}
Zammit-Mangion, A.; Cressie, N.
\newblock {FRK: An R} package for spatial and spatio-temporal prediction with
 large datasets.
\newblock {\em arXiv} {\bf 2017}, arXiv:1705.08105.

\bibitem[Lindgren \em{et~al.}(2011)Lindgren, Rue, and
 Lindstr{\"o}m]{Lindgren_2011}
Lindgren, F.; Rue, H.; Lindstr{\"o}m, J.
\newblock An explicit link between {G}aussian fields and {Gaussian Markov
 random fields: The stochastic partial} differential equation approach.
\newblock {\em J. R. Stat. Soc. Ser. B} {\bf 2011},
 {\em 73},~423--498.

\bibitem[Nychka \em{et~al.}(2015)Nychka, Bandyopadhyay, Hammerling, Lindgren,
 and Sain]{Nychka_2015}
Nychka, D.; Bandyopadhyay, S.; Hammerling, D.; Lindgren, F.; Sain, S.
\newblock A multiresolution Gaussian process model for the analysis of large
 spatial datasets.
\newblock {\em J. Comput. Graph. Stat.} {\bf 2015},
 {\em 24},~579--599.

\bibitem[Stein(2014)]{Stein_2014}
Stein, M.L.
\newblock Limitations on low rank approximations for covariance matrices of
 spatial data.
\newblock {\em Spat. Stat.} {\bf 2014}, {\em 8},~1--19.

\bibitem[Ma and Kang(2017)]{Ma_2017}
Ma, P.; Kang, E.L.
\newblock Fused {G}aussian process for very large spatial data.
\newblock {\em arXiv} {\bf 2017}, arXiv:1702.08797.

\bibitem[Zammit-Mangion \em{et~al.}(2012)Zammit-Mangion, Sanguinetti, and
 Kadirkamanathan]{Zammit_2012}
Zammit-Mangion, A.; Sanguinetti, G.; Kadirkamanathan, V.
\newblock Variational estimation in spatiotemporal systems from continuous and
 point-process observations.
\newblock {\em IEEE Trans. Signal Process.} {\bf 2012}, {\em
 60},~3449--3459.

\bibitem[Katzfuss(2017)]{Katzfuss_2017}
Katzfuss, M.
\newblock A multi-resolution approximation for massive spatial datasets.
\newblock {\em J. Am. Stat. Assoc.} {\bf 2017},
 {\em 112},~201--214.

\bibitem[Wikle and Cressie(1999)]{Wikle_1999}
Wikle, C.K.; Cressie, N.
\newblock A dimension-reduced approach to space-time Kalman filtering.
\newblock {\em Biometrika} {\bf 1999}, {\em 86},~815--829.

\bibitem[Stroud \em{et~al.}(2001)Stroud, M{\"u}ller, and
 Sans{\'o}]{Stroud_2001}
Stroud, J.R.; M{\"u}ller, P.; Sans{\'o}, B.
\newblock Dynamic models for spatiotemporal data.
\newblock {\em J. R. Stat. Soc. Ser. B} {\bf 2001},
 {\em 63},~673--689.

\bibitem[Watanabe \em{et~al.}(2015)Watanabe, Hayashi, Saeki, Maksyutov, Nasuno,
 Shimono, Hirose, Takaichi, Kanekon, Ajiro, Matsumoto, and
 Yokota]{Watanabe_2015}
Watanabe, H.; Hayashi, K.; Saeki, T.; Maksyutov, S.; Nasuno, I.; Shimono, Y.;
 Hirose, Y.; Takaichi, K.; Kanekon,~S.; Ajiro, M.; et al.
\newblock Global mapping of greenhouse gases retrieved from GOSAT Level 2
 products by using a kriging method.
\newblock {\em Int. J. Remote Sens.} {\bf 2015}, {\em
 36},~1509--1528.

\bibitem[Wunch \em{et~al.}(2017)Wunch, Wennberg, Osterman, Fisher, Naylor,
 Roehl, O'Dell, Mandrake, Viatte, Kiel, Griffith, Deutscher, Velazco, Notholt,
 Warneke, Petri, De~Maziere, Sha, Sussmann, Rettinger, Pollard, Robinson,
 Morino, Uchino, Hase, Blumenstock, Feist, Arnold, Strong, Mendonca, Kivi,
 Heikkinen, Iraci, Podolske, Hillyard, Kawakami, Dubey, Parker, Sepulveda,
 Garc\'{\i}a, Te, Jeseck, Gunson, Crisp, and Eldering]{Wunch_2017}
Wunch, D.; Wennberg, P.O.; Osterman, G.; Fisher, B.; Naylor, B.; Roehl, C.M.;
 O'Dell, C.; Mandrake, L.; Viatte,~C.; Kiel, M.; et al.
\newblock Comparisons of the {Orbiting Carbon Observatory-2 (OCO-2) $XCO_2$
 measurements with TCCON}.
\newblock {\em Atmos. Meas. Tech.} {\bf 2017}, {\em
 10},~2209--2238.

\bibitem[Hobbs \em{et~al.}(2017)Hobbs, Braverman, Cressie, Granat, and
 Gunson]{Hobbs_2017}
Hobbs, J.; Braverman, A.; Cressie, N.; Granat, R.; Gunson, M.
\newblock Simulation-based uncertainty quantification for estimating
 atmospheric CO$_2$ from satellite data.
\newblock {\em SIAM/ASA J. Uncertain. Quantif.} {\bf 2017}, {\em
 5},~956--985.

\bibitem[Nguyen \em{et~al.}(2017)Nguyen, Cressie, and Braverman]{Nguyen_2017}
Nguyen, H.; Cressie, N.; Braverman, A.
\newblock Multivariate spatial data fusion for very large remote sensing
 datasets.
\newblock {\em Remote Sens.} {\bf 2017}, {\em 9},~142, doi:10.3390/rs9020142.

\bibitem[Sherlock \em{et~al.}(2014)Sherlock, Connor, Robinson, Shiona, Smale,
 and Pollard]{Sherlock2014ll}
Sherlock, V.; Connor, B.; Robinson, J.; Shiona, H.; Smale, D.; Pollard, D.
\newblock {{TCCON Data from Lauder, New~Zealand, 125HR, Release GGG2014R0}}; Carbon Dioxide Information Analysis Center, Oak Ridge National Laboratory: Oak Ridge, TN, USA, 2014.
\newblock Available online: \url{http://dx.doi.org/10.14291/tccon.ggg2014.lauder02.R0/1149298} (accessed on \blue{20 January 2018}). 

\bibitem[Griffith \em{et~al.}(2014)Griffith, Velazco, Deutscher, Murphy, Jones,
 Wilson, Macatangay, Kettlewell, Buchholz, and Riggenbach]{Griffith2014wg}
Griffith, D.W.T.; Velazco, V.A.; Deutscher, N.; Murphy, C.; Jones, N.; Wilson,
 S.; Macatangay, R.; Kettlewell,~G.; Buchholz, R.R.; Riggenbach, M.
\newblock {{TCCON Data from Wollongong, Australia, Release GGG2014R0}}; Carbon Dioxide Information
 Analysis Center, Oak Ridge National Laboratory: Oak Ridge, TN, USA, 2014.
\newblock Available online: \url{http://dx.doi.org/10.14291/tccon.ggg2014.wollongong01.R0/1149291} (accessed on \blue{20 January 2018}). 

\bibitem[{De Maziere} \em{et~al.}(2014){De Maziere}, Sha, Desmet, Hermans,
 Scolas, Kumps, Metzger, Duflot, and Cammas]{DeMaziere2014ra}
{De Maziere}, M.; Sha, M.K.; Desmet, F.; Hermans, C.; Scolas, F.; Kumps, N.;
 Metzger, J.M.; Duflot, V.; Cammas, J.P.
\newblock {{TCCON Data from Reunion Island (La Reunion), France, Release
 GGG2014R0}}; Carbon Dioxide Information
 Analysis Center, Oak Ridge National Laboratory: Oak Ridge, TN, USA, 2014.
\newblock Available online: \url{http://dx.doi.org/10.14291/tccon.ggg2014.reunion01.R0/1149288} (accessed on \blue{20 January 2018}). 

\bibitem[Griffith \em{et~al.}(2014)Griffith, Deutscher, Velazco, Wennberg,
 Yavin, Aleks, Washenfelder, Toon, Blavier, Murphy, Jones, Kettlewell, Connor,
 Macatangay, Roehl, Ryczek, Glowacki, Culgan, and Bryant]{Griffith2014db}
Griffith, D.W.T.; Deutscher, N.; Velazco, V.A.; Wennberg, P.O.; Yavin, Y.;
 Aleks, G.K.; Washenfelder, R.; Toon,~G.C.; Blavier, J.F.; Murphy, C.; et al.
\newblock {{TCCON Data from Darwin, Australia, Release GGG2014R0}}; Carbon~Dioxide Information
 Analysis Center, Oak Ridge National Laboratory: Oak Ridge, TN, USA, 2014.
\newblock Available online: \url{http://dx.doi.org/10.14291/tccon.ggg2014.darwin01.R0/1149290} (accessed on \blue{20 January 2018}). 

\bibitem[Feist \em{et~al.}(2014)Feist, Arnold, John, and Geibel]{Feist2014ae}
Feist, D.G.; Arnold, S.G.; John, N.; Geibel, M.C.
\newblock {{TCCON Data from Ascension Island, Saint Helena, Ascension~and
 Tristan da Cunha, Release GGG2014R0}}; Carbon Dioxide Information
 Analysis Center, Oak Ridge National Laboratory: Oak Ridge, TN, USA, 2014.
\newblock Available online: \url{http://dx.doi.org/10.14291/tccon.ggg2014.ascension01.R0/1149285} (accessed on \blue{20 January 2018}). 

\bibitem[Dubey \em{et~al.}(2014)Dubey, Henderson, Green, Butterfield,
 Keppel-Aleks, Allen, Blavier, Roehl, Wunch, and Lindenmaier]{Dubey2014ma}
Dubey, M.; Henderson, B.; Green, D.; Butterfield, Z.; Keppel-Aleks, G.; Allen,
 N.; Blavier, J.F.; Roehl,~C.; Wunch, D.; Lindenmaier, R.
\newblock {{TCCON Data from Manaus, Brazil, Release GGG2014R0}}; Carbon Dioxide Information
 Analysis Center, Oak Ridge National Laboratory: Oak Ridge, TN, USA, 2014.
\newblock Available~online: \url{http://dx.doi.org/10.14291/tccon.ggg2014.manaus01.R0/1149274} (accessed on \blue{20 January 2018}). 

\bibitem[Blumenstock \em{et~al.}(2014)Blumenstock, Hase, Schneider, Garcia, and
 Sepulveda]{Blumenstock2014iz}
Blumenstock, T.; Hase, F.; Schneider, M.; Garcia, O.; Sepulveda, E.
\newblock {{TCCON Data from Izana, Tenerife, Spain, Release GGG2014R0}}; Carbon Dioxide Information
 Analysis Center, Oak Ridge National Laboratory: Oak~Ridge, TN, USA, 2014.
\newblock Available online: \url{http://dx.doi.org/10.14291/tccon.ggg2014.izana01.R0/1149295} (accessed on \blue{20 January 2018}). 

\bibitem[Kawakami \em{et~al.}(2014)Kawakami, Ohyama, Arai, Okumura, Taura,
 Fukamachi, and Sakashita]{Kawakami2014js}
Kawakami, S.; Ohyama, H.; Arai, K.; Okumura, H.; Taura, C.; Fukamachi, T.;
 Sakashita, M.
\newblock {{TCCON Data from Saga, Japan, Release GGG2014R0}}; Carbon Dioxide Information
 Analysis Center, Oak Ridge National Laboratory: Oak Ridge, TN, USA, 2014.
\newblock Available online: \url{http://dx.doi.org/10.14291/tccon.ggg2014.saga01.R0/1149283} (accessed on \blue{20 January 2018}). 


\bibitem[Wennberg \em{et~al.}(2014)Wennberg, Wunch, Roehl, Blavier, Toon, and
 Allen]{Wennberg2014ci}
Wennberg, P.O.; Wunch, D.; Roehl, C.; Blavier, J.F.L.; Toon, G.C.; Allen, N.
\newblock {{TCCON Data from California Institute of Technology, Pasadena,
 California, USA, Release GGG2014R1}}; Carbon Dioxide Information
 Analysis Center, Oak Ridge National Laboratory: Oak Ridge, TN, USA, 2014.
\newblock Available online: \url{http://dx.doi.org/10.14291/tccon.ggg2014.pasadena01.R1/1182415} (accessed on \blue{20 January 2018}). 
\newpage
\bibitem[Iraci \em{et~al.}(2014)Iraci, Podolske, Hillyard, Roehl, Wennberg,
 Blavier, Landeros, Allen, Wunch, Zavaleta, Quigley, Osterman, Albertson,
 Dunwoody, and Boyden]{Iraci2014df}
Iraci, L.; Podolske, J.; Hillyard, P.; Roehl, C.; Wennberg, P.O.; Blavier,
 J.F.; Landeros, J.; Allen, N.; Wunch, D.; Zavaleta, J.; et al.
\newblock {{TCCON Data from Armstrong Flight Research Center, Edwards, CA, USA,
 Release GGG2014R0}}; Carbon Dioxide Information
 Analysis Center, Oak Ridge National Laboratory: Oak Ridge, TN, USA, 2014.
\newblock Available online: \url{http://dx.doi.org/10.14291/tccon.ggg2014.edwards01.R0/1149289} (accessed on \blue{20 January 2018}). 

\bibitem[Morino \em{et~al.}(2014)Morino, Matsuzaki, and Shishime]{Morino2014tk}
Morino, I.; Matsuzaki, T.; Shishime, A.
\newblock {{TCCON Data from Tsukuba, Ibaraki, Japan, 125HR, Release GGG2014R1}}; Carbon Dioxide Information
 Analysis Center, Oak Ridge National Laboratory: Oak Ridge, TN, USA, 2014.
\newblock Available online: \url{http://dx.doi.org/10.14291/tccon.ggg2014.tsukuba02.R1/1241486} (accessed on \blue{20 January 2018}). 

\bibitem[Goo \em{et~al.}(2014)Goo, Oh, and Velazco]{Goo2014ay}
Goo, T.Y.; Oh, Y.S.; Velazco, V.A.
\newblock {{TCCON Data from Anmeyondo, South Korea, Release GGG2014R0}}; Carbon~Dioxide Information
 Analysis Center, Oak Ridge National Laboratory: Oak Ridge, TN, USA, 2014.
\newblock Available online: \url{http://dx.doi.org/10.14291/tccon.ggg2014.anmeyondo01.R0/1149284} (accessed on \blue{20 January 2018}). 

\bibitem[Wennberg \em{et~al.}(2014)Wennberg, Wunch, Roehl, Blavier, Toon,
 Allen, Dowell, Teske, Martin, and Martin.]{Wennberg2014oc}
Wennberg, P.O.; Wunch, D.; Roehl, C.; Blavier, J.F.; Toon, G.C.; Allen, N.;
 Dowell, P.; Teske, K.; Martin, C.; Martin., J.
\newblock {{TCCON Data from Lamont, Oklahoma, USA, Release GGG2014R0}}; Carbon Dioxide Information
 Analysis Center, Oak Ridge National Laboratory: Oak Ridge, TN, USA, 2014.
\newblock Available~online:\mbox{ \url{http://dx.doi.org/10.14291/tccon.ggg2014.lamont01.R0/1149159}} (accessed on \blue{20 January 2018}). 

\bibitem[Morino \em{et~al.}(2014)Morino, Yokozeki, Matzuzaki, and
 Shishime]{Morino2014rj}
Morino, I.; Yokozeki, N.; Matzuzaki, T.; Shishime, A.
\newblock {{TCCON Data from Rikubetsu, Hokkaido, Japan, Release~GGG2014R1}}; Carbon Dioxide Information
 Analysis Center, Oak Ridge National Laboratory: Oak~Ridge, TN, USA, 2014.
\newblock Available online: \url{http://dx.doi.org/10.14291/tccon.ggg2014.rikubetsu01.R1/1242265} (accessed on \blue{20 January 2018}). 

\bibitem[Wennberg \em{et~al.}(2014)Wennberg, Roehl, Wunch, Toon, Blavier,
 Washenfelder, Keppel-Aleks, Allen, and Ayers]{Wennberg2014pa}
Wennberg, P.O.; Roehl, C.; Wunch, D.; Toon, G.C.; Blavier, J.F.; Washenfelder,
 R.; Keppel-Aleks, G.; Allen,~N.; Ayers, J.
\newblock {{TCCON Data from Park Falls, Wisconsin, USA, Release GGG2014R0}}; Carbon Dioxide Information
 Analysis Center, Oak Ridge National Laboratory: Oak Ridge, TN, USA, 2014.
\newblock Available online: \mbox{\url{http://dx.doi.org/10.14291/tccon.ggg2014.parkfalls01.R0/1149161}} (accessed on \blue{20 January 2018}). 

\bibitem[Sussmann and Rettinger(2014)]{Sussmann2014gm}
Sussmann, R.; Rettinger, M.
\newblock {{TCCON Data from Garmisch, Germany, Release GGG2014R0}}; Carbon Dioxide Information
 Analysis Center, Oak Ridge National Laboratory: Oak Ridge, TN, USA, 2014.
\newblock Available online: \url{http://dx.doi.org/10.14291/tccon.ggg2014.garmisch01.R0/1149299 } (accessed on \blue{20 January 2018}). 

\bibitem[Warneke \em{et~al.}(2014)Warneke, Messerschmidt, Notholt, Weinzierl,
 Deutscher, Petri, Grupe, Vuillemin, Truong, Schmidt, Ramonet, and
 Parmentier]{Warneke2014or}
Warneke, T.; Messerschmidt, J.; Notholt, J.; Weinzierl, C.; Deutscher, N.;
 Petri, C.; Grupe, P.; Vuillemin,~C.; Truong, F.; Schmidt, M.; et al.
\newblock {{TCCON Data from Orleans, France, Release GGG2014R0}}; Carbon Dioxide Information
 Analysis Center, Oak Ridge National Laboratory: Oak Ridge, TN, USA, 2014.
\newblock Available online: \url{http://dx.doi.org/10.14291/tccon.ggg2014.orleans01.R0/1149276} (accessed on \blue{20 January 2018}). 

\bibitem[Te \em{et~al.}(2014)Te, Jeseck, and Janssen]{Te2014pr}
Te, Y.; Jeseck, P.; Janssen, C.
\newblock {{TCCON Data from Paris, France, Release GGG2014R0}}; Carbon Dioxide Information
 Analysis Center, Oak Ridge National Laboratory: Oak Ridge, TN, USA, 2014.
\newblock Available~online: \mbox{\url{http://dx.doi.org/10.14291/tccon.ggg2014.paris01.R0/1149279}} (accessed on \blue{20 January 2018}). 

\bibitem[Hase \em{et~al.}(2014)Hase, Blumenstock, Dohe, Gross, and
 Kiel]{Hase2014ka}
Hase, F.; Blumenstock, T.; Dohe, S.; Gross, J.; Kiel, M.
\newblock {{TCCON Data from Karlsruhe, Germany, Release~GGG2014R1}}; Carbon Dioxide Information
 Analysis Center, Oak Ridge National Laboratory: Oak~Ridge, TN, USA, 2014.
\newblock Available online: \url{http://dx.doi.org/10.14291/tccon.ggg2014.karlsruhe01.R1/1182416} (accessed on \blue{20 January 2018}). 

\bibitem[Notholt \em{et~al.}(2014)Notholt, Petri, Warneke, Deutscher,
 Buschmann, Weinzierl, Macatangay, and Grupe]{Notholt2014br}
Notholt, J.; Petri, C.; Warneke, T.; Deutscher, N.; Buschmann, M.; Weinzierl,
 C.; Macatangay, R.; Grupe,~P.
\newblock {{TCCON Data from Bremen, Germany, Release GGG2014R0}}; Carbon Dioxide Information
 Analysis Center, Oak~Ridge National Laboratory: Oak Ridge, TN, USA, 2014.
\newblock Available online: \url{http://dx.doi.org/10.14291/tccon.ggg2014.bremen01.R0/1149275} (accessed on \blue{20 January 2018}). 

\bibitem[Deutscher \em{et~al.}(2014)Deutscher, Notholt, Messerschmidt,
 Weinzierl, Warneke, Petri, Grupe, and Katrynski]{Deutscher2014bi}
Deutscher, N.; Notholt, J.; Messerschmidt, J.; Weinzierl, C.; Warneke, T.;
 Petri, C.; Grupe, P.; Katrynski,~K.
\newblock {{TCCON Data from Bialystok, Poland, Release GGG2014R1}}; Carbon Dioxide Information
 Analysis Center, Oak~Ridge National Laboratory: Oak Ridge, TN, USA, 2014.
\newblock Available online: \url{http://dx.doi.org/10.14291/tccon.ggg2014.bialystok01.R1/1183984} (accessed on \blue{20 January 2018}). 

\bibitem[Kivi \em{et~al.}(2014)Kivi, Heikkinen, and Kyro]{Kivi2014so}
Kivi, R.; Heikkinen, P.; Kyro, E.
\newblock {{TCCON Data from \blue{Sodankyl{\"a}}, Finland, Release GGG2014R0}}; Carbon Dioxide Information
 Analysis Center, Oak Ridge National Laboratory: Oak Ridge, TN, USA, 2014.
\newblock Available online: \url{http://dx.doi.org/10.14291/tccon.ggg2014.sodankyla01.R0/1149280}
(accessed on \blue{20 January 2018}). 

\newpage

\bibitem[Notholt \em{et~al.}(2014)Notholt, Warneke, Petri, Deutscher, 
 Weinzierl, Palm, and Buschmann]{Notholt2014ny}
Notholt, J.; Warneke, T.; Petri, C.; Deutscher, N.M.; Weinzierl, C.; Palm, M.;
 Buschmann, M.
\newblock {{TCCON Data from Ny {\AA}lesund, Norway, Release GGG2014R0}}; Carbon Dioxide Information
 Analysis Center, Oak Ridge National Laboratory: Oak Ridge, TN, USA, 2014.
\newblock Available online: \url{http://dx.doi.org/10.14291/tccon.ggg2014.nyalesund01.R0/1149278}
(accessed on \blue{20 January 2018}). 

\bibitem[Strong \em{et~al.}(2014)Strong, Mendonca, Weaver, Fogal, Drummond,
 Batchelor, and Lindenmaier]{Strong2014eu}
Strong, K.; Mendonca, J.; Weaver, D.; Fogal, P.; Drummond, J.; Batchelor, R.;
 Lindenmaier, R.
\newblock {{TCCON Data from Eureka, Canada, Release GGG2014R0}}; Carbon Dioxide Information
 Analysis Center, Oak Ridge National Laboratory: Oak Ridge, TN, USA, 2014.
\newblock Available online: \url{http://dx.doi.org/10.14291/tccon.ggg2014.eureka01.R0/1149271}
(accessed on \blue{20 January 2018}). 

\bibitem[Liang \em{et~al.}(2017)Liang, Gong, Han, and Xiang]{Liang_2017}
Liang, A.; Gong, W.; Han, G.; Xiang, C.
\newblock Comparison of satellite-observed {XCO$_2$ from GOSAT, OCO-2, and~ground-based TCCON}.
\newblock {\em Remote Sens.} {\bf 2017}, {\em 9},~1033, doi:10.3390/rs9101033.

\bibitem[{R Core Team}(2017)]{R}
{R Core Team}.
\newblock {\em R: A Language and Environment for Statistical Computing};
\newblock R Foundation for Statistical Computing: Vienna, Austria, 2017.

\bibitem[Banerjee \em{et~al.}(2008)Banerjee, Gelfand, Finley, and
 Sang]{Banerjee_2008}
Banerjee, S.; Gelfand, A.E.; Finley, A.O.; Sang, H.
\newblock Gaussian predictive process models for large spatial data sets.
\newblock {\em J. R. Stat. Soc. Ser. B} {\bf 2008},
 {\em 70},~825--848.

\bibitem[Rasmussen and Williams(2006)]{Rasmussen_2006}
Rasmussen, C.E.; Williams, C.K.I.
\newblock {\em {Gaussian Processes for Machine Learning}}; The MIT Press:
 Cambridge, MA, USA, 2006.

\end{thebibliography}

\end{document}